\documentclass[11pt]{article}
\usepackage[margin=1in]{geometry}
\usepackage{lmodern}

\usepackage{amssymb}
\usepackage{amsmath}
\usepackage{amsthm}
\usepackage{bm}
\usepackage{bbm}
\usepackage{mathrsfs}
\usepackage{longtable}
\usepackage{multirow}
\usepackage{graphicx}
\usepackage{authblk}
\usepackage[authoryear]{natbib}
\setcitestyle{aysep={}} 
\usepackage[pdfpagemode=UseNone,bookmarksopen=false,colorlinks=true]{hyperref}
\hypersetup{linkcolor=blue, citecolor=blue,
  filecolor=blue, urlcolor= black}

\newtheorem{theorem}{Theorem}[section]
\newtheorem{corollary}{Corollary}[section]
\newtheorem{result}{Result}[section]
\newtheorem{lemma}{Lemma}
\newtheorem{assumption}{Assumption}
\newtheorem{definition}{Definition}
\newtheorem{remark}{Remark}
\newtheorem{example}{Example}

\begin{document}

\title{The Elephant in the Cluster: Robust Inference with Dominant Groups}

\author[a]{Subhodeep Dey\thanks{Corresponding author. \textit{Email address:} \href{mailto:subhodeepdey920@gmail.com}{\texttt{subhodeepdey920@gmail.com}}}}
\author[a]{Gopal K Basak}
\author[b]{Samarjit Das}

\affil[a]{Theoretical Statistics and Mathematics Unit, Indian Statistical Institute, 203 B.T. Road, Kolkata 700108, India}
\affil[b]{Economic Research Unit, Indian Statistical Institute, 203 B.T. Road, Kolkata 700108, India}

\date{}
\maketitle

\begin{abstract}
We re-examine the asymptotic properties of the traditional pooled ordinary least squares (POLS) estimator under different within-cluster dependence regimes (strong, semi-strong and weak). This study explores several scenarios characterized by different restrictions on cluster sizes and the number of clusters, showing that POLS can be inconsistent in many realistic contexts with strong within-cluster dependence. To address this issue, we propose a simple estimator based on data averaging. We investigate the asymptotic behavior of the proposed estimator, demonstrating its consistency even in settings where POLS fails. Furthermore, we show that the proposed estimator is more efficient than POLS in many practical applications. As an advantage of its averaging structure, the proposed estimator also retains consistency under classical measurement error, thus avoiding reliance on instrumental variables (IV). A comprehensive simulation study illustrates the efficacy of the proposed estimator, and empirical findings suggest that it delivers improved goodness of fit.
\end{abstract}



\newpage
 
\section{Introduction}
\label{sec:intro}

The least squares estimator (LSE) derived from the Gauss–Markov (GM) regression model is known to be the best linear unbiased estimator (BLUE). When the assumption of normality is additionally satisfied, the LSE becomes the most efficient estimator, attaining the Rao–Cramér lower bound. However, violations of any GM assumptions may result in inefficiency or even inconsistency of the estimator. In particular, error dependence can cause the LSE to become inconsistent, and the central limit theorem (CLT) may no longer hold, complicating hypothesis testing for regression parameters. However, when data are structured in groups or clusters, valid inference may still be attainable using robust standard errors. \cite{moulton1986random, moulton1990illustration} provide several canonical examples illustrating how clustering naturally arises in empirical research. It often occurs when the sampling design involves selecting random groups, such as schools, households, or towns, and surveying all units within each selected group. Clustering may also result from cost-reduction strategies in data collection, such as interviewing multiple households located on the same block of the city. Furthermore, even under a simple random sampling scheme, dependence can emerge due to unobserved effects shared across units, for instance, a latent factor common to all households within a given state.

A plethora of literature has been developed to facilitate robust inference through appropriate correction of standard errors \citep[see][]{wooldridge2003cluster, cameron2008bootstrap, cameron2015practitioner, abadie2020sampling, canay2021wild, abadie2023should, chiang2023genuinely, mackinnon2023cluster}. \citet{cameron2011robust} and \citet{mackinnon2021wild} extend cluster-robust and wild bootstrap inference, respectively, to non-nested multiway clustering; \citet{chiang2022multiway} further extend multiway cluster-robust inference to the double/debiased machine learning framework; \citet{tabord2019inference} establishes analogous asymptotic theory for dyadic (pairwise-indexed) data. The broader literature studies various scenarios for cluster size ($N_g,$ say) and the number of clusters ($G,$ say). In practice, three different asymptotic regimes may arise: (1) $G$ is finite, and $N_g \rightarrow \infty $, (2)  $N_g \rightarrow \infty, $ and  $ G \rightarrow \infty,$ (3) $N_g$ is finite, but $ G \rightarrow \infty.$
Scenario (1) typically leads to inconsistency of the LSE and invalidity of the central limit theorem. \footnote{\cite{andrews2005cross} demonstrates that the LSE is inconsistent when sample size grows with a fixed number of clusters.} \footnote{\cite{bester2011inference} study asymptotic properties under fixed $G$ and large $N_g$, invoking a “limited memory” assumption under which dependence vanishes with distance, resulting in sparse error covariance matrices.} 
\footnote{\citet{ibragimov2010t, ibragimov2016inference} propose inference strategies for small numbers of heterogeneous clusters, requiring the cluster-level parameter estimators  to be Gaussian with possibly heterogeneous variances.}
The same inferential challenges arise when both $N_g\to\infty$ and $G\to\infty$, without additional restrictions on their relative magnitudes. Scenario (3), where the cluster size is fixed and the number of clusters increases, has been extensively studied in the above references.

However, the literature remains limited for scenario (2), with some notable exceptions. Both \cite{hansen2019asymptotic} and \cite{djogbenou2019asymptotic} rely on non-dominance conditions that assume a degree of homogeneity in cluster sizes. Consequently, their frameworks exclude settings where a few clusters are disproportionately large relative to the rest of the sample. In such heavily unbalanced contexts, the pooled ordinary least squares (POLS) estimator can perform poorly. For instance, \cite{djogbenou2019asymptotic} demonstrate that standard asymptotic approximations break down, leading to severe inferential distortions, when a single cluster is dominant, having half of the total observations. \cite{kojevnikov2023some} demonstrate that under error normality conditions, the variance of the POLS estimator cannot be consistently estimated in the presence of at least one large cluster.  The possibility of dominating cluster is also explored in \cite{sasaki2022non}; and \cite{chiang2023genuinely}. They used a different estimator and a different set of assumptions, e.g., an i.i.d. setup when $N_g$ is a random variable with finite expectation. $N_g$ is i.i.d. with finite expectation implies that $\sum_{g=1}^G N_g = O_P(G).$ This restriction rules out many real-life scenarios.

As an illustration of disproportionately large clusters, studies examining the effects of state laws on corporate governance in the United States often cluster at the state level, where Delaware alone accounts for nearly half of all incorporated firms \citep{spamann2019inference}. To illustrate further, we consider India's Annual Survey of Industries (ASI) for the years 2017-18 and 2019-20, where 4-digit NIC codes serve as natural cluster identifiers. In highly industrialized states like Maharashtra and moderately industrialized ones like West Bengal, a few clusters contain over 500 units while dozens of others contain fewer than 25. Even in less industrialized states like Tripura, a single cluster can account for over 90\% of the observations. These examples illustrate instances where a finite number of clusters are disproportionately large compared to the others.

In this paper, we propose a simple averaging-based estimator, which may be termed as the between estimator. We establish that the proposed estimator remains consistent under both asymptotic regimes: possibility (2), where $N_g\to\infty,$ $G\to\infty$; and possibility (3), where $N_g$ is fixed and $G\to\infty.$ In particular, consistency is ensured under any kinds of within-cluster dependence as long as $G \to \infty$, irrespective of cluster sizes including dominant clusters. The asymptotic normality, consistency of the proposed standard error, and the validity of the Wald-type test based on the proposed estimator are also established under the different degrees of within-cluster dependence.

We further derive the conditions under which the POLS estimator will be inconsistent. We study the asymptotic normality results and the validity of the Wald-type test for the POLS estimator under different dependence regimes. 

We also establish that the proposed estimator offers several advantages over the standard POLS estimator. In particular, the proposed estimator is shown to be asymptotically more efficient in many realistic scenarios. As a by-product of averaging, the proposed estimator retains consistency in the presence of classical measurement error in the regressors, provided the cluster sizes are sufficiently large. Consequently, this approach  circumvents the need for valid instrumental variables. We further discuss realistic empirical scenarios in which the proposed estimator is more efficient than the conventional POLS estimator.

The remainder of this paper is organized as follows. Section \ref{Section:model} outlines the model and its underlying assumptions. Section \ref{Section:POLS} reviews the pooled ordinary least squares method and presents asymptotic results, highlighting the inconsistency of the estimator in the presence of dominant clusters and strong dependence. Section \ref{Section:proposed} introduces the proposed estimator along with its asymptotic properties under various forms of error dependence, including strong, semi-strong, and weak dependence. Section \ref{Section:analysis} presents a simulation study and empirical analysis using real-world data. Section \ref{Section:conclusion} concludes the paper. All formal proofs are provided in the Appendix.

\section{The Model} \label{Section:model}

Let $Y_{gi}$ be the observation on the $i^{th}$ individual of the $g^{th}$ cluster, and suppose that it is generated according to the linear model for clustered data as 
\begin{equation} \label{Eq:model}
Y_{gi}=X_{gi}^{\prime} \beta +\epsilon_{gi}, \quad i=1,2,\ldots,N_g; \; g=1,2,\ldots,G, 
\end{equation} 
where $X_{gi}$ is a $k\times1$ vector of observed regressors (including categorical variables) for the $i^{th}$ individual in the $g^{th}$ cluster, and $\epsilon_{gi}$ is the corresponding random error.
We consider one-way clustering only. Stacking observations within a cluster yields the following model.  
\begin{equation} \label{Eq:model1}
 Y_g=X_g\beta +\epsilon_g ,\quad  g=1,2,\ldots ,G,
\end{equation}
  where $Y_g$  is a $N_g\times 1$  vector, $X_g$  is a  $N_g \times k$  matrix and $\beta$  is a $k\times 1$  vector.  Note that for every $g=1,2,\ldots,G,$ the variance-covariance matrix of ${\epsilon}_g$ is defined as $\Omega_g,$  which is an $N_g\times N_g$ positive definite matrix with uniformly bounded diagonal elements and is independent of $X=\left(X_1^\prime,X_2^\prime,\ldots,X_G^\prime\right)^\prime$.\footnote{ Conditional heteroskedasticity may be allowed with some assumptions on $X.$} Clearly, $\lambda_{\min}(\Omega_g) \geq c,$ for some $0<c<\infty.$ Before presenting the main results, we establish some fundamental assumptions and make key observations.

\begin{assumption}\label{assump:exogeneity}
    We assume that $X_g$ are exogenous with respect to $\epsilon_{g}$, $\forall g,$ i.e., 
    \[
        \mathbb{E}\left[\epsilon_{g} \mid X_g \right] = 0, \quad  \forall g.
    \]
\end{assumption}
This assumption states that the error term within each cluster is mean independent of the corresponding regressors, thereby ruling out within-cluster endogeneity.
Furthermore, if the sequence $\{(X_g,\epsilon_g)\}_{g=1}^G$ is independent 
across $g$, then the above condition implies
$\mathbb{E}\left[\epsilon_g \mid X \right] = 0,$ for all $g,$ since conditioning on the full regressor array does not introduce additional information about $\epsilon_g$ 
beyond $X_g$ under cross-cluster independence.

\begin{definition} \label{def:exact order and order uniformly}
    (i)     Let $\left\{z_n\right\}$ be a sequence of real numbers and $\left\{a_n\right\}$ be a sequence of positive real numbers.
    
    (a) By $z_n=O_e\left(a_n\right),$ we mean that for some $0<c_1\leq c_2<\infty,$
    \[ c_1 \leq \left| \frac{z_n}{a_n} \right| \leq c_2, \; \text{ for all } n. \]
    
    (b)  By $z_n=O\left(a_n\right),$ we mean that for some $0<c<\infty,$
    \[
     \left| \frac{z_n}{a_n} \right| \leq c, \; \text{ for all } n.
    \]  
    
    (ii) Let $\left\{z_{N_g,g}\right\}$ be a doubly indexed sequence of real numbers and $\left\{a_{N_g,g}\right\}$ be a doubly indexed sequence of positive real numbers, where $N_g$ is the cluster size for the $g^{th}$ cluster, which may increase to $+\infty.$
    
    (a) By $z_{N_g,g}=O_e\left(a_{N_g,g}\right)$ uniformly in $g,$ we mean that for some $0<c_1\leq c_2<\infty,$
    \[ c_1 \leq \inf_{g,N_g}\;\left| \frac{z_{N_g,g}}{a_{N_g,g}} \right| \leq \sup_{g,N_g}\;\left| \frac{z_{N_g,g}}{a_{N_g,g}} \right| \leq c_2. \]

     (b) By $z_{N_g,g}=O\left(a_{N_g,g}\right)$ uniformly in $g,$ we mean that for some $0<c<\infty,$
    \[ \sup_{g,N_g}\;\left| \frac{z_{N_g,g}}{a_{N_g,g}} \right| \leq c. \]
\end{definition}

\begin{definition} \label{def:stochastically bounded uniformly}
    (i) Let $\left\{Z_n\right\}$ be a sequence of random variables and $\left\{a_n\right\}$ be a sequence of positive real numbers.
    By $Z_n=O_P\left(a_n\right),$ we mean that for every $\varepsilon > 0$, there exist $M_\varepsilon > 0$ and ${L_\varepsilon} \in \mathbb{N}$ such that 
     \[ \mathbb{P}\left(\left|\frac{Z_n}{a_n}\right| > M_\varepsilon\right) < \varepsilon, \; \text{ for all } n \geq {L_\varepsilon}. \]

    (ii)    
    Let $\left\{Z_{N_g,g}\right\}$ be a doubly indexed sequence of random variables and $\left\{a_{N_g,g}\right\}$ be a doubly indexed sequence of positive real numbers, where $N_g$ is the cluster size for the $g^{th}$ cluster, which may increase to $+\infty.$
     By $Z_{N_g,g}=O_P\left(a_{N_g,g}\right)$ uniformly in $g,$ we mean that for every $\varepsilon > 0$, there exists $M_\varepsilon > 0$ such that 
     \[ \sup_{g,N_g}\;\mathbb{P}\left(\left|\frac{Z_{N_g,g}}{a_{N_g,g}}\right| > M_\varepsilon\right) < \varepsilon. \] 
    
\end{definition}

Definition \ref{def:stochastically bounded uniformly}(i) means that for every $\varepsilon > 0$, there exists $M_\varepsilon > 0$ such that $\underset{n}{\operatorname{sup}}\;\mathbb{P}\left(\left|\frac{Z_{n}}{a_{n}}\right| > M_\varepsilon\right) < \varepsilon.$

When applying asymptotic order notation to a matrix of finite dimension, the big-$O$ operator is defined through a matrix norm. Let $\{A_n\}$ be a sequence of $k \times m$ matrices with finite and fixed dimensions $k$ and $m$, and let $\{a_n\}$ be a sequence of positive real numbers. We write $A_n = O(a_n)$ if its matrix norm satisfies $\|A_n\| = O(a_n)$. Because all matrix norms are equivalent in finite-dimensional spaces, this asymptotic bound holds irrespective of the specific choice of norm. The same logic extends to stochastic orders, where $A_n = O_p(a_n)$ implies that the matrix norm is bounded in probability at the rate of $a_n$, such that $\|A_n\| = O_p(a_n)$.

We now characterize the dependence structure of the error terms, which is fundamental to our analysis.

\begin{definition} \label{def:dependency rates}
We define three possible kinds of cross-sectional dependence. Let us denote $\Omega_g=\mathbb{E}\left[\epsilon_g \epsilon_g^\prime \big | X_g\right]$ for a fixed cluster $g,$ and the $g^{th}$ cluster contains $N_g$ observations.
\begin{enumerate}
    \item \textbf{ Strong Dependence:} Dependence across the $N_g$ individuals is said to be strong, when $\lambda_{\max}(\Omega_{g})=O_e(N_{g}).$ 
    
    \item \textbf{ Semi-strong Dependence:} Dependence across the $N_g$ individuals is said to be semi-strong or moderate, when the following condition is met: $\lambda_{\max}(\Omega_{g})=O_e(h(N_{g})),$ where $h(N_{g})\uparrow\infty,$
    as $N_{g}\uparrow\infty,$ but $\frac{h(N_{g})}{N_{g}}\rightarrow0,$
    as $N_{g}\rightarrow\infty.$

    \item \textbf{ Weak Dependence:} Dependence across $N_g$ individuals is said
    to be weak, when $\lambda_{\max}(\Omega_{g})=O_e(1).$
\end{enumerate}
\end{definition}

Strong dependence implies that (almost) all individuals are correlated or interconnected. Semi-strong dependence implies that the number of dependent pairs increases with sample size. Weak dependence implies that all the eigenvalues of $\Omega_{g}$ are finite. Independence is regarded as weak dependence. Weak dependence may also hold for dependence that decays sufficiently fast as observations become more distant according to some measure. \citet{basak2018understanding} provide further details and examples of these types of cross-sectional dependence. Consider a scenario where we have clustered data with $G$ clusters, where the $g^{th}$ cluster contains $N_g$ observations. If the $g^{th}$ cluster exhibits strong or semi-strong dependence, then $N_g$ must diverge to $+\infty,$ along with $G\to\infty.$ Otherwise, if $N_g$ is finite, and $G\to\infty,$ then one will always be left with weakly dependent observations.

\begin{remark}
    For simplicity, we have assumed that the rate of semi-strong dependence is uniform across clusters, i.e., $h_g(N_g)\simeq h(N_g), \; \forall g,$ although it can be generalized.
\end{remark}

\begin{assumption}\label{assump:error independence}
(i) The sequence $\{(X_g,\epsilon_g)\}_{g=1}^G$ is independent across $g$. 
Moreover, for each $g = 1, \ldots, G$, the conditional covariance structure of the error terms satisfies
\begin{align*}
    \mathbb{E}\left[\epsilon_g \epsilon_g^\prime \mid X_g \right] &= \Omega_g, \\
    \mathbb{E}\left[\epsilon_g \epsilon_{g_1}^\prime \mid (X_g, X_{g_1}) \right] &= 0, \quad  g_1 \neq g,
\end{align*}
where $\Omega_g$ is positive definite. 

(ii) As per Definition \ref{def:dependency rates}, further assume that 
\[
\lambda_{\max}\left(\Omega_{g}\right)=\begin{cases}
O_{e}\left(N_{g}\right), & \text{for strong dependence,}\\
O_{e}\left(h(N_{g})\right), & \text{for semi-strong dependence,}\\
O_{e}\left(1\right), & \text{for weak dependence,}
\end{cases}
\]
uniformly in $g,$ as in Definition \ref{def:exact order and order uniformly}.
\end{assumption}

Our use of precise definitions and order notation is designed to systematically track the numerous fixed constants that arise in the assumptions, theorems, and their corresponding proofs. Assumption \ref{assump:error independence} allows for arbitrary heteroskedasticity and correlation among observations within the same cluster, as captured by the unrestricted positive definite matrix $\Omega_g$, while imposing independence across clusters. Moreover, the conditioning set $(X_g, X_{g_1})$ can equivalently be replaced by the full regressor array $X$ without changing the meaning of the assumption. Under the independence of $\{(X_g,\epsilon_g)\}_{g=1}^G$ across $g$, conditioning on $X$ does not introduce additional information beyond $(X_g, X_{g_1})$ for the pair $(\epsilon_g,\epsilon_{g_1})$. The conditions on $\lambda_{\max}(\Omega_g)$ characterize the strength of within-cluster dependence through the growth rate of the largest eigenvalue relative to cluster size $N_g$. In particular, under strong dependence, the maximum eigenvalue grows proportionally to the cluster size, indicating substantial within-cluster correlation. Under semi-strong dependence, the growth rate is governed by $h(N_g)$, allowing for intermediate forms of dependence. Under weak dependence, the maximal eigenvalue remains bounded, implying that within-cluster correlation does not grow with the cluster size. The uniformity condition ensures that these rates hold uniformly across clusters.

\begin{definition} \label{def:norms}
    We use two matrix norms: the spectral (maximum eigenvalue) norm and the Frobenius norm. The spectral norm of a matrix $A$ is defined as $\|A\|_e = \max_{z \neq 0} \frac{\|Az\|}{\|z\|},$ 
    where for any vector $z \in \mathbb{R}^n$, $\|z\| = (z'z)^{1/2}$. In particular, if $A$ is symmetric and nonnegative definite, then $\|A\|_e = \max_{z'z=1} z'Az = \lambda_{\max}(A).$
    The Frobenius norm is defined as $\|A\| = \left[ \operatorname{tr}(A'A) \right]^{1/2}.$
\end{definition}

\begin{definition} \label{def:balanced and unbalanced}
Let $\{N_g\}_{g=1}^G$ denote the cluster sizes, where $G$ is the total number of clusters. We classify the clusters according to their degree of balance as follows:

\begin{enumerate}
    \item \textbf{Nearly Balanced:} The clusters are said to be \emph{nearly balanced} if the ratio of the minimum to the maximum cluster size is bounded away from zero asymptotically. That is,
    \[
    \liminf_{G \to \infty} \frac{\min_{1 \leq g \leq G} N_g}{\max_{1 \leq g \leq G} N_g} > 0.
    \]

    \item \textbf{Unbalanced:} The clusters are said to be \emph{unbalanced} if the ratio of the minimum to the maximum cluster size vanishes asymptotically. That is, as $G\to\infty,$
    \[
    \frac{\min_{1 \leq g \leq G} N_g}{\max_{1 \leq g \leq G} N_g} = o(1).
    \]
\end{enumerate}
\end{definition}

Under nearly balanced clusters, the smallest and largest cluster sizes are of the same order, so no single cluster dominates asymptotically. In contrast, under unbalanced clusters, the smallest cluster size becomes negligible relative to the largest, allowing substantial heterogeneity in cluster sizes that may affect the asymptotic behavior of estimators and test statistics.

In the next section, we discuss the pooled ordinary least squares method and derive some asymptotic results under three types of cross-sectional dependence.

\section{ The Pooled Ordinary Least Squares (POLS)} \label{Section:POLS}

The most common estimator for clustering is the POLS (Pooled Ordinary Least Squares) estimator.
From the model described in equation \eqref{Eq:model1}, stacking over clusters yields
\[
Y=X \beta +\epsilon,
\]
where $Y$ and $\epsilon$ are $\left(\sum_{g=1}^G N_g\right)\times 1$ vectors and $X$ is a $\left(\sum_{g=1}^G N_g\right) \times k$ matrix. The traditional POLS estimator of $\beta$ is given by
$$\hat{\beta}_{POLS}=(X^{\prime}X)^{-1}X^{\prime}Y.$$ 
The variance-covariance matrix of the POLS estimator is given by
$$V_{POLS}=\mathbb{E}\left[ \left(X^{\prime}X\right)^{-1}X^{\prime}\Omega X\left(X^{\prime}X\right)^{-1} \right].$$

\cite{hansen2019asymptotic} show that the POLS estimator is consistent provided the non-dominance condition $\frac{\max_{g \le G} N_g}{\sum_{g=1}^G N_g} \to 0$ holds as $G \to \infty$. Here, we shall discuss the asymptotic properties of the POLS estimator under the three kinds of dependence. To establish the consistency of the POLS estimator, we require the following regularity conditions on the design matrix.

\begin{assumption}\label{assump: Xg'Xg=O(Ng)}
(i) Assume that $C_{G}=\mathbb{E}\left[\frac{X^{\prime}X}{\sum_{g=1}^{G}N_{g}}\right]$
is finite and uniformly positive definite over $G.$ Also, assume
that 
\[
\frac{X^{\prime}X}{\sum_{g=1}^{G}N_{g}}-C_{G}=o_{P}(1).
\]

(ii) Assume that for some $q>1,$
\[
\mathbb{E}\left[\left(\operatorname{tr}\left(X_{g}^{\prime}X_{g}\right)\right)^{q}\right]=O\left(N_{g}^{q}\right) \; \text{ uniformly in } g,
\]
as in Definition \ref{def:exact order and order uniformly}.
\end{assumption}

Part (i) of this assumption ensures that the scaled design matrix converges in probability to a finite, non-singular deterministic limit. Assumption \ref{assump: Xg'Xg=O(Ng)}(ii) bounds the growth rate of regressor moments across clusters. This directly implies that $\left\Vert{X}_g^\prime {X}_g\right\Vert=O_P\left(N_g\right),$ uniformly across clusters.

The following lemmas are well-known in the literature and are included here for completeness.

\begin{lemma} \label{lemma:A_n=B_n+o_P(1) implies A_n = O_P(1)}
    Let $\left\{A_n\right\}$ be a sequence of finite-dimensional matrices such that $A_n=B_n+o_P(1),$ where $B_n$ is finite and uniformly positive definite over $n.$ Then, $A_n=O_P(1).$
\end{lemma}
The proof of this lemma is provided in the Appendix for completeness. 
\begin{lemma}\label{lemma:Exact order summability}
    Let $\left\{z_{N_g,g}\right\}$ and $\left\{a_{N_g,g}\right\}$ be doubly indexed sequences of positive real numbers such that $z_{N_g,g}=O_e\left(a_{N_g,g}\right)$ uniformly in $g,$ where $N_g$ is the cluster size for the $g^{th}$ cluster. Then for any $G\geq 1,$  $\sum_{g=1}^G z_{N_g,g}=O_e\left(\sum_{g=1}^G a_{N_g,g}\right).$
\end{lemma}
The proof of this lemma is given in the Appendix. Analogous results hold if we replace $O_e(\cdot)$ by $O(\cdot).$ 

\begin{theorem}
\label{thm:ols consistency}
Under Assumptions \ref{assump:exogeneity}, \ref{assump:error independence} and \ref{assump: Xg'Xg=O(Ng)}, $\hat{\beta}_{POLS}$ is consistent for $\beta$ (i) for nearly balanced clusters under any kind of dependence, (ii) for unbalanced clusters with any kind of weak or semi-strong dependence structures,   (iii) for unbalanced clusters with at least one strongly dependent cluster, if $\frac{\underset{g\leq G}{\max} N_g}{\sum_{g=1}^G N_g} \to 0.$
\end{theorem}

The proof of Theorem \ref{thm:ols consistency} is given in the Appendix.  Although the non-dominance condition imposed in this theorem is the same as Assumption 1 of \cite{hansen2019asymptotic}, decomposing it across different dependence structures allows us to establish the consistency of the estimator under weak and semi-strong dependence even in the unbalanced case, i.e., without requiring the non-dominance condition. For strong dependence, the non-dominance condition is indeed required for the consistency of POLS.

\begin{remark}\label{rem:explanation of strong, semi-strong, weak dependence}
    From now on, unless otherwise stated, by strong dependence, we refer to a scenario in which $O(G)$ clusters exhibit strong dependency, while other clusters may be semi-strongly or weakly dependent. Note that the number of strongly dependent clusters may be, say, $\frac{G}{2}$ or $\frac{G}{3},$ which are also included in the above case. Similarly, unless otherwise stated, semi-strong dependence means that $O(G)$ clusters are semi-strongly dependent, with no strongly dependent clusters present, although weakly dependent clusters may still exist. Weak dependence, on the other hand, describes a situation where all clusters are weakly dependent, with no strong or semi-strongly dependent clusters involved.
\end{remark}

\begin{assumption}\label{assump:additional assumption on errors}
  (i) Assume that for some $0 < c_1 < \infty$ and some $p>1$, $\mathbb{E}\left[ | \epsilon_{gi} |^{2p} \mid X_g \right] \le c_1$ for all $i,g$ and uniformly in $X_g$.
   
 (ii)  Further assume that the following conditional expectations hold uniformly in $g$ and $X_g$ for strong, semi-strong, and weak dependence, respectively: 
\begin{align*}
 \mathbb{E}\left[ \sum_{\underset{i\neq j \neq s}{i,j,s=1}}^{N_g}\epsilon_{gi}^{2}\epsilon_{gj}\epsilon_{gs} \Bigg| X_g\right] &= O\left(N_g^3\right), \quad O\left(N_g^2{h(N_g)}\right), \quad \text{and} \quad O\left(N_g^2\right); \\
\mathbb{E}\left[ \sum_{\underset{i\neq j \neq s \neq t}{i,j,s,t=1}}^{N_g}\epsilon_{g_{i}}\epsilon_{g_{j}}\epsilon_{g_{s}}\epsilon_{g_{t}}  \Bigg |  X_g\right]
 &= O\left(N_g^4\right), \quad O\left({N^2_g}{h^2(N_g)}\right), \quad \text{and} \quad O\left(N_g^2\right). 
\end{align*}
\end{assumption}

Part (i) in the above assumption essentially ensures the existence of moments beyond the second order, thereby controlling the tail behavior of the error terms $\epsilon_{gi}.$ It also establishes the necessary regularity conditions for deriving the limiting distribution of the estimator. Part (ii) of the assumption specifies the orders of the higher-order joint moments of the errors under different within-cluster dependence structures. The following discussion is based on one of several sufficient conditions that provide intuition and justification for the moment conditions imposed in Assumption \ref{assump:additional assumption on errors}.

Under strong dependence, the error terms within a cluster exhibit substantial correlation, so that cross-products of distinct errors do not average out. As a result, the triple sum 
$\sum_{i\neq j\neq s}\epsilon_{gi}^{2}\epsilon_{gj}\epsilon_{gs}$ accumulates at rate $N_g^{3}$, while the quadruple sum 
$\sum_{i\neq j\neq s\neq t}\epsilon_{gi}\epsilon_{gj}\epsilon_{gs}\epsilon_{gt}$ grows at rate $N_g^{4}$, reflecting the total number of non-negligible interacting terms. 

To justify the orders in Assumption \ref{assump:additional assumption on errors} under semi-strong dependence, consider a block-diagonal dependence structure within the $g$th cluster. Suppose that the error covariance matrix $\Omega_g$ consists of $O(N_g / h(N_g))$ blocks, each of size $O\left(h(N_g)\right)$. 
For the triple sum, nonzero contributions arise only when at least two of the indices belong to the same block. If all three error terms fall within the same block, then the number of such terms is of order $N_g / h(N_g) \times h^3(N_g)$, yielding an overall order of $N_g h^2(N_g)$. If two of the first moment error terms belong to the same block and the remaining squared term belongs to a different block, then the number of such terms is of order $N_g^2 / h^2(N_g) \times h^3(N_g)$, which leads to the dominant order $N_g^2 h(N_g)$. All other index configurations yield zero contribution due to cross-block independence.
A similar argument applies to the quadruple sum. If all four error terms fall within the same block, the resulting order is $N_g / h(N_g) \times h^4(N_g) = N_g h^3(N_g)$. If the indices split across two blocks with two terms in each block, the number of contributing terms is of order $N_g^2 / h^2(N_g) \times h^4(N_g)$, yielding $N_g^2 h^2(N_g)$, which dominates. All remaining configurations vanish by independence across blocks.
Therefore, under semi-strong dependence, the leading orders of the triple and quadruple sums are $O\left(N_g^2 h(N_g)\right)$ and $O\left(N_g^2 h^2(N_g)\right)$, respectively, as stated in Assumption \ref{assump:additional assumption on errors}.

For weak cross-sectional dependence, consider a block-diagonal structure in which the $g$th cluster consists of $N_g$ blocks of uniformly bounded size. In this case, dependence is confined to finite-sized blocks.
For the triple sum, if all three error terms fall within the same block, the number of such terms is of order $N_g$. If two of the first moment error terms belong to the same block and the remaining squared term belongs to a different block, the number of such terms is of order $N_g^2$, which dominates. All other index configurations yield zero contribution due to independence across blocks.
For the quadruple sum, if all four error terms fall within the same block, the contribution is of order $N_g$. If the indices split across two blocks with two terms in each block, the number of contributing terms is of order $N_g^2$, which again dominates. All remaining configurations vanish by cross-block independence.

\begin{theorem}\label{thm:ols inconsistency}
     Assume that there exists at least one strongly dependent cluster such that the cluster size of the maximum of such clusters, say, $N^*$ satisfy the condition $\frac{N^*}{\sum_{g=1}^G N_g} \nrightarrow 0.$
     Then, under Assumptions \ref{assump:exogeneity}, \ref{assump:error independence},  \ref{assump: Xg'Xg=O(Ng)} with $q=2,$ \ref{assump:additional assumption on errors}(i) with $p=2$, and  $\lambda_{\min}\left(\sum_{g=1}^G \mathbb{E}\left[X_{g}^{\prime}\Omega_{g}X_{g}\right]\right) \geq C \sum_{g\in \mathcal{SD}} N_g^2$, for some positive constant $C$, the POLS estimator $\hat{\beta}_{POLS}$ is inconsistent, where $\mathcal{SD}$ is the set of strongly dependent clusters.
\end{theorem}

The proof of Theorem \ref{thm:ols inconsistency} is given in the Appendix. 
If the clusters are highly unbalanced and exhibit strong within-cluster dependence, the standard POLS estimator may yield inconsistent results. This theorem establishes the fact that the non-dominance condition is not only a sufficient condition for consistency, but also necessary when strong within-cluster dependence is present. The assumption of the lower bound on the minimum eigenvalue of $\sum_{g=1}^G \mathbb{E}\left[X_{g}^{\prime}\Omega_{g}X_{g}\right]$ in Theorem \ref{thm:ols inconsistency} will be justified and explained in \eqref{eq: order of lambda min D_G} in a subsequent discussion.

\begin{example}      
    Let $N^*$ be the maximum size among a finite number $L$ of dominant clusters. Assume that the remaining $G-L$ clusters grow at a rate $M \to \infty$, bounded above by $c_2 M$ with $0<c_2<\infty$. Then, the ratio
    \[
     \frac{\max_{g\leq G} N_g}{\sum_{g=1}^{G}N_g} \geq \frac{N^*}{L N^* + (G-L)c_2 M} = \frac{\frac{N^*}{MG}}{L\left(\frac{N^*}{MG}\right) + \left(1 - \frac{L}{G}\right)c_2}.
    \]
    As $G \to \infty$, the term $\frac{L}{G} \to 0$. If $\frac{N^*}{MG} \to +\infty$, the lower bound converges to $\frac{1}{L}$. If instead, (i) the limit infimum of  $\frac{N^*}{MG}$ remains bounded away from $0$ and $+\infty$, then the limit infimum of the ratio has a strictly positive lower bound, and (ii) the limit supremum of $\frac{N^*}{MG}$ is bounded away from zero (limit infimum may still converge to $0$), then there exists a subsequence of $\frac{N^*}{MG}$ which is bounded away from $0$ and $+\infty$, which further deduces that the ratio has a strictly positive lower bound.  Thus, $\hat{\beta}_{POLS}$ is inconsistent if $\frac{N^*}{MG} \nrightarrow 0$ as $G \to \infty$.
\end{example}

\begin{remark}
  Note that for the inconsistency of $\hat{\beta}_{POLS}$, the boundedness of the regressors and errors up to the fourth moment can be relaxed and replaced by the boundedness for the $(2+\delta)^{\text{th}}$ moment for some $0<\delta<\infty$.
\end{remark}

Next, we intend to study the asymptotic normality of $\hat{\beta}_{POLS}$ under different degrees of within-cluster dependence. This requires the analysis of the matrix: 
\[
D_{G}=\frac{1}{\sum_{g=1}^{G}N_{g}^{2}}\sum_{g=1}^{G}\mathbb{E}\left[X_{g}^{\prime}\Omega_{g}X_{g}\right].
\]
The $k\times k$ matrix measures the average variability contributed by the clusters, after accounting for within-cluster dependence. Using the convexity of $\lambda_{\max}(\cdot)$ and the inequality
$\lambda_{\max}(A B A') \le \lambda_{\max}(A A') \lambda_{\max}(B)$
for $B \succeq 0$, we obtain
\[
\lambda_{\max}(D_G)
\le
\frac{1}{\sum_{g=1}^G N_g^2}
\sum_{g=1}^{G}
\mathbb{E}\left[\lambda_{\max}(X_g'X_g)\right]
\lambda_{\max}(\Omega_g).
\]
Using the fact that $\mathbb{E}\left[\lambda_{\max}\left({X}_{g}^{\prime}{X}_{g}\right)\right]=O(N_g)$ by Assumption \ref{assump: Xg'Xg=O(Ng)}, and applying Assumption \ref{assump:error independence}, it follows from Lemma \ref{lemma:Exact order summability} that
\begin{equation}
\label{eq: order of lambda_max D_G}
 \lambda_{\max}\left(D_{G}\right) = \begin{cases}
O\left(1\right), & \text{for strong dependence,}\\
O\left(\frac{\sum_{g=1}^{G}N_{g}h(N_{g})}{\sum_{g=1}^{G}N_{g}^{2}}\right), & \text{for semi-strong dependence,}\\
O\left(\frac{\sum_{g=1}^{G}N_{g}}{\sum_{g=1}^{G}N_{g}^{2}}\right), & \text{for weak dependence.}
\end{cases}
\end{equation}
 To determine the order of $\lambda_{\min}\left(D_G\right)$ for strong, semi-strong, and weak cross-sectional dependence, consider the following expression:
 \begin{align}
& \lambda_{\min}\left(D_G\right) \nonumber\\
 & \geq \, \frac{1}{\sum_{g=1}^G N_g^2}\cdot\lambda_{\min}\left(\sum_{g:\frac{\lambda_{\min}\left(\mathbb{E}\left[X_{g}^{\prime}\Omega_{g}X_{g}\right]\right)}{N_{g}^{2}}\geq c>0}\mathbb{E}\left[X_{g}^{\prime}\Omega_{g}X_{g}\right]\right) \label{eq: expansion of lambdamin(DG) term 1} \\
    & \quad + \frac{1}{\sum_{g=1}^G N_g^2}\cdot\lambda_{\min}\left(\sum_{\underset{\frac{\lambda_{\min}\left(\mathbb{E}\left[X_{g}^{\prime}\Omega_{g}X_{g}\right]\right)}{N^2_{g}}=o(1);\;h(N_{g})\uparrow\infty;\;\frac{h(N_{g})}{N_{g}}=o(1)}{g:\frac{\lambda_{\min}\left(\mathbb{E}\left[X_{g}^{\prime}\Omega_{g}X_{g}\right]\right)}{N_{g}h(N_{g})}\geq c_{1}>0}}\mathbb{E}\left[X_{g}^{\prime}\Omega_{g}X_{g}\right]\right) \label{eq: expansion of lambdamin(DG) term 2} \\
 & \quad + \frac{1}{\sum_{g=1}^G N_g^2}\cdot\lambda_{\min}\left( \sum_{\underset{\frac{\lambda_{\min}\left(\mathbb{E}\left[X_{g}^{\prime}\Omega_{g}X_{g}\right]\right)}{N_{g}h(N_g)}=o(1);\;h(N_{g})\uparrow\infty;\;\frac{h(N_{g})}{N_{g}}=o(1)}{g:\frac{\lambda_{\min}\left(\mathbb{E}\left[X_{g}^{\prime}\Omega_{g}X_{g}\right]\right)}{N_{g}}\geq c_{2}>0}} \mathbb{E}\left[X_{g}^{\prime}\Omega_{g}X_{g}\right]\right). \label{eq: expansion of lambdamin(DG) term 3}
\end{align}

Since $\lambda_{\min}(A+B+C) \geq \lambda_{\min}(A) + \lambda_{\min}(B) + \lambda_{\min}(C)$ for non-negative definite matrices $A$, $B$, and $C$, if one of these minimum eigenvalues dominates, it effectively determines the overall lower bound, allowing the remaining lower-order terms to be ignored. Here, if the number of strongly dependent clusters is of order $O_e(G),$ then only the term in \eqref{eq: expansion of lambdamin(DG) term 1} dominates, and we have $\lambda_{\min}\left(D_G\right)=O_e\left(1\right).$ If the number of clusters that are semi-strongly dependent is of order $O_e(G),$ and no strongly dependent clusters are present, then the term in \eqref{eq: expansion of lambdamin(DG) term 1} does not contribute, the term in \eqref{eq: expansion of lambdamin(DG) term 2} dominates and we have $\lambda_{\min}\left(D_G\right) = O_e\left({\sum_{g=1}^{G} N_g h(N_g)} / {\sum_{g=1}^{G} N_g^2}\right).$ If all the clusters are weakly dependent, then only the term in \eqref{eq: expansion of lambdamin(DG) term 3} contributes, and we have $\lambda_{\min}\left(D_G\right) = O_e\left(\sum_{g=1}^G N_g / \sum_{g=1}^{G} N_g^2\right).$ 
Thus, the splitting helps us to analyze the cases where the equality of orders of $\lambda_{\min}\left(D_{G}\right)$ and $\lambda_{\max}\left(D_{G}\right)$ for the three kinds of dependence arises.

For the preceding arguments, some of the sufficient conditions may be stated as follows: for any arbitrary index set $S_G^* \subseteq \{1,\dots,G\}$ satisfying $\vert{}S_G^*\vert{} = O_e(G)$, we assume that:
  $$\begin{aligned} \text{(i)} \quad & \sum_{g\in S_G^*}N_g^2 = O_e\left(\sum_{g=1}^G N_g^2\right), && \text{under strong dependence,} \\ \text{(ii)} \quad & \sum_{g\in S_G^*}N_g h(N_g) = O_e\left(\sum_{g=1}^G N_g h(N_g)\right), && \text{under semi-strong dependence,} \\ \text{(iii)} \quad & \sum_{g\in S_G^*}N_g = O_e\left(\sum_{g=1}^G N_g\right), && \text{under weak dependence.} \end{aligned}$$
Notably, these conditions are naturally satisfied in the nearly balanced case. Then, it follows that 
\begin{equation}
\label{eq: order of lambda min D_G}
     \lambda_{\min}(D_G) \,= \,
 \begin{cases}
O_{e}\left(1\right), & \text{for strong dependence,}\\
O_{e}\left(\frac{\sum_{g=1}^{G}N_{g}h(N_{g})}{\sum_{g=1}^{G}N_{g}^{2}}\right), & \text{for semi-strong dependence,}\\
O_{e}\left(\frac{\sum_{g=1}^{G}N_{g}}{\sum_{g=1}^{G}N_{g}^{2}}\right), & \text{for weak dependence.}
\end{cases}
\end{equation}
Thus, \eqref{eq: order of lambda_max D_G} and \eqref{eq: order of lambda min D_G} provide a convenient framework for analyzing the equality of the orders of the minimum and maximum eigenvalues of the matrix $D_G$. 
To summarize the preceding discussion, we next impose the following 
condition.

\begin{assumption}\label{assump:lamba_max=lambda_min for OLS}
Assume that $D_G$ is finite and positive definite, and that both $\lambda_{\min}(D_G)$ and $\lambda_{\max}(D_G)$ are of the exact order:
\[
\begin{cases}
O_{e}\left(1\right), & \text{for strong dependence,}\\
O_{e}\left(\frac{\sum_{g=1}^{G}N_{g}h(N_{g})}{\sum_{g=1}^{G}N_{g}^{2}}\right), & \text{for semi-strong dependence,}\\
O_{e}\left(\frac{\sum_{g=1}^{G}N_{g}}{\sum_{g=1}^{G}N_{g}^{2}}\right), & \text{for weak dependence.}
\end{cases}
\]
\end{assumption} 

This assumption is justified in the preceding discussion, and it follows from Assumption \ref{assump:lamba_max=lambda_min for OLS} that the $k\times k$ matrix ${D_{G}}/{\left\Vert D_{G}\right\Vert}$ is uniformly positive definite for any kind of dependence.

\begin{theorem}
\label{thm:ols normality}
Let $W_G=C_G^{-1} D_G C_G^{-1}$. Under Assumptions \ref{assump:exogeneity}, \ref{assump:error independence}, \ref{assump: Xg'Xg=O(Ng)} with $q=p$, \ref{assump:additional assumption on errors}(i) and \ref{assump:lamba_max=lambda_min for OLS},
\[
    \frac{\sum_{g=1}^G N_g}{\sqrt{\sum_{g=1}^{G}N_{g}^{2}}}W_{G}^{-1/2}\left(\hat{\beta}_{POLS}-\beta\right)\xrightarrow{d}N_{k}\left(0,I_{k}\right), \quad \text{ as } G\to\infty,
\]
(i) for strong dependence if $\frac{\sum_{g=1}^G N_g^{2p}}{\left(\sum_{g=1}^G N_g^2\right)^p}\to 0$, (ii) for semi-strong dependence if $\frac{\sum_{g=1}^G N_g^{2p}}{\left(\sum_{g=1}^G N_g h(N_g)\right)^p}\to 0$, (iii) for weak dependence if $\frac{\sum_{g=1}^G N_g^{2p}}{\left(\sum_{g=1}^G N_g\right)^p}\to 0$. 
\end{theorem}

The proof of Theorem \ref{thm:ols normality} is given in the Appendix.

\begin{remark}
The conditions imposed on cluster sizes in the above theorem require further discussion. For the asymptotic normality of the POLS, \cite{hansen2019asymptotic} assumed that for some $2\leq r \leq s <\infty,$ $\frac{\left(\sum_{g=1}^G N_g^r\right)^{2/r}}{n}\leq C<\infty$ and $\max_{g\leq G}\frac{N_g^2}{n}\to 0,$ as $n\to\infty,$ where $n=\sum_{g=1}^G N_g,$ and $\sup_i \mathbb{E}[|y_i|^{2s}]<\infty, \, \sup_i \mathbb{E}[\Vert x_i\Vert^{2s}]<\infty,$ with either $r<s,$ or $(y_i,x_i)$ having identical marginal distributions. \cite{djogbenou2019asymptotic} also assume similar orders (See Assumptions 1, 2, 3 in their paper) to establish the asymptotic normality of the POLS, consistency of the CRVE, and asymptotic normality of the test statistic (their Theorem 2.1), under local alternatives. The conditions imposed on the cluster sizes in \cite{hansen2019asymptotic} reduce to 
\begin{equation} \label{eq: hansen asympnom conditions}
    \frac{\sum_{g=1}^G N_g^{2p}}{\left(\sum_{g=1}^G N_g \right)^p} \leq C < \infty, \quad \text{ and } \quad \frac{\max_{g\leq G} N_g^2}{\sum_{g=1}^G N_g}\to 0.
\end{equation}
However, from Theorem \ref{thm:ols normality}, the equivalent condition for the asymptotic normality of the POLS to hold under strong dependence is $\frac{\max_{g\leq G}N_g^2}{\sum_{g=1}^G N_g^2}\to0,$ as $G\to\infty$, since
\[
\left(\frac{\max_{g\leq G}N_g^2}{\sum_{g=1}^G N_g^2}\right)^p\leq \frac{\sum_{g=1}^G N_g^{2p}}{\left(\sum_{g=1}^G N_g^2\right)^p} \leq \frac{\max_{g\leq G}N_g^{2(p-1)}}{\left(\sum_{g=1}^G N_g^2\right)^{p-1}}.
\]
Therefore, the asymptotic normality in Theorem \ref{thm:ols normality} is established under less restrictive conditions on cluster sizes than those in \eqref{eq: hansen asympnom conditions} in the presence of strong dependence. Note that under the nearly balanced clusters case with $N_g\simeq N$, for strong dependence, the asymptotic normality holds generally, since the required condition reduces to  $O\left(\frac{\sum_{g=1}^{G}N_{g}^{2p}}{\left(\sum_{g=1}^{G}N_{g}^{2}\right)^{p}}\right)=O\left(G^{1-p}\right),$ which vanishes as $G\to\infty.$

For semi-strong dependence, the condition $\frac{\sum_{g=1}^G N_g^{2p}}{\left(\sum_{g=1}^G N_g h(N_g)\right)^p}\to 0$ is less restrictive than the first condition in \eqref{eq: hansen asympnom conditions}. Also, for the nearly balanced clusters with $N_g\simeq N$, our condition reduces to $\left(\frac{N}{h(N)}\right)^p \frac{1}{G^{p-1}}\to 0,$ which is less restrictive than the second condition in \eqref{eq: hansen asympnom conditions} as long as $h(N)\geq \sqrt{N}$ and $p>2.$

For weak dependence, our condition is more restrictive than those in \eqref{eq: hansen asympnom conditions}. Under the nearly balanced clusters with $N\to\infty,$ the condition reduces to $\frac{N^p}{G^{p-1}}\to 0.$
\end{remark}

The conventional cluster-robust variance estimator (CRVE) takes the form:
$$\hat{V}_{POLS}=\left(X^{\prime}X\right)^{-1}\left(\sum_{g=1}^{G}X_{g}^{\prime}u_{g}u_{g}^{\prime}X_{g}\right)\left(X^{\prime}X\right)^{-1}$$ 
with cluster-specific residuals $u_{g}=Y_{g}-X_{g}\hat{\beta}_{POLS},$  for $g=1,2,\ldots,G.$ Initial derivations of this estimator, including \cite{white1984asymptotic} for balanced clusters and \cite{liang1986longitudinal} for unbalanced settings, assume a finite number of observations per cluster. Subsequently, \cite{hansen2007asymptotic} established that the CRVE remains asymptotically valid when the cluster sizes $N_g \to \infty$ alongside the number of clusters $G \to \infty$. More recently, \cite{hansen2019asymptotic} and \cite{djogbenou2019asymptotic} demonstrated the asymptotic normality of the POLS estimator, consistency of the CRVE, and asymptotic validity of Wald-type tests under specific non-dominance regularity conditions under a general within-cluster dependence framework. 

\begin{theorem}
\label{thm:Vols_consistent}
Let us define
\[
\hat{W}_{G}=\left(\frac{X^{\prime}X}{\sum_{g=1}^{G}N_{g}}\right)^{-1}\hat{D}_{G}\left(\frac{X^{\prime}X}{\sum_{g=1}^{G}N_{g}}\right)^{-1},\qquad \hat{D}_{G}=\frac{1}{\sum_{g=1}^{G}N_{g}^{2}}\sum_{g=1}^{G}X_{g}^{\prime}u_{g}u_{g}^{\prime}X_{g}.
\]
Then, under Assumptions \ref{assump:exogeneity}, \ref{assump:error independence}, \ref{assump: Xg'Xg=O(Ng)} with $q=2,$ \ref{assump:additional assumption on errors}(i) with $p=2$, and \ref{assump:lamba_max=lambda_min for OLS}, $\hat{W}_{G}$ is consistent for $W_{G}$ as $G\to\infty,$ (i) for strong dependence if $\frac{\sum_{g=1}^G N_g^{4}}{\left(\sum_{g=1}^G N_g^2\right)^2}\to 0$, (ii) for semi-strong dependence if $\frac{\sum_{g=1}^G N_g^{4}}{\left(\sum_{g=1}^G N_g h(N_g)\right)^2}\to 0$, (iii) for weak dependence if $\frac{\sum_{g=1}^G N_g^{4}}{\left(\sum_{g=1}^G N_g\right)^2}\to 0$. 
\end{theorem}

The proof of Theorem \ref{thm:Vols_consistent} is provided in the Appendix. We establish the consistency of the CRVE under suitable non-dominance conditions across three distinct dependence regimes. This classification into different dependence regimes enables us to state the required conditions separately for each type of dependence structure. In particular, as noted in the previous remark, under strong dependence, our restriction on cluster sizes is less restrictive than those in \cite{hansen2019asymptotic}. Under semi-strong dependence, the condition in Theorem \ref{thm:Vols_consistent} is comparatively less restrictive, if $h(N)\geq N^\delta,$ for any $\delta>1/2.$ In contrast, under weak dependence, our conditions are comparatively more restrictive, as we use the maximum bound $O\left(N_g^4\right)$ for the term  $\mathbb{E}\left[\left\Vert X_g'(\epsilon_g \epsilon_g' - \Omega_g) X_g \right\Vert^2\right]$,
rather than using a sharper bound for the weak dependence structure.

Let us consider the testing problem of testing $H_{0}:R^{l\times k}\beta^{k\times1}=r^{l\times1}$ against $H_{A}:R^{l\times k}\beta^{k\times1}\ne r^{l\times1}$
, where $R^{l\times k}$ and $r^{l\times1}$ are known matrix and
vector, respectively, with $\operatorname{rank}(R^{l\times k})=l\leq k.$ 

\begin{theorem}\label{thm:testing for OLS}
    Let $\Pi_{G}^{l\times l}=RW_{G}R^{\prime}$
and $\hat{\Pi}_{G}^{l\times l}=R\hat{W}_{G}R^{\prime}$. Then, under
$H_{0}:R\beta=r,$ under Assumptions \ref{assump:exogeneity}, \ref{assump:error independence}, \ref{assump: Xg'Xg=O(Ng)} with $q=2,$ \ref{assump:additional assumption on errors}(i) with $p=2$, and \ref{assump:lamba_max=lambda_min for OLS},
\[
\frac{\left(\sum_{g=1}^{G}N_{g}\right)^{2}}{\sum_{g=1}^{G}N_{g}^{2}}\left(R\hat{\beta}_{POLS}-r\right)^{\prime}\hat{\Pi}_{G}^{-1}\left(R\hat{\beta}_{POLS}-r\right)\xrightarrow{d}\chi_{l}^{2},\quad \text{ as } G\to\infty
\]
(i) for strong dependence if $\frac{\sum_{g=1}^G N_g^{4}}{\left(\sum_{g=1}^G N_g^2\right)^2}\to 0$, (ii) for semi-strong dependence if $\frac{\sum_{g=1}^G N_g^{4}}{\left(\sum_{g=1}^G N_g h(N_g)\right)^2}\to 0$, (iii) for weak dependence if $\frac{\sum_{g=1}^G N_g^{4}}{\left(\sum_{g=1}^G N_g\right)^2}\to 0$. 
\end{theorem}

The proof of Theorem \ref{thm:testing for OLS} is given in the Appendix. 

We introduce the proposed estimator in the next section. Therein, we establish its asymptotic properties and derive a corresponding test statistic.

\section{The Proposed Estimator} \label{Section:proposed}

To study the proposed estimator, we return to the model in \eqref{Eq:model}. Taking the cluster averages of the dependent variables, explanatory variables, and random errors yields $\bar{Y}_g,$ $\bar{X}^{1\times k}_g,$ and $\bar{\epsilon}_g,$ respectively. Then the model \eqref{Eq:model1} reduces to 
\begin{equation} \label{Eq:model3}
\bar{Y}_{g}=\bar{X}_{g}\beta +\bar{\epsilon}_{g},\;\;  g=1,2,\ldots,G.
\end{equation}

By stacking all the group averages, the  model can be written in matrix form as
\begin{equation}\label{Eq:model4}
\bar{Y}=\bar{X}\beta +\bar{\epsilon},
\end{equation}
where $\bar{Y} $ is a $G\times 1$ vector, $\bar{X} $ is a $G\times k$ matrix. As we have mentioned earlier in Section \ref{Section:model}, $X_g$ may contain categorical variables as well.


 \begin{remark}\label{rem:intercept allow}
     We can allow the intercept to vary across clusters, assuming a classical random coefficient model for intercepts, as follows.
     \begin{equation*} 
         Y_g=\alpha_g+X_g\beta +\epsilon_g ,\;\;  g=1,2,\ldots ,G.
    \end{equation*}
    However, this generalization is not immediate for the POLS.
 \end{remark}

\begin{remark}
\label{Rem:heteroskedastic_model}
The averaged model in \eqref{Eq:model3} is heteroskedastic and free of cross-sectional dependence. Here, we transform our clustered data into heteroskedastic data due to averaging.
\end{remark}

The OLS estimator for this transformed (average) model is
\[
\hat{\beta}_A=\left(\bar{X}'\bar{X}\right)^{-1} \bar{X}'\bar{Y}.
\]
This is our proposed estimator. We establish the asymptotic properties of this estimator and examine its comparative advantages. The variance-covariance matrix of the proposed estimator is given by $V=\mathbb{E}[(\hat{\beta}_A-\beta)(\hat{\beta}_A-\beta)^\prime],$ which reduces to 
\[
V =\mathbb{E}\left[ (\bar{X}'\bar{X})^{-1} \left(\sum_{g=1}^G  \bar{X}_g'{\bar{X}_g} \sigma_g^2 \right) (\bar{X}'\bar{X})^{-1} \right],
\]
where $\sigma_g^2 =  \operatorname{Var}\left(\bar{\epsilon}_g \mid X_g \right) =  \frac{\mathbf{1}'\Omega_g \mathbf{1}}{N_g^2},$ for $g=1,2,\ldots, G.$

\begin{remark}\label{rem:order of sigma_g^2}
    Since $\sigma_{g}^{2} \leq \frac{1}{N_g} \lambda_{\max}(\Omega_g),$ it follows from Assumption \ref{assump:error independence} that 
    \begin{equation}\label{eq:order of sigma_g^2}
    \sigma_{g}^{2} = \begin{cases}
    O(1), & \text{for strong dependence,}\\
    O\left(\frac{h(N_{g})}{N_{g}}\right), & \text{for semi-strong dependence,}\\
    O_e\left(\frac{1}{N_{g}}\right), & \text{for weak dependence.}
    \end{cases}    
    \end{equation}
    uniformly in $g$. The exact order $O_e(1/N_g)$ under weak dependence follows from the following bounds: $\frac{1}{N_{g}}\lambda_{\min}(\Omega_{g}) \leq \sigma_{g}^{2} \leq \frac{1}{N_{g}}\lambda_{\max}(\Omega_{g})$, as both the eigenvalues are $O_e(1)$ uniformly in $g$.
\end{remark}

For valid statistical inference, we require an appropriate standard error. The natural choice is any standard error robust to heteroskedasticity.  Here we adapt White's heteroskedasticity consistent covariance matrix (HCCM) estimator (see \cite{white1980heteroskedasticity}), which provides consistent covariance matrix estimates under general conditions.
 Define $ e =\bar{Y}-\bar{X}\hat{\beta}_A. $ Note that $e=[e_1, e_2, \ldots , e_G]^\prime.$
The natural estimator of the variance is :
\begin{equation}\label{Eq:v_hat}  
\hat{V} = \widehat{\operatorname{Var}(\hat{\beta}_A) }=\left(\bar{X}'\bar{X}\right)^{-1} \left(\sum_{g=1}^{G} {\bar{X}_g}'{\bar{X}_g}e_g^2\right) \left(\bar{X}'\bar{X}\right)^{-1}.
\end{equation}

The asymptotic properties of $\hat{\beta}_A$ and $\hat{V}$ implicitly depend on the dependence structure of the error terms. The following subsection examines the asymptotic properties of the proposed estimator.

\subsection{Asymptotic Results and Hypothesis Testing}

We now analyze the asymptotic properties of the proposed estimator under the three dependence regimes defined. Furthermore, we develop a Wald-type test based on our estimator.  First, we impose a White-type (see \cite[p.~818]{white1980heteroskedasticity}) condition on the averaged regressors.

\begin{assumption}\label{assump: Xbar'Xbar=O(G)}
(i)    Assume that  $A_{G}=\mathbb{E}\left[\frac{\bar{X}^{\prime}\bar{X}}{G}\right]$
    is finite and uniformly positive definite over $G,$ where $\bar{X}$ is obtained by stacking the cluster averages $\bar{X}_g=\frac{1}{N_g} \sum_{i=1}^{N_g}X_{gi}, $ for all explanatory variables. Also, assume that 
    \[
    \frac{\bar{X}^{\prime}\bar{X}}{G}-A_{G}=o_{P}(1), \quad \text{ as } G\to\infty.
    \]

(ii)  For some fixed $0<c_0<\infty,$ assume that 
\[
\sup_g \mathbb{E}\left[\left(\operatorname{tr}\left(\bar{X}_{g}^{\prime}\bar{X}_{g}\right)\right)^{2}\right]\leq c_0.
\]

\end{assumption}

Part (i) of this assumption establishes a standard regularity condition requiring the sample average of the cluster-level cross products to converge in probability to a finite and positive definite population matrix, which guarantees the asymptotic invertibility of the design matrix. Assumption \ref{assump: Xbar'Xbar=O(G)}(ii) implies that $\left|\bar{X}_g\bar{X}^\prime_g\right|=O_P(1),$ uniformly in $g,$ since for any $\delta>0,$ and for some $0<c_0^*<\infty, \; M>0,$ by Markov inequality, 
\[
\sup_g \mathbb{P}\left(\left|\bar{X}_g\bar{X}^\prime_g\right|>M\right) \leq \frac{\underset{g}{\text{sup}} \; \mathbb{E}\left[\left| \bar{X}_g\bar{X}^\prime_g  \right|\right]}{M} \leq \frac{c_0^*}{M} < \delta.
\] 

We now proceed to establish the consistency of the estimator.
\begin{theorem}
\label{thm:consistent}
Under Assumptions \ref{assump:exogeneity}, \ref{assump:error independence}(i), and \ref{assump: Xbar'Xbar=O(G)}(i), for any kinds of within-cluster dependence, and for any cluster sizes,  $\hat{\beta}_A$ is consistent for $\beta$.
\end{theorem}

The proof of Theorem \ref{thm:consistent} is provided in the Appendix. This theorem establishes the consistency of the proposed averaging estimator under general dependence regimes, without imposing restrictions on cluster sizes. Note that the non-dominance condition required for the consistency of the POLS estimator is not needed for the consistency of the proposed estimator. In particular, $\hat{\beta}_A$ remains consistent even when this condition is violated. Real-life examples of dominating clusters have been mentioned in the Introduction. Moreover, the presence of $O(G)$, or $o(G)$ number of strongly, semi-strongly, or weakly dependent clusters does not affect the consistency of the estimator. 

Note that $\hat{\beta}_A - \beta = \left(\frac{\bar{X}'\bar{X}}{G}\right)^{-1} \frac{\bar{X}'\bar{\epsilon}}{G}.$ Clearly, $\hat{\beta}_A$ is an unbiased estimator of $\beta.$
Also, $\left(\frac{\bar{X}^{\prime}\bar{X}}{G}\right)^{-1}=O_P(1),$ which follows from Lemma \ref{lemma:A_n=B_n+o_P(1) implies A_n = O_P(1)},  given that $\left(\frac{\bar{X}^{\prime}\bar{X}}{G}\right)^{-1}-A_{G}^{-1}=o_P(1),$ a result followed from Assumption \ref{assump: Xbar'Xbar=O(G)}(i), where $A_{G}$ is finite and uniformly positive definite over $G.$ Applying Assumptions \ref{assump:exogeneity} and \ref{assump:error independence}, the variance-covariance matrix of $\frac{\bar{X}^{\prime}\bar{\epsilon}}{G}$ reduces to 
\[
\operatorname{Var}\left(\frac{\bar{X}^{\prime}\bar{\epsilon}}{G}\right) = \frac{1}{G^2}\sum_{g=1}^{G} \mathbb{E}\left[\bar{X}_{g}^{\prime}\bar{X}_{g}\right]\sigma_{g}^{2}.
\]
If we additionally include Assumption \ref{assump: Xbar'Xbar=O(G)}(ii), then, applying the triangular inequality for norms, substituting the order of $\sigma_g^2$  provided in equation \eqref{eq:order of sigma_g^2} of Remark \ref{rem:order of sigma_g^2}, which follows from Assumption \ref{assump:error independence}, and using Lemma \ref{lemma:Exact order summability}, we obtain
$$\left\Vert \operatorname{Var}\left(\frac{\bar{X}^{\prime}\bar{\epsilon}}{G}\right)\right\Vert=
\begin{cases}
O\left(\frac{1}{G}\right), \text{ for strong dependence }\\
O\left(\frac{1}{G^2}\sum_{g=1}^G \frac{h(N_g)}{N_g}\right), \text{ for semi-strong dependence }\\
O\left(\frac{1}{G^2}\sum_{g=1}^G \frac{1}{N_g}\right), \text{ for weak dependence }
\end{cases}$$ 
The above discussion yields the following result.

\begin{corollary}
     Under Assumptions \ref{assump:exogeneity}, \ref{assump:error independence} and \ref{assump: Xbar'Xbar=O(G)}, $\hat{\beta}_A$ is $\sqrt{G}$ consistent for strong dependence. $\hat{\beta}_A$ is $\sqrt{G}\cdot \bar{h}^{-\frac{1}{2}}_1$ consistent for semi-strong dependence, and $\sqrt{G}\cdot \bar{m}^{-\frac{1}{2}}$ consistent for weak dependence, where $\bar{h}_{1}=\frac{1}{G}\sum_{g=1}^G \frac{h(N_g)}{N_g}$ and $\bar{m}=\frac{1}{G}\sum_{g=1}^G \frac{1}{N_g}.$  
\end{corollary}

Next, to establish the asymptotic normality of $\hat{\beta}_A,$ we study the matrix: 
\[
B_G = \frac{1}{G}\sum_{g=1}^{G}\mathbb{E}\left[\bar{X}_{g}^{\prime}\bar{X}_{g}\right]\sigma_{g}^{2}.
\]
The $k\times k$ matrix $B_G$ represents the aggregate contribution of the clusters to the asymptotic variance, combining the second moment of the averaged regressors with the variance of the averaged errors. The differences in the resulting orders under various dependence structures arise from the behavior of $\sigma_g^2$.

Note that 
\[
\lambda_{\max}\left(B_G\right)\,\leq \,\frac{1}{G} \sum_{g=1}^{G}\lambda_{\max}\left(\mathbb{E}\left[\bar{X}_{g}^{\prime}\bar{X}_{g}\right]\sigma_{g}^{2}\right).
\]
Applying Lemma \ref{lemma:Exact order summability}, together with the order of $\sigma_g^2$ given in \eqref{eq:order of sigma_g^2} from Remark \ref{rem:order of sigma_g^2}, and using the fact that $\mathbb{E}\left[\bar{X}_{g}^{\prime}\bar{X}_{g}\right] = O(1)$ uniformly in $g$ by Assumption \ref{assump: Xbar'Xbar=O(G)}(ii), it follows that
\begin{equation}
\label{eq: order of lambda_max B_G}
   \lambda_{\max}\left(B_G\right) \, = \, \begin{cases}
O\left(1\right), & \text{for strong dependence,}\\
O\left(\frac{1}{G}\sum_{g=1}^{G}\frac{h(N_{g})}{N_{g}}\right), & \text{for semi-strong dependence,}\\
O\left(\frac{1}{G}\sum_{g=1}^{G}\frac{1}{N_{g}}\right), & \text{for weak dependence.}
\end{cases} 
\end{equation}
To determine the order of $\lambda_{\min}\left(B_G\right)$ for strong, semi-strong, and weak cross-sectional dependence, consider the following expression:
\begin{align}
\lambda_{\min}\left(B_G\right) \; \geq & \; \frac{1}{G}\cdot\lambda_{\min}\left(\sum_{g:\sigma_{g}^2\geq c>0}\mathbb{E}\left[\bar{X}_{g}^{\prime}\bar{X}_{g}\right]\sigma_{g}^{2}\right) \label{eq: expansion of lambdamin(BG) term 0}\\
& + \; \frac{1}{G}\cdot\lambda_{\min}\left(\sum_{\underset{\sigma_{g}^2=o(1);\;h(N_{g})\uparrow\infty;\;h(N_g)=o(N_g)}{g:\frac{N_g}{h(N_{g})}\sigma^2_{g}\geq c_{1}>0}}\mathbb{E}\left[\bar{X}_{g}^{\prime}\bar{X}_{g}\right]\sigma_{g}^{2}\right) \label{eq: expansion of lambdamin(BG) term 1}\\
  & +\;\frac{1}{G}\cdot\lambda_{\min}\left(\sum_{\underset{\frac{N_g}{h(N_g)}\sigma^2_{g}=o(1);\;h(N_{g})\uparrow\infty;\;h(N_{g})=o(N_g)}{g:N_{g}\sigma^2_{g}\geq c_{2}>0}}\mathbb{E}\left[\bar{X}_{g}^{\prime}\bar{X}_{g}\right]\sigma_{g}^{2}\right). \label{eq: expansion of lambdamin(BG) term 2}
\end{align}
For non-negative definite matrices $A, B$, and $C$, the inequality $\lambda_{\min}(A+B+C) \geq \lambda_{\min}(A) + \lambda_{\min}(B) + \lambda_{\min}(C)$ holds. Consequently, if one of these minimum eigenvalues asymptotically dominates the others, the lower bound is effectively determined by this leading term, ignoring the lower-order terms. Here, if the number of strongly dependent clusters is of order $O_e(G),$ then only the term on the right-hand side in \eqref{eq: expansion of lambdamin(BG) term 0} dominates, and we have $\lambda_{\min}\left(B_G\right)=O_e\left(1\right).$ If the number of clusters that are semi-strongly dependent is of order $O_e(G),$ and no strongly dependent clusters are present, then the term in \eqref{eq: expansion of lambdamin(BG) term 0} does not contribute, the term in \eqref{eq: expansion of lambdamin(BG) term 1} dominates, and we have $\lambda_{\min}\left(B_G\right)=O_e\left(\frac{1}{G}\sum_{g=1}^{G}\frac{h(N_{g})}{N_{g}}\right).$ If all the clusters are weakly dependent, then only the term in \eqref{eq: expansion of lambdamin(BG) term 2} contributes, and we have $\lambda_{\min}\left(B_G\right)=O_e\left(\frac{1}{G}\sum_{g=1}^{G}\frac{1}{N_{g}}\right).$ In the preceding arguments, a sufficient condition for the orders to hold is that the average of  $\mathbb{E}\left[\bar{X}_g^{\prime}\bar{X}_g\right]$ over any collection of $O_e(G)$ elements is positive definite. More precisely, for any index set $S_G \subseteq \{1,\dots,G\}$ such that $|S_G| = O_e(G)$, we assume that $\frac{1}{|S_G|} \sum_{g \in S_G} \mathbb{E}\left[\bar{X}_g^{\prime}\bar{X}_g\right]$ is positive definite.
 Therefore, it follows that 
\begin{equation}
\label{eq: order of lambda min B_G}
     \lambda_{\min}(B_G) \,= \,
\begin{cases}
O_{e}\left(1\right), & \text{for strong dependence,}\\
O_{e}\left(\frac{1}{G}\sum_{g=1}^{G}\frac{h(N_{g})}{N_{g}}\right), & \text{for semi-strong dependence,}\\
O_{e}\left(\frac{1}{G}\sum_{g=1}^{G}\frac{1}{N_{g}}\right), & \text{for weak dependence.}
\end{cases}
\end{equation}
Thus, \eqref{eq: order of lambda_max B_G} and \eqref{eq: order of lambda min B_G} provide a convenient framework for analyzing the equality of the orders of the minimum and maximum eigenvalues of the matrix $B_G$. 
To summarize the preceding discussion, we next impose the following condition.

\begin{assumption}\label{assump:lamba_max=lambda_min} 
Assume that $B_G$ is finite and positive definite, and that both $\lambda_{\min}(B_{G})$ and $\lambda_{\max}(B_{G})$ are of the exact order:
\[
\begin{cases}
O_{e}\left(1\right), & \text{for strong dependence,}\\
O_{e}\left(\frac{1}{G}\sum_{g=1}^{G}\frac{h(N_{g})}{N_{g}}\right), & \text{for semi-strong dependence,}\\
O_{e}\left(\frac{1}{G}\sum_{g=1}^{G}\frac{1}{N_{g}}\right), & \text{for weak dependence.}
\end{cases}
\]
\end{assumption}

It follows from Assumption \ref{assump:lamba_max=lambda_min} that
the $k\times k$ matrix ${B_{G}}/{\left\Vert B_{G}\right\Vert}$
is uniformly positive definite for any kind of dependence.

\begin{lemma}\label{lemma: sum h^2/N^2 to 0}
 Under semi-strong and weak dependence, as $G\to\infty,$ the following hold, respectively, 
\begin{equation} \label{eq: sum h^2/N^2 to 0}
    \frac{\sum_{g=1}^{G} \frac{h^{2}(N_{g})}{N_{g}^{2}}}{\left( \sum_{g=1}^{G} \frac{h(N_{g})}{N_{g}} \right)^{2}} \to 0 \quad \text{ and } \quad \frac{\sum_{g=1}^{G} \frac{1}{N_{g}^{2}}}{\left( \sum_{g=1}^{G} \frac{1}{N_{g}} \right)^{2}} \to 0
\end{equation}
for (i) the nearly balanced case and (ii) the unbalanced case with $O(G)$ clusters being nearly balanced.
\end{lemma}
The proof of Lemma \ref{lemma: sum h^2/N^2 to 0} is given in the Appendix. The following theorem establishes the asymptotic distribution of the estimator $\hat{\beta}_A$ under general dependence conditions.

\begin{theorem}
\label{thm:normal}
Under Assumptions \ref{assump:exogeneity}, \ref{assump:error independence}, \ref{assump:additional assumption on errors} with $p=2$,  \ref{assump: Xbar'Xbar=O(G)}, and \ref{assump:lamba_max=lambda_min}, for strong, semi-strong and weak dependence, for nearly balanced clusters, as $G\to\infty$, $V_{G}^{-1/2}\sqrt{G}\left(\hat{\beta}_{A}-\beta\right)\xrightarrow{d}N_{k}\left(0,I_{k}\right),$ where $V_G=A^{-1}_G B_G A^{-1}_G.$ For the unbalanced case, the asymptotic normality result holds for: (i) strong dependence generally, (ii) weak or semi-strong dependence, if $O(G)$ clusters are nearly balanced.
\end{theorem}

The proof of Theorem \ref{thm:normal} is given in the Appendix.
The asymptotic normality result remains valid under nearly balanced clusters without requiring Assumption \ref{assump:additional assumption on errors}(ii). Note that for strong dependence, the asymptotic normality holds without additional assumptions on cluster sizes. However, for weak or semi-strong dependence, the additional restrictions follow from Lemma \ref{lemma: sum h^2/N^2 to 0}. Once the asymptotic normality of the proposed estimator has been established, we now address the consistency of the variance estimator.

\begin{theorem}
\label{thm:consistency of V}
Under Assumptions \ref{assump:exogeneity}, \ref{assump:error independence}, \ref{assump:additional assumption on errors} with $p=2$,  \ref{assump: Xbar'Xbar=O(G)}, and \ref{assump:lamba_max=lambda_min}, for nearly balanced clusters, under strong, semi-strong and weak dependence, $\hat{V}_G $ is consistent for $V_G$, where $$\hat{V}_G=\left(\frac{\bar{X}'\bar{X}}{G}\right)^{-1} \hat{B}_G \left(\frac{\bar{X}'\bar{X}}{G}\right)^{-1}, \qquad \hat{B}_{G}=\frac{1}{G}\sum_{g=1}^{G}\bar{X_{g}}^{\prime}\bar{X_{g}}e_{g}^{2},$$  and $e_{g}=\bar{Y}_{g}-\bar{X}_{g}\hat{\beta}_{A}.$  For the unbalanced case, the consistency holds for: (i) strong dependence generally, (ii) weak or semi-strong dependence, if $O(G)$ clusters are nearly balanced.

\end{theorem}

The proof of Theorem \ref{thm:consistency of V} is given in the Appendix.


\begin{remark}
    The asymptotic normality of $\hat{\beta}_A$ and the consistency of $\hat{V}_G$ hold for weak or semi-strong dependence for unbalanced clusters, even if $O(G)$ clusters (say, $G/2$ clusters) are extremely large and $O(G)$ clusters (say, $G/2$ clusters) are nearly balanced.
\end{remark}

Note that the Lyapunov condition for the asymptotic normality of $\hat{\beta}_{A}$ reduces to
\[
\begin{cases}
O\left(G^{-1}\right), & \text{for strong dependence,}\\
O\left(\frac{\sum_{g=1}^{G}\frac{h^{2}(N_{g})}{N_{g}^{2}}}{\left(\sum_{g=1}^{G}\frac{h(N_{g})}{N_{g}}\right)^{2}}\right), & \text{for semi-strong dependence,}\\
O\left(\frac{\sum_{g=1}^{G}\frac{1}{N_{g}^{2}}}{\left(\sum_{g=1}^{G}\frac{1}{N_{g}}\right)^{2}}\right), & \text{for weak dependence,}
\end{cases}
\]
as in \eqref{eq: Lyapunov condition orders for beta hat} given in the proof of Theorem \ref{thm:normal} provided in the Appendix. Also, for the consistency of $\hat{V}_G$, note that $\frac{1}{\left\Vert B_{G}\right\Vert}\left\Vert \hat{B}_{G}-B_{G} \right\Vert$ reduces to
\[
\frac{1}{\left\Vert B_{G}\right\Vert}\left\Vert \hat{B}_{G}-B_{G} \right\Vert=\begin{cases}
O_{P}\left(\frac{1}{\sqrt{G}}\right), & \text{for strong dependence,}\\
O_{P}\left(\frac{\sqrt{\sum_{g=1}^{G}\frac{h^{2}(N_{g})}{N_{g}^{2}}}}{\sum_{g=1}^{G}\frac{h(N_{g})}{N_{g}}}\right), & \text{for semi-strong dependence,}\\
O_{P}\left(\frac{\sqrt{\sum_{g=1}^{G}\frac{1}{N_{g}^{2}}}}{\sum_{g=1}^{G}\frac{1}{N_{g}}}\right), & \text{for weak dependence,}
\end{cases}
\]
as explained in \eqref{eq: consistency of V orders} from the proof of Theorem \ref{thm:consistency of V} given in the Appendix. The above discussion is summarized in the following remark.

\begin{remark}\label{Rem:asympnorm and consistency of V for unbalanced}
\begin{enumerate}
    \item   The asymptotic normality of $\hat{\beta}_A$ and the consistency of $\hat{V}_G$ are preserved in the unbalanced case under strong dependence.
    \item   For the unbalanced case, if all the clusters are semi-strongly dependent, then it requires $\frac{\sum_{g=1}^G  \frac{h^2(N_g)}{N_g^2}}{\left(\sum_{g=1}^G  \frac{h(N_g)}{N_g}\right)^2}\rightarrow 0,$ for the asymptotic normality of $\hat{\beta}_A$ and the consistency of $\hat{V}_G.$
    \item   For the unbalanced case, if all the clusters are weakly dependent, then $\frac{\sum_{g=1}^G  \frac{1}{N_g^2}}{\left(\sum_{g=1}^G  \frac{1}{N_g}\right)^2}\rightarrow 0$ is required for the asymptotic normality of $\hat{\beta}_A$ and the consistency of $\hat{V}_G.$
\end{enumerate}
\end{remark}

The condition ${\sum_{g=1}^G  \frac{h^2(N_g)}{N_g^2}} \big/ {\left(\sum_{g=1}^G  \frac{h(N_g)}{N_g}\right)^2}\rightarrow 0$ is equivalent to  ${\max_{g\leq G} \frac{h(N_g)}{N_g}} \big/ {\sum_{g=1}^G  \frac{h(N_g)}{N_g}}\rightarrow 0,$ since 
\[
\left(\frac{\max_{g\leq G} \frac{h(N_g)}{N_g}}{\sum_{g=1}^G  \frac{h(N_g)}{N_g}}\right)^2 \leq \frac{\max_{g\leq G} \frac{h^2(N_g)}{N^2_g}}{\left(\sum_{g=1}^G  \frac{h(N_g)}{N_g}\right)^2} \leq \frac{\sum_{g=1}^G  \frac{h^2(N_g)}{N_g^2}}{\left(\sum_{g=1}^G  \frac{h(N_g)}{N_g}\right)^2} \leq \frac{\max_{g\leq G} \frac{h(N_g)}{N_g}}{\sum_{g=1}^G  \frac{h(N_g)}{N_g}}.
\]
In particular, setting $h(N_g)=1$ yields the analogous condition for weak dependence.

\begin{remark}
Consider the case of semi-strong dependence and the unbalanced
clusters case, where there are $L$ extremely large clusters. If the type of semi-strong dependence changes with clusters, i.e., instead of $h(N_{g}),$ we have $h_{g}(N_{g}),$ and
without loss of generality, we assume that $\frac{h_{1}(N_{1})}{N_{1}}<\frac{h_{2}(N_{2})}{N_{2}}<\cdots<\frac{h_{L}(N_{L})}{N_{L}}$ for the extremely large clusters, and $h_{g}(N)\simeq h(N)$ for the remaining clusters, then for the asymptotic normality of $\hat{\beta}_A$ and the consistency of $\hat{V}_G$ to hold, the condition boils down to $\frac{L}{G-L}\frac{h_{L}^{2}(N_{L})N^{2}}{N_{L}^{2}h^{2}(N)}=O(1).$
Therefore, if $L$ is finite or $O(G)$, then one just needs $\frac{h_{L}(N_{L})N}{N_{L}h(N)}=O(1),$
since $\frac{L}{G-L}=O(1)$ and $\frac{N}{N_{L}}=o(1).$
\end{remark}

Theorem \ref{thm:normal} and Theorem \ref{thm:consistency of V} provide a framework for developing a Wald-type test using our proposed estimator in the general linear hypothesis problem. The following theorem states our proposed test.

\begin{theorem}
\label{thm:testing}
Let $\Delta_{G}^{l\times l}=RV_{G}R^{\prime}$
and $\hat{\Delta}_{G}^{l\times l}=R\hat{V}_{G}R^{\prime}$. Then,
under $H_{0}:R\beta=r,$ under Assumptions \ref{assump:exogeneity}, \ref{assump:error independence}, \ref{assump:additional assumption on errors} with $p=2$,  \ref{assump: Xbar'Xbar=O(G)}, and \ref{assump:lamba_max=lambda_min}, for nearly balanced clusters, 
\[
G\left(R\hat{\beta}_{A}-r\right)^{\prime}\hat{\Delta}_{G}^{-1}\left(R\hat{\beta}_{A}-r\right)\xrightarrow{d}\chi_{l}^{2},\quad \text{ as } G\to\infty.
\]

For the unbalanced case, the above holds for: (i) strong dependence generally, (ii) weak or semi-strong dependence, if $O(G)$ clusters are nearly balanced.
\end{theorem}

The proof of Theorem \ref{thm:testing} is given in the Appendix.
The test described in Theorem \ref{thm:testing} can be used to determine whether explanatory variables in a model are significant, for cross-sectional data with cluster dependence.

\begin{remark} \label{rem:bounded_clusters}
    Our theoretical framework also accommodates scenarios with bounded cluster sizes, that is, $N_g$ is bounded above by a finite constant for all $g$. Bounded cluster sizes prevent the covariance matrix eigenvalues from diverging, so these clusters fall strictly under the regime of weak dependence. As a result, all of our main theoretical results, including parameter consistency, asymptotic normality, the consistency of the variance estimator, and the validity of the test statistic, remain valid.

    However, there are two important differences regarding the asymptotic behavior in this setting. Firstly, the rate of convergence for consistency is faster under weak dependence when cluster sizes grow large compared to the bounded cluster size case. Secondly, establishing asymptotic normality and consistency of the variance estimator for bounded clusters under weak dependence is mathematically less restrictive, since the conditions in \eqref{eq: sum h^2/N^2 to 0} reduce to $O(1/G),$ which vanishes as $G\to\infty$. Consequently, similar arguments ensure the validity of the Wald-type test established in Theorem \ref{thm:testing}.
\end{remark}

The following subsection presents a comparison between the proposed estimator and the POLS estimator.

\subsection{Efficiency}

In addition to consistency and asymptotic normality, it is of interest to compare the efficiency of the proposed estimator, $\hat{\beta}_A$, with that of the standard pooled ordinary least squares (POLS) estimator. 

To illustrate the relative efficiency properties, consider a data-generating process characterized by an asymptotically dominant cluster. Suppose that for each cluster $g=1,\dots,G$, the error covariance matrix exhibits equicorrelation, given by
\[
\Omega_g = (a-b) I_{N_g} + b \mathbf{1}\mathbf{1}^{\prime},
\]
where $0 < b < a < \infty$. This structure introduces strong within-cluster dependence, as all observations within a given cluster are positively correlated. Furthermore, assume the first cluster is asymptotically dominant in size, such that $N_1 \to \infty$, while the remaining clusters are of size $N_g = n$ (not necessarily fixed) for $g=2,3,\dots,G$, with $\frac{n}{N_1} \to 0.$ Thus, one cluster is substantially larger than the rest.

Under this specified structure, the variance of the POLS estimator depends on the standard design matrix sum $\sum_{g=1}^G X_g^\prime X_g = O_e(N_1+nG)$ and the cross-product term driven by within-cluster correlation, $\sum_{g=1}^G X^\prime_g \mathbf{1} \mathbf{1}^\prime X_g = O_e(N_1^2+n^2G)$. Combining these terms yields the scalar convergence rate:
\[
\operatorname{Var}(\hat{\beta}_{POLS}) = O_e\left(\frac{N_1^2+n^2G}{(N_1+nG)^2}\right).
\]

In contrast, the variance of the proposed averaged estimator depends on cluster-level averages. Since the averaged design matrix sum is $\sum_{g=1}^G \bar{X}_g^\prime \bar{X}_g = O_e(G)$, and $\sum_{g=1}^G \bar{X}_g^\prime \bar{X}_g \left[\frac{a-b}{N_g}+b\right] = O_e(G)$, we have
\[
\operatorname{Var}(\hat{\beta}_A) = O_e\left(\frac{1}{G}\right).
\]

\begin{result}\label{result:efficiency}
Under the specified equicorrelated error structure with a dominant cluster, $\hat{\beta}_A$ is asymptotically more efficient than $\hat{\beta}_{POLS}$. The two estimators are asymptotically equally efficient if and only if $\frac{N_1^2}{n^2 G} \to 0$.
\end{result}

This result is formalized by analyzing the asymptotic efficiency ratio, $\rho$, of their scalar convergence rates. Substituting the derived orders yields:
$$
\rho \, = \, O\left( \frac{N_{1}^{2}G+n^2 G^{2}}{(N_{1}+nG)^{2}} \right) \, = \, O\left( \frac{1+\frac{N_1^2}{n^2 G}}{1+\frac{N_1^2}{n^2 G^2}+\frac{2N_1}{nG}} \right).
$$

Evaluating the limit of $\rho$ as $G \to \infty$ and $N_1 \to \infty$ reveals three distinct efficiency regimes:
$$
\rho \to 
\begin{cases}
1, & \text{if } \displaystyle \frac{N_1^2}{n^2 G} \to 0, \\[10pt]
1 + c, & \text{if } \displaystyle \frac{N_1^2}{n^2 G} \to c \in (0,\infty), \\[10pt]
\infty, & \text{if } \displaystyle \frac{N_1^2}{n^2 G} \to \infty.
\end{cases} 
$$

In the first regime, where $\frac{N_1^2}{n^2 G} \to 0$, the influence of the dominant cluster vanishes asymptotically, resulting in equal efficiency ($\rho \to 1$) between the two estimators. In the second regime, the dominant cluster grows at a proportional rate to the remaining sample such that $\frac{N_1^2}{n^2 G} \to c$. Here, $\hat{\beta}_A$ achieves strictly greater asymptotic precision by a constant factor. Finally, consider the third regime where $\frac{N_1^2}{n^2 G} \to \infty.$ In this case, the dominant cluster becomes sufficiently large to determine the overall variance, making the POLS estimator highly inefficient, with $\rho \to \infty$. Since $\rho \geq 1$ in all three regimes, the proposed estimator $\hat{\beta}_A$ is always at least as efficient as $\hat{\beta}_{POLS}$, and in many cases it is strictly more efficient.

\subsection{Results under classical measurement error}

Many economic data sets are susceptible to being contaminated with measurement error.  The presence of measurement errors is known to produce biased and inconsistent estimates, with a possibility of a high amount of bias (attenuation bias). The whole inference may break down. The standard solution is to find valid instruments. Finding valid instruments is a painful and debatable task. However, simple averaging circumvents the search for IVs.

Let us consider the model where independent errors are present in the observations of the regressors. Let $X_{g}^{*}$ be the true regressor variable and $X_{g}$ be the observed regressor with some measurement error $\Gamma_{g},$ where $\Gamma_g=\left(\left(\gamma_{gij}:i=1,2,\ldots, N_g; \; j=1,2,\ldots ,k\right)\right)^{N_g\times k}=\left[\gamma_{g\cdot 1}, \gamma_{g\cdot 2}, \ldots, \gamma_{g\cdot k}\right].$  Then we have the following model:
\begin{align}
Y_{g} & =X_{g}^{*}\beta+\epsilon_{g}, \text{ where } X_{g}=X_{g}^{*}+\Gamma_{g} \nonumber \\ 
 & =X_{g}\beta+\eta_{g},\mbox{ where \ensuremath{\eta_{g}=\epsilon_{g}- \Gamma_{g}  \beta}} \label{eq:endogeneity model}
\end{align}

In the above model, the rows of $\Gamma_{g}$ are $ k$-dimensional random errors, which are independent between clusters and are independently distributed with $\epsilon_{gi}$. Now, let us consider some assumptions on $X^*_{gi}$ and $\gamma_{gi}$, which are somewhat similar to the previous assumptions for $X_{gi}$ and $\epsilon_{gi}$.

\begin{assumption}\label{assump:exogeneity for measurement error}    
    We assume that $X_g^*$ is exogenous 
    with respect to both the measurement error $\gamma_{g\cdot j}$ and 
    the structural error $\epsilon_g$, for all $j$ and $g$. That is,
    \[
        \mathbb{E}\left[\gamma_{g\cdot j} \mid X_g^* \right] = 0,
        \qquad
        \mathbb{E}\left[\epsilon_g \mid X_g^* \right] = 0,
        \qquad \forall j,g.
    \]
\end{assumption}

This assumption requires that both the measurement errors and the 
structural disturbances are mean independent of the true regressors 
within each cluster. In particular, the measurement error is assumed to be classical in the sense of having zero conditional mean given the true covariates, and the structural error contains no systematic component predictable from $X_g^*$. Furthermore, if the sequence $\{(X_g^*,\Gamma_g,\epsilon_g)\}_{g=1}^G$ 
is independent across $g$, then the above conditions are equivalent to $ \mathbb{E}\left[\gamma_{g\cdot j} \mid X^* \right] = 0,$ and $ \mathbb{E}\left[\epsilon_g \mid X^* \right] = 0, $ for all $j,g,$ where $X^*=\left(X^{*\prime}_{1},X^{*\prime}_{2},\ldots,X^{*\prime}_{G}\right)^\prime$, since conditioning on the full matrix 
$X^*$ does not provide additional information beyond $X_g^*$ under 
cross-cluster independence.

\begin{assumption}\label{assump:error independence for measurement error}
Assume that the sequence $\{(X_g^*,\Gamma_g,\epsilon_g)\}_{g=1}^G$ 
is independent across $g$.

(i) For each $g = 1, \ldots, G$, the conditional covariance of the structural error satisfies
\begin{align*}
    \mathbb{E}\left[\epsilon_g \epsilon_g^\prime \mid X_g^* \right] &= \Omega_g, \\
    \mathbb{E}\left[\epsilon_g \epsilon_{g_1}^\prime \mid (X_g^*, X_{g_1}^*) \right] &= 0, \quad  g_1 \neq g,
\end{align*}
where $\Omega_g$ is an $N_g \times N_g$ positive definite matrix, 
independent of $X_g^*$. Moreover, uniformly in $g$, the maximum eigenvalue 
of $\Omega_g$ satisfies
\[
\lambda_{\max}(\Omega_g) =
\begin{cases}
O_e(N_g), & \text{under strong dependence}, \\
O_e\left(h(N_g)\right), & \text{under semi-strong dependence}, \\
O_e(1), & \text{under weak dependence},
\end{cases}
\]
as defined in Definition \ref{def:dependency rates} and in the sense of 
Definition \ref{def:exact order and order uniformly}.

(ii) For each $g = 1, \ldots, G$, and $j,l=1,\ldots,k$, the conditional cross-covariance of the measurement errors satisfies
\begin{align*}
    \mathbb{E}\left[\gamma_{g\cdot j}\gamma_{g\cdot l}' \mid X_g^* \right] &= \Lambda_{g,jl}, \\
    \mathbb{E}\left[\gamma_{g\cdot j}\gamma_{g_1\cdot l}' \mid (X_g^*, X_{g_1}^*) \right] &= 0, \quad  g_1 \neq g.
\end{align*}
For $j=l$, the marginal covariance matrix $\Lambda_{g,jj}$ is an $N_g \times N_g$ positive definite matrix, independent of $X_g^*$, with uniformly bounded diagonal elements. Furthermore, the maximum eigenvalue of $\Lambda_{g,jj}$ satisfies \footnote{From the practitioner's point of view, it is somewhat unlikely that $\gamma_{g\cdot j}$ is strongly dependent. $\gamma_{g\cdot j }$ to be strongly dependent, (almost) all the individuals unitedly have to misreport at the time of the survey.}  
\[
\lambda_{\max}(\Lambda_{g,jj})
=
\begin{cases}
O_e(N_g), & \text{under strong dependence}, \\
O_e\left(h_\gamma(N_g,j)\right), & \text{under semi-strong dependence}, \\
O_e(1), & \text{under weak dependence},
\end{cases}
\]
uniformly in $j$ and $g$, with $h_\gamma(N_g,j)/N_g \to 0$ as $N_g \to \infty$ for each $j$.
\end{assumption}

This assumption specifies the second-moment structure of both the 
structural errors and the measurement errors in the clustered setting. 
Part (i) allows for arbitrary heteroskedasticity and within-cluster 
dependence on the structural disturbances, as summarized by the 
positive definite matrix $\Omega_g$, while imposing independence across clusters. The growth rate of $\lambda_{\max}(\Omega_g)$ characterizes the strength of within-cluster dependence according to Definition \ref{def:dependency rates}. Part (ii) imposes an analogous structure on the measurement errors. Within each cluster, the measurement errors may exhibit arbitrary dependence summarized by $\Lambda_{g,jj}$, while remaining conditionally uncorrelated across clusters. The boundedness of the diagonal elements ensures that individual-level measurement error variances remain controlled, and the order of 
$\lambda_{\max}(\Lambda_{g,jj})$ determines the degree of within-cluster 
dependence in the measurement errors. Note that $\left\Vert \Lambda_{g,jl}\right\Vert \leq \left\Vert \Lambda_{g,jj}\right\Vert^{1/2} \left\Vert \Lambda_{g,ll}\right\Vert^{1/2}$. Therefore, the spectral norm of the cross-covariance matrix $\Lambda_{g,jl}$ is bounded by the maximum eigenvalues of the corresponding marginal covariance matrices. Finally, under the independence of 
$\{(X_g^*,\Gamma_g,\epsilon_g)\}_{g=1}^G$ across $g$, conditioning on the 
entire regressor matrix $X^*$ instead of 
$(X_g^*,X_{g_1}^*)$ does not alter the meaning of the assumption.

\begin{assumption}\label{assump: Xg'Xg=O(Ng) for measurement error}
(i) Assume that $C_{G}^{*}=\mathbb{E}\left[\frac{X^{*\prime}X^{*}}{\sum_{g=1}^{G}N_{g}}\right]$
is finite and uniformly positive definite over $G.$ Also, assume
that 
\[
\frac{X^{*\prime}X^{*}}{\sum_{g=1}^{G}N_{g}}-C_{G}^{*}=o_{P}(1).
\]

(ii) Assume that
\[
\mathbb{E}\left[\left(\operatorname{tr}\left(X_g^{*\prime}X^*_{g}\right)\right)\right]=O\left(N_{g}\right) \; \text{ uniformly in } g,
\]
as in Definition \ref{def:exact order and order uniformly}.
\end{assumption}


For the model \eqref{eq:endogeneity model}, the pooled ordinary least squares estimate for $\beta$ is 
\begin{eqnarray}   
    \hat{\beta}_{POLS}=(X^{\prime}X)^{-1}\sum_{g=1}^G  X_g^{\prime}Y_g=\beta+(X^{\prime}X)^{-1}\sum_{g=1}^G  X_g^{\prime}\eta_g. \label{eq:beta hat ols for measurement error}
\end{eqnarray}

Let us consider another additional assumption on $\Gamma_g$, which will be necessary for the probability convergence of $\Gamma_g^\prime \Gamma_g.$
\begin{assumption}\label{assump: convergence of Gamma'Gamma}
    Assume that as $G \to \infty$,
\begin{enumerate}
    \item[(i)] $\frac{1}{\sum_{g=1}^G N_g}
        \sum_{g=1}^G \mathbb{E}\left[\Gamma_g^\prime \Gamma_g \big| X^*_g
    \right]\rightarrow C_0,$ where $C_0$ is a diagonal and positive definite matrix.

    \item[(ii)] $\frac{1}{\left(\sum_{g=1}^G N_g\right)^2}
    \sum_{g=1}^G
    \mathbb{E}\left[
        \operatorname{tr}
        \Big(
        \Gamma_g^\prime \Gamma_g
        -
        \mathbb{E}[\Gamma_g^\prime \Gamma_g]
        \Big)^2
        \,\Big|\, X^*_g
    \right]
    \;\rightarrow\; 0.$
\end{enumerate}
\end{assumption}
Assumption \ref{assump: convergence of Gamma'Gamma} imposes a conditional law-of-large-numbers 
requirement on the aggregated measurement error quadratic form. 
Part (i) ensures that the normalized second moment of the measurement 
errors converges to a fixed diagonal positive definite matrix, 
preventing degeneracy or explosive growth asymptotically. 
Part (ii) guarantees that the conditional variance of the normalized sum 
vanishes, so that $\frac{1}{\sum_{g=1}^G N_g}\sum_{g=1}^G \Gamma_g^\prime \Gamma_g$ converges in probability (conditionally on $X^*$) to $C_0$ as $G \to \infty$.

\begin{result}\label{result:Inconsistent_ols_endogeneity}
Under Assumptions \ref{assump:exogeneity for measurement error}, \ref{assump:error independence for measurement error},  \ref{assump: Xg'Xg=O(Ng) for measurement error} and \ref{assump: convergence of Gamma'Gamma}, in the presence
of classical measurement error, $\hat{\beta}_{POLS}$ is biased and inconsistent for $\beta$, under
any kind of dependence for both $\epsilon_{g}$ and $\gamma_{g\cdot j}$.
\end{result}

The proof of this result is given in the Appendix.

Since POLS produces biased and inconsistent parameter estimates in
the presence of classical measurement error, hypothesis tests based on it also become seriously misleading. On the other hand, taking the cluster average for each cluster yields the following model:
\begin{equation} \label{eq:endogeneity model for average}
\bar{Y}_{g} = \bar{X}_{g}\beta + \bar{\eta}_{g},
\end{equation}
where $\bar{X}_{g}=\bar{X}_{g}^{*}+\bar{\gamma}_{g}$ and $\bar{\eta}_{g}=\bar{\epsilon}_{g}-\bar{\gamma}_{g}\beta$, with $\bar{\gamma}_g$ being a $1 \times k$ vector whose $j$-th element is $\frac{1}{N_g}\sum_{i=1}^{N_g} \gamma_{gij}$. Now, let us consider the standard assumptions for $\bar{X}^*_g$ and $\bar{\gamma}_g$.
\begin{assumption}\label{assump: Xbar'Xbar=O(G) for measurement error}
 Assume that $A_{G}^{*}=\mathbb{E}\left[\frac{\bar{X}^{*\prime}\bar{X}^{*}}{G}\right]$
is finite and uniformly positive definite over $G,$ where $\bar{X}^*=\left(\bar{X}^{*\prime}_1,\bar{X}^{*\prime}_2,\ldots,\bar{X}^{*\prime}_G\right)^\prime.$ Also, assume
that
\[
\frac{\bar{X}^{*\prime}\bar{X}^{*}}{G}-A_{G}^{*}=o_{P}(1).
\]
\end{assumption}

Note that for all $j,g,$ $\mathbb{E}\left[\bar{\gamma}_{gj} \big|  X^*_g\right]=0$ and $V\left( \bar{\gamma}_{gj} \big|  X^*_g\right)=\frac{\mathbf{1}^{\prime}\Lambda_{g,jj}\mathbf{1}}{N_{g}^{2}},$ where $\bar\gamma_{gj}$ is the $j^{th}$ component of $\bar \gamma_g$. 
In addition, from the Assumption \ref{assump:error independence for measurement error}, we have 
\begin{eqnarray}
\operatorname{Var}\left( \bar{\gamma}_{gj} \mid  X^*_g\right)=\begin{cases}
    O\left(1\right), & \text{ for strong dependence of } \gamma_{g\cdot j} \\
    O\left(\frac{h_{\gamma}(N_{g},j)}{N_{g}}\right), & \text{ for semi-strong dependence of } \gamma_{g\cdot j} \\
    O\left(\frac{1}{N_{g}}\right), & \text{ for weak dependence of } \gamma_{g\cdot j}
\end{cases} \label{eq:order of gamma gj bar square}
\end{eqnarray}
uniformly in $j,g$ and $X^*_g.$

\begin{assumption}\label{assump:existence of moment for gamma} 
Assume that there exists a constant $0 < c_2 < \infty$ such that
\[
\mathbb{E}\left[
    \left| \gamma_{gij} \right|^{4}
    \,\big|\, X_g^*
\right]
\le c_2,
\qquad
\text{for all } i,j,g,
\]
uniformly in $X_g^*$.
\end{assumption}
This assumption requires that the measurement errors possess uniformly 
bounded conditional fourth moments. In particular, it guarantees that higher-order 
fluctuations of the measurement error terms remain asymptotically controlled.

\begin{assumption}\label{assump: convergence of Gamma bar'Gamma bar for average model}
Suppose that the measurement errors exhibit strong dependence within clusters. Assume that as $G \to \infty$,
\begin{enumerate}
    \item[(i)]
    $\frac{1}{G}
        \sum_{g=1}^G  \mathbb{E}\left[\bar{\gamma}_g^{\prime}\bar{\gamma}_g
        \,\Big|\, X^*_g
    \right]\rightarrow C^*,$
    where $C^*$ is a diagonal and positive definite matrix.

    \item[(ii)] $\frac{1}{G^{2}}
    \sum_{g=1}^G
    \mathbb{E}\left[
        \operatorname{tr}
        \Big(
        \bar{\gamma}_g^{\prime}\bar{\gamma}_g
        -
\mathbb{E}\left[\bar{\gamma}_g^{\prime}\bar{\gamma}_g\right]
        \Big)^{2}
        \,\Big|\, X^*_g
    \right]
    \;\rightarrow\; 0.$
\end{enumerate}
\end{assumption}

This assumption imposes a conditional law-of-large-numbers requirement 
on the cluster-averaged measurement errors under strong within-cluster 
dependence. Part (i) ensures that the cross-cluster average of 
$\bar{\gamma}_g^{\prime}\bar{\gamma}_g$ converges to a fixed diagonal 
positive definite matrix, preventing degeneracy in the limit. 
Part (ii) guarantees that the conditional variance of the normalized sum 
vanishes asymptotically, implying convergence in probability (conditionally 
on $X^*$) of $\frac{1}{G}\sum_{g=1}^G \bar{\gamma}_g^{\prime}\bar{\gamma}_g$ to $C^*$ as $G \to \infty$.

As discussed in Section \ref{Section:proposed}, our proposed estimator is the ordinary least squares estimator on the model \eqref{eq:endogeneity model for average}, which can be written as
\begin{eqnarray}
\hat{\beta}_A=(\bar X^{\prime}\bar X)^{-1}\sum_{g=1}^G  \bar X_g^{\prime}\bar Y_g=\beta+(\bar X^{\prime}\bar X)^{-1}\sum_{g=1}^G  \bar X_g^{\prime}\bar \eta_g. \label{eq:beta hat for measurement error}
\end{eqnarray}

\begin{lemma} \label{lemma: m' error consistency N_g large condition}
(i) For semi-strong dependence, let $f(N_g)$ be a function of $N_g$ such that $f(N_g) = O_e(h(N_g)/N_g)$ for $O(G)$ many clusters, and $f(N_g) = O_e(1/N_g)$ for the remaining clusters. Also, suppose that the number of bounded clusters is $o(G)$, while $N_g \to \infty$ for all remaining clusters. Then, as $G \to \infty$,
\[
O\left(\frac{1}{G} \sum_{g=1}^G f(N_g)\right) \to 0.
\]

(ii) For weak dependence, suppose that the number of bounded clusters is $o(G)$, while $N_g \to \infty$ for all remaining clusters. Then, as $G \to \infty$,
\[
O\left(\frac{1}{G} \sum_{g=1}^G \frac{1}{N_g}\right) \to 0.
\]
\end{lemma}

The proof of this lemma is given in the Appendix.

\begin{result}
\label{result:consistent_our_estimate_endogeneity}
Under Assumptions \ref{assump:exogeneity for measurement error}, \ref{assump:error independence for measurement error}, \ref{assump: Xbar'Xbar=O(G) for measurement error}, \ref{assump:existence of moment for gamma} and \ref{assump: convergence of Gamma bar'Gamma bar for average model}, in the presence of classical measurement error,  regardless of the dependence structure of $\epsilon_{g},$ the estimator $\hat{\beta}_A$ is consistent under weak or semi-strong dependence of the measurement errors provided that the number of bounded clusters is $o(G)$, while $N_g \to \infty$ for all remaining clusters; conversely, $\hat{\beta}_A$ is inconsistent under strong dependence of the measurement errors.
\end{result}

The proof of Result \ref{result:consistent_our_estimate_endogeneity} is given in the Appendix. Note that for this consistency to hold, it is permissible for a subset of clusters to remain finite in size. The requirement is that the total number of these bounded clusters is $o(G)$, while the size of all remaining clusters, $N_g \to \infty$, which is formalized in Lemma \ref{lemma: m' error consistency N_g large condition}. Table \ref{tab:measurement error} summarizes our claim under the classical measurement error for any kind of dependence of $\epsilon_g$.

\begin{table}[htbp]
\centering
\caption{Consistency of $\hat{\beta}_{POLS}$ and $\hat{\beta}_A$ Under Classical Measurement Error for Arbitrary Dependence in $\epsilon_g$.}
\label{tab:measurement error}
\small
\begin{tabular}{lccc}
\hline 
Estimator & Weak Dep. of $\gamma_{g\cdot j}$ & Semi-strong Dep. of $\gamma_{g\cdot j}$ & Strong Dep. of $\gamma_{g\cdot j}$ \\
\hline 
$\hat{\beta}_{POLS}$ & Inconsistent & Inconsistent & Inconsistent \\
$\hat{\beta}_A$ & Consistent & Consistent & Inconsistent \\
\hline 
\end{tabular}
\end{table}

In summary, the presence of measurement error leads to the inconsistency of $\hat{\beta}_{POLS}$ regardless of the dependence structure of $\epsilon_g$. Conversely, the estimator $\hat{\beta}_A$ maintains consistency provided the measurement errors exhibit only weak or semi-strong dependence, irrespective of the dependence in $\epsilon_g$. 

The following section evaluates the performance of both our proposed test and the POLS-based test through applications to simulated and empirical datasets.







\section{Simulation and Empirical Illustration} \label{Section:analysis}
 
This section evaluates the finite-sample performance of the proposed estimator relative to the pooled ordinary least squares (POLS) estimator under a common data-generating process (DGP). We first present a series of
Monte Carlo simulations to assess estimation accuracy and inferential validity under nearly balanced and unbalanced cluster structures, and then turn to an empirical illustration using real-world data.
 
\subsection{Simulation Study}

To examine the finite-sample properties of the proposed methods, we conduct Monte Carlo simulations for a linear model. We generate $R^* = 10{,}000$ replications for each scenario, with each replication drawing a sample from the DGP consisting of $G$ clusters with $N_g$ individuals per cluster. The simulated data are generated as $Y_{gi} = X_{gi}^{\prime}\beta + \epsilon_{gi}$, where $X_{gi}$ is a $(k+1) \times 1$ vector of regressors comprising an intercept and $k$ explanatory variables, and $\beta$ is the corresponding
$(k+1) \times 1$ parameter vector. We consider both $k=1$ and $k=4$ below. The error terms are generated as
$\epsilon_g \sim N_{N_g}(0, \Omega_g)$, with $\Omega_g = M_g M_g^{\prime}$, where $M_g$ is an $N_g \times N_g$ matrix whose entries are drawn independently from $U(-5, 10)$. By construction, $\lambda_{\max}(\Omega_g)$
is of order $N_g$, ensuring strong within-cluster dependence. Cluster averages $\bar{Y}_g$ and $\bar{X}_g$ are computed in each replication, and
the DGP is repeated $R^*$ times while keeping the true parameters fixed.
 
We test the general linear hypothesis $H_0: R\beta = r$ against $H_1: R\beta \neq r$ at a significance level of $\alpha = 0.05$. For the proposed estimator, we compute the cluster-level residuals $e_g = \bar{Y}_g - \bar{X}_g \hat{\beta}_A$, and construct the test
statistic 
\begin{equation}
    T = \left(R\hat{\beta}_{A} - r\right)^{\prime} \left(R\hat{V}R^{\prime}\right)^{-1} \left(R\hat{\beta}_{A} - r\right),
\end{equation}
where the variance estimator is
$\hat{V} = \left(\bar{X}^{\prime}\bar{X}\right)^{-1} \sum_{g=1}^G \bar{X}_{g}^{\prime}\bar{X}_{g}e_{g}^{2}\left(\bar{X}^{\prime}\bar{X}\right)^{-1}$.
For comparison, the standard test statistic based on POLS is
\begin{equation}
    T_{POLS} = \left(R\hat{\beta}_{POLS} - r\right)^{\prime} \left(R\hat{V}_{POLS}R^{\prime}\right)^{-1} \left(R\hat{\beta}_{POLS} - r\right),
\end{equation}
where $\hat{V}_{POLS} = ({X}^{\prime}{X})^{-1} \left( \sum_{g=1}^G {X}_{g}^{\prime}u_{g}u_{g}^{\prime}{X}_{g} \right) ({X}^{\prime}{X})^{-1}$, with residuals $u_g = Y_g - X_g \hat{\beta}_{POLS}$. The empirical size is calculated as the proportion of test statistics falling in the rejection region under $H_0$, and the empirical power is defined analogously under $H_1$. To ensure a meaningful comparison of power between the two estimators, we apply a size correction. The size-corrected critical value is defined as the 95th percentile of the test statistic values simulated under the null hypothesis, and is subsequently used to evaluate the size-corrected empirical power.
 
\subsubsection{Nearly Balanced Clusters}
 
In the first scenario, $N_g$ is drawn uniformly from $\{25, \dots, 50\}$ for all $g$. We begin with a single explanatory variable ($k = 1$), so that the regressors $X_{gi}$ are two-dimensional (an intercept and one slope). The design matrix $X_g$ is constructed with a vector of ones in the first column and $N_g$ independent draws from $N(\mu_g, \omega_g^2)$ in the second column, where $\mu_g \sim U(10, 100)$ and $\omega_g^2 \sim U(200, 300)$.
 
Table~\ref{tab:Balanced} reports results for the unscaled covariance matrix $\Omega_g = M_g M_g^{\prime}$, while Table~\ref{tab:Balanced with scaled
Omega_g} presents results for a scaled covariance matrix that guarantees strong dependence. Here, $\Omega_g$ is defined as $\Omega_g = M_g M_g^{\prime}\, N_g / \lambda_{\max}(M_g M_g^{\prime})$; by construction, the maximum eigenvalue of this $\Omega_g$ is exactly $N_g$.

\begin{table}[htbp] 
\centering
\setlength{\tabcolsep}{4pt}
\caption{Comparison of the POLS and proposed estimator under strong dependence for nearly balanced clusters.}
\vspace{.1cm}
\label{tab:Balanced}
\resizebox{\linewidth}{!}{%
\begin{tabular}{lccccc}
\hline\hline 
$G$ & 25 & 50 & 100 & 200 & 500 \\
\hline 
MSE (Intercept) & 13.44 & 9.5 & 6.08 & 2.637 & 1.12 \\
MSE (Intercept) (POLS) & 14.38 & 10.9 & 6.61 & 3.005 & 1.24 \\
MSE (Slope) & $6.21\times 10^{-5}$ & $2.69\times 10^{-5}$ & $1.71\times 10^{-5}$ & $6.71\times 10^{-6}$ & $3.31\times 10^{-6}$ \\
MSE (Slope) (POLS) & $6.72\times 10^{-5}$ & $3.01\times 10^{-5}$ & $1.83\times 10^{-5}$ & $7.62\times 10^{-6}$ & $3.62\times 10^{-6}$ \\
Size & 0.1128 & 0.0807 & 0.0688 & 0.0559 & 0.0548 \\
Size (POLS) & 0.1247 & 0.0760 & 0.0684 & 0.0563 & 0.0523 \\
Size-corr. crit. value & 6.42 & 4.905 & 4.446 & 4.042 & 4.005 \\
Size-corr. crit. value (POLS) & 6.67 & 4.685 & 4.456 & 4.034 & 3.911 \\
Power & 0.3098 & 0.3211 & 0.4803 & 0.7488 & 0.9905 \\
Power (POLS) & 0.2799 & 0.3019 & 0.4397 & 0.7176 & 0.9839 \\
Size-corr. power & 0.1839 & 0.2214 & 0.4295 & 0.7342 & 0.9884 \\
Size-corr. power (POLS) & 0.1486 & 0.2251 & 0.3867 & 0.7021 & 0.9832 \\
\hline
\end{tabular}%
}
\vspace{1ex}
\par
\raggedright
\small
\textit{Note:} The number of replications is 10,000. The MSE is calculated at $\beta_{0}=(1,0)$. Size is calculated at $\beta_{0}=(1,0)$ and power is calculated at $\beta_{1}=(1,0.08)$.
\end{table}

\begin{table}[htbp] 
\centering
\setlength{\tabcolsep}{4pt}
\caption{Comparison of the POLS and proposed estimator under strong dependence for nearly balanced clusters with scaled dispersion matrix.}
\label{tab:Balanced with scaled Omega_g}
\resizebox{\linewidth}{!}{%
\begin{tabular}{lccccc}
\hline\hline 
$G$ & 25 & 50 & 100 & 200 & 500 \\
\hline 
MSE (Intercept) & 0.167 & 0.1235 & 0.0553 & 0.0276 & 0.0104 \\
MSE (Intercept) (POLS) & 0.168 & 0.1360 & 0.0563 & 0.0277 & 0.0105 \\
MSE (Slope) & $4.12\times 10^{-5}$ & $3.41\times 10^{-5}$ & $1.33\times 10^{-5}$ & $7.66\times 10^{-6}$ & $2.81\times 10^{-6}$ \\
MSE (Slope) (POLS) & $4.17\times 10^{-5}$ & $3.42\times 10^{-5}$ & $1.34\times 10^{-5}$ & $7.63\times 10^{-6}$ & $2.82\times 10^{-6}$ \\
Size & 0.0806 & 0.0712 & 0.0591 & 0.0552 & 0.0503 \\
Size (POLS) & 0.0816 & 0.0743 & 0.0583 & 0.0550 & 0.0499 \\
Size-corr. crit. value & 4.99 & 4.490 & 4.179 & 4.002 & 3.852 \\
Size-corr. crit. value (POLS) & 4.96 & 4.670 & 4.131 & 3.991 & 3.828 \\
Power & 0.1072 & 0.1090 & 0.1393 & 0.1972 & 0.4368 \\
Power (POLS) & 0.1068 & 0.1110 & 0.1398 & 0.1976 & 0.4359 \\
Size-corr. power & 0.0697 & 0.0829 & 0.1232 & 0.1856 & 0.4355 \\
Size-corr. power (POLS) & 0.0701 & 0.0777 & 0.1246 & 0.1855 & 0.4371 \\
\hline
\end{tabular}%
}
\vspace{1ex}
\par
\small
\raggedright
\textit{Note:} The number of replications is 10,000. The MSE is calculated at $\beta_{0}=(1,0)$. Size is calculated at $\beta_{0}=(1,0)$ and power is calculated at $\beta_{1}=(1,0.003)$.
\end{table}

We next extend the design to $k = 4$ explanatory variables, with cluster sizes $N_g$ again drawn uniformly from $\{25, \dots, 50\}$ for all $g$. The 4-dimensional non-constant component of the regressors, denoted $\tilde{X}_{gi}$, is generated from a multivariate normal distribution, $\tilde{X}_{gi} \sim N_k(\mu_g \mathbf{1}, \omega_g^2 I_k)$, where $k = 4$, $\mu_g \sim U(10, 100)$, and $\omega_g^2 \sim U(200, 300)$. The full $5 \times 1$ regressor vector is then constructed by placing an intercept in the first coordinate, yielding $X_{gi} = (1, \tilde{X}_{gi}^{\prime})^{\prime}$. We test the joint null hypothesis $H_0: R\beta = r$ against $H_1: R\beta \neq r$, where the restriction matrix is $R = [0_{4\times 1}, I_4]$ and $r = (0.1, 0.2, 0.3, 0.4)^{\prime}$.

\begin{table}[ht]
\centering
\caption{Empirical Size and Power of the Proposed and POLS Tests under Strong Dependence for Nearly Balanced Clusters ($k=4$).}
\label{tab:Balanced with k=4}
\begin{tabular}{lccc}
\hline 
G & $100$ & $200$ & $500$ \\
\hline 
Size & 0.090 & 0.073 & 0.073 \\
Size (POLS) & 0.074 & 0.066 & 0.074 \\
Size-corrected critical value & 11.6 & 10.3 & 10.7 \\
Size-corrected critical value (POLS) & 10.6 & 10.5 & 10.2 \\
Power & 0.260 & 0.476 & 0.858 \\
Power (POLS) & 0.260 & 0.485 & 0.882 \\
Size-corrected power & 0.178 & 0.431 & 0.804 \\
Size-corrected power (POLS) & 0.201 & 0.418 & 0.852 \\
\hline 
\end{tabular}
\vspace{1ex}
\par
\raggedright
\small \textit{Note:} The number of replications is $10,000$. The empirical size is calculated under the null hypothesis with $\beta_{0}=(1, 0.1, 0.2, 0.3, 0.4)^{\prime}$, and the empirical power is calculated under the alternative hypothesis with $\beta_{1}=(1, 0.2, 0.2, 0.3, 0.4)^{\prime}$.
\end{table}

Table~\ref{tab:Balanced with k=4} shows that, for nearly balanced clusters, the empirical size, power, and size-corrected power of the proposed test are comparable to those of the traditional POLS-based test, and that both size and power converge toward their theoretical target values as the number of clusters increases. The mean squared errors of the proposed and POLS estimators are likewise similar in this setting, indicating that the two approaches perform equivalently when clusters are of comparable size. Next, we explore the results for scenarios featuring highly unbalanced data, specifically where a single cluster is exceptionally large.

\subsubsection{Unbalanced Clusters With One Large Cluster}

In the second scenario, the data are partitioned so that the first cluster is substantially larger than the others. For $g = 2, 3, \dots, G$, the cluster sizes $N_g$ are drawn uniformly from $\{4, \dots, 10\}$.
 
The non-constant component of the regressors for the dominant cluster is constructed using the leading eigenvector of its error covariance matrix. Let $p_1^{(\max)}$ denote the eigenvector corresponding to $\lambda_{\max}(\Omega_1)$, and define the base vector $X_1^* = \text{sgn}(p_1^{(\max)})$. The regressor matrix for the first cluster is then $X_1 = [\mathbf{1}, c_1 X_1^*]$, where each element $c_{1j} \sim U(2, 10)$ for $j = 1, \dots, N_1$. For the remaining clusters ($g \geq 2$), the regressor matrices $X_g$ are constructed as in the balanced case, using independent draws from $N(\mu_g, \omega_g^2)$ with $\mu_g \sim U(10, 100)$ and $\omega_g^2 \sim U(200, 300)$. The model retains a single explanatory variable ($k = 1$).
 
Tables~\ref{tab:Unbalanced_1} and~\ref{tab:Unbalanced_2} compare the performance of the proposed estimator and test against the POLS estimator and its associated Wald test. We report the mean squared error (MSE) for both estimators, along with the empirical size and power of the corresponding tests, across a range of values of $G$ and $N_1$.\footnote{We also considered an alternative data-generating process that introduces intra-cluster dependence in the regressors, $X_g$, using a random coefficients structure. Because this specification yields similar conclusions, the corresponding simulation results are omitted for brevity.}

\begin{table}[htbp]
\centering
\setlength{\tabcolsep}{4pt}
\caption{Comparison of the POLS and proposed estimator under strong dependence for one large cluster.}
\label{tab:Unbalanced_1}
\resizebox{\linewidth}{!}{%
\begin{tabular}{lccccc}
\hline\hline 
$(G, N_1)$ & (25, 500) & (25, 1500) & (50, 500) & (50, 1500) & (100, 500) \\
\hline 
MSE (Intercept) & 159.46 & 317.22 & 39.17 & 122.85 & 9.745 \\
MSE (Intercept) (POLS) & 2421.32 & 7594.11 & 2181.78 & 7257.25 & 1666.94 \\
MSE (Slope) & 0.0620 & 0.0702 & 0.0088 & 0.0290 & 0.0025 \\
MSE (Slope) (POLS) & 0.4637 & 1.515 & 0.4520 & 1.590 & 0.3940 \\
Size & 0.0755 & 0.0447 & 0.0480 & 0.0420 & 0.0439 \\
Size (POLS) & 0.9419 & 0.9683 & 0.9530 & 0.9690 & 0.9542 \\
Size-corr. crit. value & 4.691 & 3.675 & 3.678 & 3.571 & 3.638 \\
Size-corr. crit. value (POLS) & 202.36 & 278.99 & 95.92 & 375.27 & 28.44 \\
Power & 0.2078 & 0.1429 & 0.2930 & 0.2110 & 0.5299 \\
Power (POLS) & 0.9379 & 0.9528 & 0.9420 & 0.9750 & 0.9398 \\
Size-corr. power & 0.1582 & 0.1529 & 0.3070 & 0.2450 & 0.5489 \\
Size-corr. power (POLS) & 0.1358 & 0.0073 & 0.1820 & 0.1060 & 0.3377 \\
\hline 
\end{tabular}%
}
\vspace{1ex}
\par
\raggedright
\small
\textit{Note:} The number of replications is 10,000. The MSE is calculated at $\beta_{0}=(1,0)$. Size is calculated at $\beta_{0}=(1,0)$ and power is calculated at $\beta_{1}=(1,0.08)$.
\end{table}

\begin{table}[htbp]
\centering
\setlength{\tabcolsep}{4pt}
\caption{Comparison of the POLS and proposed estimator under strong dependence for one large cluster (extended sample sizes).}
\label{tab:Unbalanced_2}
\resizebox{\linewidth}{!}{%
\begin{tabular}{lccccc}
\hline\hline 
$(G, N_1)$ & (100, 1500) & (200, 500) & (200, 1500) & (500, 500) & (500, 1500) \\
\hline 
MSE (Intercept) & 29.33 & 3.99 & 8.30 & 1.155 & 1.97 \\
MSE (Intercept) (POLS) & 6592.31 & 1217.71 & 5665.02 & 618.57 & 4214.31 \\
MSE (Slope) & 0.0057 & 0.0011 & 0.0021 & $2.9\times 10^{-4}$ & $4.7\times 10^{-4}$ \\
MSE (Slope) (POLS) & 1.588 & 0.284 & 1.391 & 0.140 & 0.945 \\
Size & 0.0390 & 0.0453 & 0.0316 & 0.0468 & 0.0350 \\
Size (POLS) & 0.9780 & 0.9449 & 0.9836 & 0.3853 & 0.9873 \\
Size-corr. crit. value & 3.321 & 3.703 & 3.239 & 3.753 & 3.336 \\
Size-corr. crit. value (POLS) & 209.50 & 13.156 & 57.10 & 5.119 & 16.30 \\
Power & 0.3940 & 0.8090 & 0.6299 & 0.9864 & 0.9089 \\
Power (POLS) & 0.9780 & 0.8539 & 0.9798 & 0.5382 & 0.9563 \\
Size-corr. power & 0.4460 & 0.8200 & 0.6719 & 0.9869 & 0.9193 \\
Size-corr. power (POLS) & 0.1350 & 0.2232 & 0.6065 & 0.5026 & 0.4323 \\
\hline 
\end{tabular}%
}
\vspace{1ex}
\par
\raggedright
\small
\textit{Note:} The number of replications is 10,000. The MSE is calculated at $\beta_{0}=(1,0)$. Size is calculated at $\beta_{0}=(1,0)$ and power is calculated at $\beta_{1}=(1,0.08)$.
\end{table}

We again consider a scaled version of $\Omega_g$ that guarantees strong dependence. Here, $\Omega_g$ is generated as $\Omega_g = M_g M_g^{\prime}\, N_g / \lambda_{\max}(M_g M_g^{\prime})$, under which $\lambda_{\max}(\Omega_g) = N_g$ in Tables \ref{tab:Unbalanced_1 scaled Omega_g} and \ref{tab:Unbalanced_2_scaled Omega_g}. We evaluate
power at a smaller alternative, $\beta_1 = (1, 0.003)^{\prime}$, keeping all other parameters identical to the previous setup.

\begin{table}[htbp]
\centering
\setlength{\tabcolsep}{4pt}
\caption{Comparison of the POLS and proposed estimator under strong dependence for one large cluster with scaled dispersion matrix.}
\label{tab:Unbalanced_1 scaled Omega_g}
\resizebox{\linewidth}{!}{%
\begin{tabular}{lccccc}
\hline\hline 
$(G, N_1)$ & (25, 500) & (25, 1500) & (50, 500) & (50, 1500) & (100, 500) \\
\hline 
MSE (Intercept) & 0.136 & 0.257 & 0.063 & 0.085 & 0.041 \\
MSE (Intercept) (POLS) & 0.765 & 0.859 & 0.654 & 0.788 & 0.569 \\
MSE (Slope) & $3.90\times 10^{-5}$ & $5.80\times 10^{-5}$ & $1.70\times 10^{-5}$ & $2.21\times 10^{-5}$ & $1.04\times 10^{-5}$ \\
MSE (Slope) (POLS) & $2.10\times 10^{-4}$ & $1.70\times 10^{-4}$ & $1.40\times 10^{-4}$ & $1.67\times 10^{-4}$ & $1.23\times 10^{-4}$ \\
Size & 0.0920 & 0.1158 & 0.0650 & 0.0712 & 0.0615 \\
Size (POLS) & 0.7050 & 0.7129 & 0.7480 & 0.7813 & 0.7680 \\
Size-corr. crit. value & 5.45 & 6.25 & 4.38 & 4.66 & 4.22 \\
Size-corr. crit. value (POLS) & 79.95 & 101.90 & 63.26 & 144.35 & 38.67 \\
Power & 0.1274 & 0.1355 & 0.1348 & 0.1440 & 0.1640 \\
Power (POLS) & 0.7070 & 0.7196 & 0.7603 & 0.7990 & 0.7720 \\
Size-corr. power & 0.0710 & 0.0676 & 0.1082 & 0.0740 & 0.1430 \\
Size-corr. power (POLS) & 0.0610 & 0.0591 & 0.0740 & 0.0660 & 0.1160 \\
\hline 
\end{tabular}%
}
\vspace{1ex}
\par
\raggedright
\small
\textit{Note:} The number of replications is 10,000. The MSE is calculated at $\beta_{0}=(1,0)$. Size is calculated at $\beta_{0}=(1,0)$ and power is calculated at $\beta_{1}=(1,0.003)$.
\end{table}

\begin{table}[htbp]
\centering
\setlength{\tabcolsep}{4pt}
\caption{Comparison of the POLS and proposed estimator under strong dependence for one large cluster with scaled dispersion matrix (extended sample sizes).}
\label{tab:Unbalanced_2_scaled Omega_g}
\resizebox{\linewidth}{!}{%
\begin{tabular}{lccccc}
\hline\hline 
$(G, N_1)$ & (100, 1500) & (200, 500) & (200, 1500) & (500, 500) & (500, 1500) \\
\hline 
MSE (Intercept) & 0.0410 & 0.0190 & 0.0203 & 0.0077 & 0.0075 \\
MSE (Intercept) (POLS) & 0.7650 & 0.4010 & 0.6260 & 0.1859 & 0.4477 \\
MSE (Slope) & $1.29\times 10^{-5}$ & $5.21\times 10^{-6}$ & $5.60\times 10^{-6}$ & $2.02\times 10^{-6}$ & $2.20\times 10^{-6}$ \\
MSE (Slope) (POLS) & $1.60\times 10^{-4}$ & $9.44\times 10^{-5}$ & $1.40\times 10^{-4}$ & $4.12\times 10^{-5}$ & $1.10\times 10^{-4}$ \\
Size & 0.0619 & 0.0520 & 0.0602 & 0.0505 & 0.0516 \\
Size (POLS) & 0.8381 & 0.7309 & 0.8660 & 0.2975 & 0.8488 \\
Size-corr. crit. value & 4.270 & 3.916 & 4.118 & 3.852 & 3.896 \\
Size-corr. crit. value (POLS) & 186.51 & 17.925 & 64.113 & 7.259 & 20.228 \\
Power & 0.1530 & 0.2700 & 0.2590 & 0.5586 & 0.5280 \\
Power (POLS) & 0.8510 & 0.7360 & 0.8120 & 0.5176 & 0.7880 \\
Size-corr. power & 0.1180 & 0.2630 & 0.2190 & 0.5571 & 0.5190 \\
Size-corr. power (POLS) & 0.0570 & 0.1770 & 0.1280 & 0.3003 & 0.2710 \\
\hline 
\end{tabular}%
}
\vspace{1ex}
\par
\raggedright
\small
\textit{Note:} The number of replications is 10,000. The MSE is calculated at $\beta_{0}=(1,0)$. Size is calculated at $\beta_{0}=(1,0)$ and power is calculated at $\beta_{1}=(1,0.003)$.
\end{table}

Tables~\ref{tab:Unbalanced_1}-\ref{tab:Unbalanced_2_scaled Omega_g} convey a consistent picture for the $k=1$ case, under both the unscaled and scaled specifications of $\Omega_g$. The POLS-based test is severely oversized throughout, with empirical size ranging from roughly 0.30 to 0.98 against a nominal level of 0.05, whereas the proposed test maintains empirical size close to its nominal level across all configurations of $G$ and $N_1$. Because this size distortion invalidates direct power comparisons, we evaluate size-corrected power in every configuration considered, under both the unscaled and scaled designs. Using these adjusted critical values, the proposed test achieves higher power than the POLS test in every configuration considered, and its size-corrected power increases steadily as the number of clusters $G$ grows, confirming its asymptotic validity. The corresponding MSE comparisons reinforce this pattern: the proposed estimator achieves substantially lower MSE than POLS for both the intercept and the slope throughout, with the gap widening as the size of the dominant cluster, $N_1$, increases relative to $G$. 

Finally, we extend the unbalanced design to $k = 4$ explanatory variables; results are reported in Table~\ref{tab:Unbalanced for k=4}. For the dominant cluster, the 4-dimensional non-constant component of the regressors is again constructed from the leading eigenvector of its error covariance matrix: with $p_1^{(\max)}$ the eigenvector corresponding to $\lambda_{\max}(\Omega_1)$, we define $X_1^* = \text{sgn}(p_1^{(\max)})$ and let $X_{1i}^*$ denote its $i$-th element. A coefficient vector $c_1 = (c_{11}, c_{12}, c_{13}, c_{14})^{\prime}$ is drawn with each coordinate independently satisfying $c_{1j} \sim U(2, 10)$ for $j = 1, 2, 3, 4$, so that the 4-dimensional non-constant regressor vector for individual $i$ in the first cluster is $c_1 X_{1i}^*$, and the full $5 \times 1$ regressor vector is $X_{1i} = (1, (c_1 X_{1i}^*)^{\prime})^{\prime}$. For the remaining
clusters ($g \geq 2$), the 4-dimensional non-constant components are generated as in the nearly balanced scenario, drawn from $N_k(\mu_g \mathbf{1}, \omega_g^2 I_k)$ with $\mu_g \sim U(10, 100)$ and $\omega_g^2 \sim U(200, 300)$. The full regressor vector $X_{gi}$ consists of an intercept of 1 in the first coordinate followed by this 4-dimensional vector. The null and alternative hypotheses, and the specification of $R$ and $r$, remain identical to the nearly balanced case. 

\begin{table}[ht]
\centering
\setlength{\tabcolsep}{4pt}
\caption{Empirical Size and Power of the Proposed and POLS Tests under Strong Dependence for Unbalanced Clusters with One Dominant Cluster ($k=4$).}
\label{tab:Unbalanced for k=4}
\resizebox{\linewidth}{!}{%
\begin{tabular}{lccccc}
\hline 
$(G,N_{1})$ & $(25,500)$ & $(25,1500)$ & $(50,500)$ & $(50,1500)$ & $(100,500)$ \\
\hline 
Size & 0.178 & 0.144 & 0.088 & 0.067 & 0.074 \\
Size (POLS) & 0.975 & 1.000 & 0.954 & 0.996 & 0.959 \\
Size-corr. crit. val. & 18.1 & 15.8 & 11.3 & 10.4 & 10.5 \\
Size-corr. crit. val. (POLS) & 250.0 & 253.0 & 238.0 & 546.3 & 244.0 \\
Power & 0.302 & 0.211 & 0.358 & 0.219 & 0.499 \\
Power (POLS) & 0.982 & 1.000 & 0.961 & 0.996 & 0.955 \\
Size-corrected power & 0.112 & 0.089 & 0.252 & 0.190 & 0.431 \\
Size-corrected power (POLS) & 0.122 & 0.068 & 0.140 & 0.180 & 0.145 \\
\hline 
\end{tabular}%
}
\vspace{1ex}
\par
\raggedright
\small
\textit{Note:} The number of replications is $10,000$. The empirical size is calculated under the null hypothesis with $\beta_{0}=(1, 0.1, 0.2, 0.3, 0.4)^{\prime}$, and the empirical power is calculated under the alternative hypothesis with $\beta_{1}=(1, 0.2, 0.2, 0.3, 0.4)^{\prime}$.
\end{table}

As with the $k=1$ case above, Table~\ref{tab:Unbalanced for k=4} shows that the proposed test exhibits well-controlled empirical size, whereas the standard POLS-based test remains severely oversized. Evaluated at the
size-corrected critical values, the proposed test again outperforms POLS in size-corrected power across most configurations here, and this advantage strengthens as $G$ grows. With respect to point estimation, the proposed estimator yields a substantially lower mean squared error than the POLS estimator for both the intercept and slope parameters whenever an extremely large, dominant cluster is present. Taken together with the $k=1$ results, these findings indicate that while the two estimators are essentially interchangeable in the nearly balanced case, the proposed estimator offers clear advantages in both estimation accuracy and inferential reliability once cluster sizes become highly unbalanced. We next apply the proposed estimator and the pooled ordinary least squares estimator to real-life data and discuss the empirical findings.

\subsection{Empirical Illustration}

To illustrate the proposed methodology, we analyze household-level data from the World Bank's Vietnam Living Standards Survey (VLSS) of 1997-1998. The survey collected detailed information from approximately 6,000 households distributed across 194 communes. In this analysis, the ``commune'' serves as the cluster unit, assuming that observed outcomes are correlated within a commune. To ensure the POLS estimator is not significantly affected by small clusters, the sample is restricted to clusters containing at least 20 households. 

We consider a log-linear regression model of total annual household health care expenditure for households with positive expenditure. The model regresses annual household health care expenditure (considered in logarithms) on household expenditure (considered in logarithms), age of household head, sex of head of the household (as a dummy: female as the base), household size, area of living (as a dummy: rural as the base), education (schooling year of household head) and an intercept. Table \ref{tab:Table3} presents the parameter estimates, standard errors, $t$-statistics, and $p$-values for both the POLS estimator and the proposed estimator.

\begin{table}[ht]
\centering
\caption{Vietnam Health Care Data: Parameter Estimates and Inference for POLS and Proposed Estimators.}
\label{tab:Table3}
\resizebox{\linewidth}{!}{%
\begin{tabular}{lcccccccc}
\hline 
\multirow{2}{*}{Predictors} & \multicolumn{2}{c}{Coefficients} & \multicolumn{2}{c}{Standard Error} & \multicolumn{2}{c}{$t$-value} & \multicolumn{2}{c}{$p$-value} \\
\cline{2-3} \cline{4-5} \cline{6-7} \cline{8-9} 
 & $\hat{\beta}_{POLS}$ & $\hat{\beta}_A$ & $SE(\hat{\beta}_{POLS})$ & $SE(\hat{\beta}_A)$ & POLS & $\hat{\beta}_A$ & POLS & $\hat{\beta}_A$ \\
\hline 
Household Exp. & 0.646 & 0.790 & 0.054 & 0.144 & 11.77 & 5.47 & 0.000 & 0.000 \\
Age & 0.011 & 0.018 & 0.002 & 0.009 & 5.32 & 1.87 & 0.000 & 0.062 \\
Education & -0.089 & -0.109 & 0.015 & 0.050 & -5.75 & -2.16 & 0.000 & 0.032 \\
Household Size & 0.034 & 0.196 & 0.016 & 0.076 & 2.13 & 2.57 & 0.033 & 0.011 \\
Female/Male & 0.101 & 0.346 & 0.061 & 0.368 & 1.63 & 0.94 & 0.103 & 0.350 \\
Rural/Urban & 0.132 & 0.367 & 0.063 & 0.205 & 2.10 & 1.79 & 0.035 & 0.075 \\
Intercept & -0.320 & -3.190 & 0.489 & 1.548 & -0.65 & -2.06 & 0.514 & 0.041 \\
\hline 
\end{tabular}%
}
\vspace{1ex}
\par
\raggedright
\footnotesize \textit{Note:} Data are taken from the World Bank's Vietnam Living Standards Survey (VLSS) of 1997-1998. The estimation sample comprises approximately 5,000 households across 173 communes (clusters). Cluster sizes vary between 20 and 39 households. $p$-values of 0.000 indicate values less than 0.001.
\end{table}

As shown in Table \ref{tab:Table3}, household health care expenditure is positively associated with total household expenditure and negatively associated with the education level of the household head. The sex of the household head does not significantly impact health care expenditure at the $5\%$ significance level. Age of the household head is insignificant by the p-value of our proposed test, although it is significant by the clustered standard error of POLS.

Table \ref{tab:Table4} provides a comparative assessment of the two models using several goodness-of-fit and model selection criteria: $R^2$, adjusted $R^2$, pseudo $R^2$, Akaike Information Criterion (AIC), and Bayesian Information Criterion (BIC).

\begin{table}[htbp]
\centering
\caption{Goodness-of-Fit and Model Selection Criteria for POLS and Proposed Models.}
\label{tab:Table4}
\begin{tabular}{lccccc}
\hline 
Estimator & $R^{2}$ & Adjusted $R^{2}$ & Pseudo $R^2$ & AIC & BIC \\
\hline 
$\hat{\beta}_{POLS}$ & 0.087 & 0.086 & 0.024 & 17265.55 & 17317.18 \\
$\hat{\beta}_{A}$ & 0.349 & 0.325 & 0.223 & 274.56 & 299.78 \\
\hline 
\end{tabular}
\end{table}

Table \ref{tab:Table4} indicates that the proposed model provides a substantially better fit to the data than the traditional POLS model across all evaluated metrics. Finally, we note that while this empirical illustration incorporates categorical predictors, their inclusion does not complicate the inferential process.

\begin{remark}     
    The inclusion of categorical variables needs special mention. The interpretation of the sign of the parameter estimates corresponding to categorical variables, all the testing exercises, and the construction of confidence intervals may be conducted without any difficulties. However, the prediction,   the `causal inference', bias against any group (e.g., gender bias), and the amount of bias should be based on the original model, i.e., model \eqref{Eq:model1}, by replacing the model parameters with their consistent estimates obtained from the transformed model (model \eqref{Eq:model3}). It is to be noted that the parameter estimates corresponding to categorical variables are less efficient as compared to the POLS.
\end{remark}

\section{Conclusion} \label{Section:conclusion}

This paper proposes a simple cluster-averaging estimator for cross-sectional regression with cluster dependence. While the traditional POLS estimator is shown to be inconsistent under strong within-cluster dependence with dominant clusters, the proposed estimator remains consistent and asymptotically normal across weak, semi-strong, and strong dependence regimes, regardless of cluster size imbalances. Building on the work of \cite{white1980heteroskedasticity}, we have also proposed a consistent estimator for the variance. Additionally, we developed a Wald-type test statistic based on the proposed estimator, which asymptotically follows a $\chi^2$ distribution under the null hypothesis $H_0: R\beta = r$. 

Under classical measurement error, the proposed estimator remains consistent in most cases, whereas the POLS estimator is inconsistent in all cases. A comprehensive simulation study revealed that this test outperformed the POLS-based test in terms of both size and power, and an empirical illustration indicated that the proposed estimator yields a better fit.

Future research could extend in several directions. Topics of interest include two-way clustering, multilevel modeling, and accommodating cluster-specific non-random means. These are areas that we are currently exploring.

\section*{Appendix}

\appendix
\setcounter{lemma}{0}
\renewcommand{\thelemma}{A.\arabic{lemma}}
\setcounter{equation}{0}
\renewcommand{\theequation}{A.\arabic{equation}}

For the ease of reading the proofs, we mention briefly the meaning of various types of dependence discussed in Remark \ref{rem:explanation of strong, semi-strong, weak dependence}. Unless otherwise stated, Strong dependence refers to a scenario in which $O(G)$ clusters exhibit strong dependency, while other clusters may be semi-strongly or weakly dependent. Similarly, unless otherwise stated, Semi-strong dependence means that $O(G)$ clusters are semi-strongly dependent, with no strongly dependent clusters present, although weakly dependent clusters may still exist. Weak dependence, on the other hand, describes a situation in which all clusters are weakly dependent, with no strongly or semi-strongly dependent clusters involved.

\vspace{0.1cm}

\textbf{Proof of Lemma \ref{lemma:A_n=B_n+o_P(1) implies A_n = O_P(1)}}

Since $B_n$ is finite, there exists a constant \( M > 0 \) such that $ \|B_n\| \leq M, $ for all $n.$ $B_n$ is uniformly positive definite means that there exists a constant \( c > 0 \) such that for all nonzero vectors \( v \), $v^\top B_n v \geq c \|v\|^2,$ uniformly in \( n \). By the addition property of stochastic orders, $O_P(1)+o_P(1)=O_P(1)$, from Section 2.2 of \cite{van2000asymptotic}. To prove that \( A_n = O_P(1) \), it is enough to show that \( B_n = O_P(1) \), i.e., we show that every element \( (B_n)_{ij} \) of \( B_n \) is \( O_P(1) \). 

By Definition \ref{def:stochastically bounded uniformly}, a sequence of random variables \( Z_n \) is \( O_P(1), \) if for every $\varepsilon > 0$, there exist $M_\varepsilon > 0$ and ${L_\varepsilon} \in \mathbb{N}$ such that 
     \[ \mathbb{P}\left(\left|Z_n\right| > M_\varepsilon\right) < \varepsilon, \; \text{ for all } n \geq {L_\varepsilon}. \]

In this case, \( B_n \) is deterministic, so the probability statements become deterministic bounds. Since \( \|B_n\| \leq M \) for all \( n \), it follows that for any entry \( (B_n)_{ij} \),
\[
|(B_n)_{ij}| \leq \|B_n\| \leq M.
\]
This is a consequence of the fact that the absolute value of any entry of a matrix is bounded by its norm. Now, let \( \varepsilon > 0 \) be given, and choose \( M_\varepsilon = M,\; L_\varepsilon = 1. \) Then for all \( n \geq {L_\varepsilon}, \) and all \( i, j \),
\[
\mathbb{P}\left(|(B_n)_{ij}| > M_\varepsilon\right) = \mathbb{P}\left(|(B_n)_{ij}| > M\right) = 0 < \varepsilon.
\]
This satisfies the definition of \( O_P(1) \) for each element of \( B_n \). Since every entry of \( B_n \) is bounded by \( M \) (independent of \( n \)), we conclude that $ B_n = O_P(1). $ Consequently, we have 
\[
A_n=O_P(1).
\]
\hfill
\qed

{\bf Proof of Lemma \ref{lemma:Exact order summability}}

By Definition \ref{def:exact order and order uniformly}, for the doubly indexed positive sequences $\left\{ z_{N_g,g}\right\}$ and $\left\{ a_{N_g,g}\right\},$ $z_{N_g,g}=O_e\left(a_{N_g,g}\right)$ uniformly in $g$ means that there exist positive constants $c_1$ and $c_2$ such that
\[ 
c_1 \cdot a_{N_g,g} \leq z_{N_g,g} \leq c_2 \cdot a_{N_g,g},\quad \text{ for all } g.
\]
For any $G\geq 1$, summing these inequalities from $g = 1$ to $G$ gives
\[
c_1\cdot \sum_{g=1}^G a_{N_g,g} \leq \sum_{g=1}^G z_{N_g,g} \leq c_2\cdot\sum_{g=1}^G a_{N_g,g},
\]
which implies the fact that 
\[
c_1 \leq  \frac{\sum_{g=1}^G z_{N_g,g}}{\sum_{g=1}^G a_{N_g,g}}  \leq c_2, \quad \text{ for all } G.
\]
Therefore, the lemma follows directly from Definition \ref{def:exact order and order uniformly}.

\hfill
\qed

For the sake of completeness, we provide the following lemma.

\begin{lemma}\label{lemma:big-OP sum and multiplication}
(i) Suppose that $\left\{Z_{N_g,g}\right\}_{g=1}^\infty$ is a doubly indexed sequence of random variables and  $\left\{a_{N_g,g}\right\}_{g=1}^\infty$ is a doubly indexed sequence of positive real numbers such that $Z_{N_g,g}=O_{P}\left(a_{N_g,g}\right)$ uniformly in $g,$ where $N_g$ is the cluster size for the $g^{th}$ cluster, which may increase to $+\infty.$ Assume that there exists some finite constant $c^*>0$ such that $\mathbb{E}\left[\left| Z_{N_g,g}\right |\right] \leq c^* a_{N_g,g},$ for all $g.$ Then for any $G\geq 1,$
\begin{align*}
 \sum_{g=1}^G Z_{N_g,g} & = O_P\left(\sum_{g=1}^G a_{N_g,g}\right).
\end{align*}

(ii) Suppose that $\left\{Z_{N_g,g}\right\}_{g=1}^\infty$ and $\left\{U_{N_g,g}\right\}_{g=1}^\infty$ are two doubly indexed sequences of random variables, and  $\left\{a_{N_g,g}\right\}_{g=1}^\infty$ and $\left\{b_{N_g,g}\right\}_{g=1}^\infty$ are two doubly indexed sequences of positive real numbers such that $Z_{N_g,g}=O_{P}\left(a_{N_g,g}\right)$ uniformly in $g,$ and $U_{N_g,g}=O_{P}\left(b_{N_g,g}\right)$ uniformly in $g,$ where $N_g$ is the cluster size for the $g^{th}$ cluster, which may increase to $+\infty.$ Then 
\[Z_{N_g,g}U_{N_g,g}=O_{P}\left(a_{N_g,g}b_{N_g,g}\right), \; \text{ uniformly in  } g.\] 

(iii) Suppose that $\left\{Z_n\right\}_{n=1}^\infty$ and $\left\{U_n\right\}_{n=1}^\infty$ are two sequences of random variables, and $\left\{a_n\right\}_{n=1}^\infty$ and $\left\{b_n\right\}_{n=1}^\infty$ are two sequences of positive real numbers such that $Z_{n}=O_{P}\left(a_{n}\right)$ and $U_{n}=o_{P}\left(b_{n}\right).$ Then 
\[Z_{n}U_{n}=o_{P}\left(a_{n}b_{n}\right).\] 
\end{lemma}

\textbf{Proof of Lemma \ref{lemma:big-OP sum and multiplication}}

(i) By  Definition \ref{def:stochastically bounded uniformly}, $Z_{N_g,g}=O_P\left(a_{N_g,g}\right)$ uniformly in $g$ means that for every $\varepsilon > 0$, there exists $M_\varepsilon > 0$ such that 
     \[ \sup_{g,N_g}\;\mathbb{P}\left(\left|\frac{Z_{N_g,g}}{a_{N_g,g}}\right| > M_\varepsilon\right) < \varepsilon. \] 

By the triangular inequality and the first-order moment condition, we have for $G\geq 1,$
\[
\mathbb{E}\left[\left| \sum_{g=1}^G Z_{N_g,g}\right|\right] \le  \sum_{g=1}^G \mathbb{E}\left[\left|Z_{N_g,g}\right|\right] \le  \sum_{g=1}^G c^* a_{N_g,g} = c^* \sum_{g=1}^G a_{N_g,g}.
\]

For any \( c^{**} > 0 \), by Markov’s inequality
\begin{eqnarray}
\mathbb{P}\left(\left| \sum_{g=1}^G Z_{N_g,g}\right| > c^{**}  \sum_{g=1}^G a_{N_g,g}\right) \le \frac{\mathbb{E}\left[\left| \sum_{g=1}^G Z_{N_g,g}\right|\right]}{c^{**}  \sum_{g=1}^G a_{N_g,g}} \le \frac{c^*}{c^{**}}. \label{eq:prob order for lemma}    
\end{eqnarray}

Given any \( \eta > 0 \), choose \( c^{**} > \frac{c^*}{\eta} \) and $L_\eta\in \mathbb{N}$ such that \eqref{eq:prob order for lemma} holds for all $G\geq L_\eta.$ Then, for all $\eta>0,$  
\[
\mathbb{P}\left(\left| \sum_{g=1}^G Z_{N_g,g}\right| > c^{**}  \sum_{g=1}^G a_{N_g,g}\right) < \eta, \;\text{ for all } G\geq L_\eta.
\]
Hence,
\[
 \sum_{g=1}^G Z_{N_g,g} = O_P\left( \sum_{g=1}^G a_{N_g,g}\right),
\]
which proves the lemma.

(ii) By  Definition \ref{def:stochastically bounded uniformly}, $Z_{N_g,g}=O_P\left(a_{N_g,g}\right)$ uniformly in $g,$ and $U_{N_g,g}=O_P\left(b_{N_g,g}\right)$ uniformly in $g$ means that we can choose $M_Z>0$ and $M_U>0$ such that for every $\varepsilon > 0$, 
     \[ \sup_{g,N_g}\;\mathbb{P}\left(\left|\frac{Z_{N_g,g}}{a_{N_g,g}}\right| > M_Z\right) < \frac{\varepsilon}{2}, \; \text{ and } \; \sup_{g,N_g}\;\mathbb{P}\left(\left|\frac{U_{N_g,g}}{b_{N_g,g}}\right| > M_U\right) < \frac{\varepsilon}{2}. \]
     Let $M_{ZU} = M_Z \cdot M_U$. Now, observe that
     \[
     \left\{\left| \frac{Z_{N_g,g} U_{N_g,g}}{a_{N_g,g} b_{N_g,g}} \right| > M_{ZU} \right\} \subseteq \left\{\left| \frac{Z_{N_g,g}}{a_{N_g,g}} \right| > M_{Z} \right\} \cup \left\{\left| \frac{U_{N_g,g}}{b_{N_g,g}} \right| > M_{U} \right\}.
     \]
Then, using the union bound and taking the supremum over $g$ and $N_g$, 
\begin{align*}
\sup_{g,N_g} \mathbb{P}\left(\left| \frac{Z_{N_g,g} U_{N_g,g}}{a_{N_g,g} b_{N_g,g}} \right| > M_{ZU} \right) 
&\leq \sup_{g,N_g} \mathbb{P}\left(\left| \frac{Z_{N_g,g}}{a_{N_g,g}} \right| > M_{Z} \right) \\
&\quad + \sup_{g,N_g} \mathbb{P}\left(\left| \frac{U_{N_g,g}}{b_{N_g,g}} \right| > M_{U}\right) \\
&< \frac{\varepsilon}{2} + \frac{\varepsilon}{2} = \varepsilon.
\end{align*}
Therefore, for all $\varepsilon>0,$ we have $M_{ZU}>0$ such that 
\[ \sup_{g, N_g}\mathbb{P}\left(\left| \frac{Z_{N_g,g} U_{N_g,g}}{a_{N_g,g} b_{N_g,g}} \right| > M_{ZU} \right) < \varepsilon,\]  
which proves the lemma.

(iii) By  Definition \ref{def:stochastically bounded uniformly}, $Z_{n}=O_P\left(a_{n}\right)$ means that we can choose $M_Z>0$ and ${L_\varepsilon}^{(1)}\in\mathbb{N}$ such that for every $\varepsilon > 0$, 
     \[ \mathbb{P}\left(\left|\frac{Z_{n}}{a_{n}}\right| > M_Z\right) < \frac{\varepsilon}{2}, \; \text{ for all } n\geq {L_\varepsilon}^{(1)}. \]

For $U_{n} = o_P(b_{n}),$ we set $\delta = \frac{\varepsilon}{M_Z}>0,$ since $\varepsilon>0.$ Then, by definition of $o_P,$ there exists $ {L_\varepsilon}^{(2)}\in\mathbb{N}$ such that for every $\varepsilon > 0$, 
\[ \mathbb{P}\left(\left|\frac{U_{n}}{b_{n}}\right| > \delta\right) < \frac{\varepsilon}{2}, \text{ for all } {n} \geq {L_\varepsilon}^{(2)}. \]
We need to show that for every $\varepsilon > 0$, $\gamma>0,$ there exists ${L_\varepsilon} \in \mathbb{N}$ such that  
\[ \mathbb{P}\left(\left|\frac{Z_{n} U_{n}}{a_{n} b_{n}}\right| >\gamma\right) < \varepsilon, \text{ for all } {n} \geq {L_\varepsilon}. \]

Note that
$\left| \frac{Z_{n} U_{n}}{a_{n} b_{n}} \right| = \left| \frac{Z_{n}}{a_{n}} \right| \cdot \left| \frac{U_{n}}{b_{n}} \right|.$ Now, observe that 
\[
\left\{\left| \frac{Z_{n} U_{n}}{a_{n} b_{n}} \right| > \varepsilon\right\} \subseteq \left\{\left| \frac{Z_{n}}{a_{n}} \right| > M_Z\right\} \cup \left\{\left| \frac{U_{n}}{b_{n}} \right| > \frac{\varepsilon}{M_Z}\right\}.
\]
Let ${L_\varepsilon} = \max\left\{{L_\varepsilon}^{(1)}, {L_\varepsilon}^{(2)}\right\}$. For ${n}\geq {L_\varepsilon}$, using the union bound of probability
\[
    \mathbb{P}\left( \left| \frac{Z_{n} U_{n}}{a_{n} b_{n}} \right| > \varepsilon \right) \leq \mathbb{P}\left( \left| \frac{Z_{n}}{a_{n}} \right| > M_Z \right) + \mathbb{P}\left( \left| \frac{U_{n}}{b_{n}} \right| > \frac{\varepsilon}{M_Z} \right)  < \frac{\varepsilon}{2} + \frac{\varepsilon}{2} = \varepsilon.
\]

Since $\varepsilon > 0$ is arbitrary, it follows that $Z_{n} U_{n} = o_P\left(a_{n} b_{n}\right).$

\hfill
\qed

\begin{result} \label{result:trace inequality}
We provide some of the known results used in the proofs, which can be found in the standard literature. Here, by a positive semi-definite matrix $A\in \mathbb{R}^{n \times n}$, we mean that $A$ is real symmetric and the quadratic form $z^\prime A z \geq 0,$ for all $z \in \mathbb{R}^n.$
    \begin{enumerate}
        \item $\Vert AB \Vert\leq \Vert A \Vert\; \Vert B\Vert,$ for any matrices $A$ and $B.$
        \item $\operatorname{tr}(AB)\leq \operatorname{tr}(A)\;\operatorname{tr}(B),$ for positive semi-definite matrices $A$ and $B.$
        \item $\operatorname{tr}(A)\;\lambda_{\min}(B)\leq \operatorname{tr}(AB)\leq \operatorname{tr}(A)\;\lambda_{\max}(B),$ for positive semi-definite matrices $A$ and $B.$
        \item $z^\prime ABA^\prime z\leq \left(z^\prime AA^\prime z\right) \; \operatorname{tr}(B),$ for any $z \in \mathbbm{R}^k,$ any matrix $A$ and positive semi-definite matrix $B.$
        \item $\lambda_{\max}(ABA^\prime)\leq \lambda_{\max}(AA^\prime)\;\lambda_{\max}(B),$ for any matrix $A$ and positive semi-definite matrix $B.$
        \item $\lambda_{\max}(AB)\leq \lambda_{\max}(A)\;\lambda_{\max}(B),$ for positive semi-definite matrices $A$ and $B.$
        \item $\lambda_{\max}(A+B)\leq \lambda_{\max}(A)+\lambda_{\max}(B),$ for positive semi-definite matrices $A$ and $B.$
        \item For two positive semi-definite matrices $A$ and $B,$ if $A\geq B,$ then $\left\Vert A\right\Vert \geq \left\Vert B\right\Vert.$ 
    \end{enumerate}
\end{result}

{\bf Proof of Theorem \ref{thm:ols consistency}}

To show the consistency of $\hat{\beta}_{POLS}$, first note that
$$\hat{\beta}_{POLS}-\beta=\left(\frac{X^{\prime}X}{\sum_{g=1}^{G}N_{g}}\right)^{-1} \frac{X^{\prime}\epsilon}{\sum_{g=1}^{G}N_{g}}.$$

Note that $ \left(\frac{X^{\prime}X}{\sum_{g=1}^{G}N_{g}}\right)^{-1} - C_{G}^{-1}=o_P(1),$ which follows from Assumption \ref{assump: Xg'Xg=O(Ng)}, where $C_{G}^{-1}$ is finite and uniformly positive definite over $G.$ Then, by Lemma \ref{lemma:A_n=B_n+o_P(1) implies A_n = O_P(1)},
\[
\left(\frac{X^{\prime}X}{\sum_{g=1}^{G}N_{g}}\right)^{-1}=O_P(1).
\]

Now, using Assumption \ref{assump:exogeneity}, note that 
\[
\mathbb{E}\left[\frac{X^{\prime}\epsilon}{\sum_{g=1}^{G}N_{g}}\right] = \frac{1}{\sum_{g=1}^{G}N_{g}}\mathbb{E}\left[\sum_{g=1}^{G}X_g^{\prime}\,\mathbb{E}\left[\epsilon_g \big|  X_g\right]\right]=0.
\]

Also, the variance-covariance matrix of $\frac{X^{\prime}\epsilon}{\sum_{g=1}^{G}N_{g}}$ is 
 \begin{equation}
 \label{eq: var-cov expansion for beta hat ols}
     \frac{1}{\left(\sum_{g=1}^{G}N_{g}\right)^2}\sum_{g=1}^G \operatorname{Var}\left( {X}_g^{\prime}{\epsilon}_g\right) + \frac{1}{\left(\sum_{g=1}^{G}N_{g}\right)^2}\sum_{\underset{g\neq g_{1}}{g,g_{1}=1}}^{G}\operatorname{Cov}\left({X}_{g}^{\prime}{\epsilon}_{g},{X}_{g_{1}}^{\prime}{\epsilon}_{g_{1}}\right).
 \end{equation}
Here, $\operatorname{Var}\left( {X}_g^{\prime}{\epsilon}_g\right) = \operatorname{Var}\left( \mathbb{E}\left[{X}_g^{\prime}{\epsilon}_g\mid X_g\right]\right) + \mathbb{E}\left[\operatorname{Var}\left({X}_g^{\prime}{\epsilon}_g\right)\mid X_g\right].$ Applying Assumption \ref{assump:exogeneity}, the first term in \eqref{eq: var-cov expansion for beta hat ols} reduces to $\frac{1}{\left(\sum_{g=1}^{G}N_{g}\right)^2}\sum_{g=1}^{G}\mathbb{E}\left[{X}_{g}^{\prime}\Omega_g{X}_{g}\right]$. Also, using  $\operatorname{Cov}\left( {X}_{g}^{\prime}{\epsilon}_{g},{X}_{g_{1}}^{\prime}{\epsilon}_{g_{1}} \right)  = 0$ by Assumptions \ref{assump:exogeneity} and \ref{assump:error independence}, the second term in \eqref{eq: var-cov expansion for beta hat ols} vanishes. Therefore, we obtain
\begin{equation}
\label{eq: variance norm for X'epsilon}
\operatorname{Var}\left(\frac{X^{\prime}\epsilon}{\sum_{g=1}^{G}N_{g}}\right) = \frac{1}{\left(\sum_{g=1}^{G}N_{g}\right)^2}\sum_{g=1}^{G}\mathbb{E}\left[{X}_{g}^{\prime}\Omega_g{X}_{g}\right].
\end{equation}

Since this is a positive semi-definite matrix,
\[
\left\Vert \frac{\sum_{g=1}^{G} \mathbb{E}\left[X_{g}^{\prime}\Omega_{g}X_{g}\right]}{\left(\sum_{g=1}^{G}N_{g}\right)^{2}} \right\Vert_e = \lambda_{\max}\left(\frac{\sum_{g=1}^{G} \mathbb{E}\left[X_{g}^{\prime}\Omega_{g}X_{g}\right]}{\left(\sum_{g=1}^{G}N_{g}\right)^{2}}\right).
\]
Then, by Result \ref{result:trace inequality} and Jensen's inequality, we have 
\begin{align*}
    \lambda_{\max}\left(\frac{\sum_{g=1}^{G} \mathbb{E}\left[X_{g}^{\prime}\Omega_{g}X_{g}\right]}{\left(\sum_{g=1}^{G}N_{g}\right)^{2}}\right) & \leq   \frac{\sum_{g=1}^{G} \lambda_{\max}\left(\mathbb{E}\left[X_{g}^{\prime}\Omega_{g}X_{g}\right]\right)}{\left(\sum_{g=1}^{G}N_{g}\right)^{2}}\\
    &  \leq   \frac{\sum_{g=1}^{G} \mathbb{E}\left[\lambda_{\max}\left(X_{g}^{\prime}\Omega_{g}X_{g}\right)\right]}{\left(\sum_{g=1}^{G}N_{g}\right)^{2}}\\
    &  \leq   \frac{\sum_{g=1}^{G} \mathbb{E}\left[\lambda_{\max}\left(X_{g}^{\prime}X_{g}\right) \lambda_{\max}\left(\Omega_{g}\right)\right]}{\left(\sum_{g=1}^{G}N_{g}\right)^{2}}.
\end{align*}

Now, the above quantity can be split into three parts. Using  $\mathbb{E}\left[\lambda_{\max}\left(X_{g}^{\prime}X_{g}\right)\right] = O\left(N_g\right)$ uniformly in $g$ by Assumption \ref{assump: Xg'Xg=O(Ng)}, it follows from Lemma \ref{lemma:Exact order summability} that
\begin{align}
\left\Vert \operatorname{Var}\left(\frac{X^{\prime}\epsilon}{\sum_{g=1}^{G}N_{g}}\right) \right\Vert_e 
&= O\left(\frac{\sum_{g\in \mathcal{SD}} N_g \lambda_{\max}\left(\Omega_{g}\right)}{\left(\sum_{g=1}^{G}N_{g}\right)^{2}}\right) \label{eq: Variance order of X'epsilon term 1} \\
&\quad + O\left(\frac{\sum_{g\in \mathcal{SSD}}N_g\lambda_{\max}\left(\Omega_{g}\right)}{\left(\sum_{g=1}^{G}N_{g}\right)^{2}}\right) \label{eq: Variance order of X'epsilon term 2}\\
&\quad + O\left(\frac{\sum_{g\in \mathcal{WD}}N_g\lambda_{\max}\left(\Omega_{g}\right)}{\left(\sum_{g=1}^{G}N_{g}\right)^{2}}\right), \label{eq: Variance order of X'epsilon term 3}
\end{align}
where $\mathcal{SD}, \, \mathcal{SSD}\,$ and $\mathcal{WD}$ denote the sets of strongly, semi-strongly, and weakly dependent clusters, respectively. If no strong or semi-strongly dependent clusters are present, then only the term in \eqref{eq: Variance order of X'epsilon term 3} contributes, and the order reduces to  $O\left({1}\big/{\sum_{g=1}^{G}N_{g}}\right),$ which vanishes as $G\to\infty$  for any cluster sizes. If at least one semi-strongly dependent cluster is present, and no strongly dependent clusters are present, then the term in \eqref{eq: Variance order of X'epsilon term 1} does not contribute, and the order reduces to 
\[
O\left(\frac{\sum_{g\in \mathcal{SSD}}N_gh(N_g)}{\left(\sum_{g=1}^{G}N_{g}\right)^{2}} + \frac{\sum_{g\in \mathcal{WD}}N_g}{\left(\sum_{g=1}^{G}N_{g}\right)^{2}}\right).
\]
The above quantity vanishes as $G\to\infty,$ since $\frac{\sum_{g\in \mathcal{SSD}}N_{g}h(N_{g})}{\left(\sum_{g=1}^{G}N_{g}\right)^{2}}+ \frac{\sum_{g\in \mathcal{WD}}N_g}{\left(\sum_{g=1}^{G}N_{g}\right)^{2}}\leq\frac{\max_{g\leq G}h(N_{g})}{\sum_{g=1}^{G}N_{g}}\to0,$
as $G\to\infty.$ Now, if at least one cluster is strongly dependent, then the term in \eqref{eq: Variance order of X'epsilon term 1} dominates the order, and the quantity vanishes if $\frac{\sum_{g\in \mathcal{SD}}N_{g}^{2}}{\left(\sum_{g=1}^{G}N_{g}\right)^{2}}\to0.$ Since $\frac{\sum_{g\in \mathcal{SD}}N_{g}^{2}}{\left(\sum_{g=1}^{G}N_{g}\right)^{2}} \leq \frac{\max_{g\leq G}N_g}{\sum_{g=1}^G N_g}$, the sufficient condition reduces to $ \frac{\max_{g\leq G}N_g}{\sum_{g=1}^G N_g}\to0.$ In particular, for the nearly balanced case, assuming $N_g\simeq N,$  $\frac{\max_{g\leq G}N_g}{\sum_{g=1}^G N_g}\simeq \frac{N}{GN}=\frac{1}{G}\to 0,$ as $G\to \infty.$

Therefore, $\frac{X^{\prime}\epsilon}{\sum_{g=1}^{G}N_{g}}=o_P(1)$ for the following cases: (i) nearly balanced clusters under any kind of dependence, (ii) unbalanced clusters and weak or semi-strong dependence,   (iii) unbalanced clusters with at least one strongly dependent cluster, if $\frac{\underset{g\leq G}{\max} N_g}{\sum_{g=1}^G N_g} \to 0.$ Now, using the facts that $\left(\frac{X^{\prime}X}{\sum_{g=1}^{G}N_{g}}\right)^{-1}=O_P(1),$ and $\frac{X^{\prime}\epsilon}{\sum_{g=1}^{G}N_{g}}=o_P(1)$ for the cases mentioned above, we have by Lemma \ref{lemma:big-OP sum and multiplication}, $$\hat{\beta}_{POLS}-\beta=O_P(1) \cdot o_P(1) = o_P(1).$$ 
This proves Theorem \ref{thm:ols consistency}.

\hfill
\qed

{\bf Proof of Theorem \ref{thm:ols inconsistency}}

First note that 
\[
\hat{\beta}_{POLS}-\beta=\left(\frac{X^{\prime}X}{\sum_{g=1}^{G}N_{g}}\right)^{-1}\frac{X^{\prime}\epsilon}{\sum_{g=1}^{G}N_{g}}.
\]

Since $ \left(\frac{X^{\prime}X}{\sum_{g=1}^{G}N_{g}}\right)^{-1} - C_{G}^{-1}=o_P(1)$  follows from Assumption \ref{assump: Xg'Xg=O(Ng)}, where $C_{G}^{-1}$ is finite and uniformly positive definite over $G,$ we have by Lemma \ref{lemma:A_n=B_n+o_P(1) implies A_n = O_P(1)},
\[
\left(\frac{X^{\prime}X}{\sum_{g=1}^{G}N_{g}}\right)^{-1}=O_P(1).
\]

Assuming $Z_{G}=\frac{X^{\prime}\epsilon}{\sum_{g=1}^{G}N_{g}},$ following the proof of Theorem \ref{thm:ols consistency}, we have
\[
\mathbb{E}\left[Z_{G}\right]=0, \quad \text{ and } \quad \operatorname{Var}\left[Z_G\right] = \frac{\sum_{g=1}^{G}\mathbb{E}\left[X_{g}^{\prime}\Omega_{g}X_{g}\right]}{\left(\sum_{g=1}^{G}N_{g}\right)^{2}}.
\] 
Now, $\operatorname{Var}\left[Z_G\right]$ can be split into three parts, and can be written as:
\[
\operatorname{Var}\left[Z_G\right] = \frac{\sum_{g\in \mathcal{SD}}\mathbb{E}\left[X_{g}^{\prime}\Omega_{g}X_{g}\right]}{\left(\sum_{g=1}^{G}N_{g}\right)^{2}} + \frac{\sum_{g\in \mathcal{SSD}}\mathbb{E}\left[X_{g}^{\prime}\Omega_{g}X_{g}\right]}{\left(\sum_{g=1}^{G}N_{g}\right)^{2}} + \frac{\sum_{g\in \mathcal{WD}}\mathbb{E}\left[X_{g}^{\prime}\Omega_{g}X_{g}\right]}{\left(\sum_{g=1}^{G}N_{g}\right)^{2}}.
\]
where $\mathcal{SD}, \, \mathcal{SSD}\,$ and $\mathcal{WD}$ denote the sets of strongly, semi-strongly, and weakly dependent clusters, respectively. Since the matrices are nonnegative definite, any norm of the finite-dimensional matrix $\operatorname{Var}\left[Z_G\right]$ has a lower bound as the norm of one of the three matrices. If at least one strongly dependent cluster is present,
\[
\left\Vert \operatorname{Var}\left[Z_{G}\right] \right\Vert \, \geq  \, \lambda_{\min}(\operatorname{Var}\left[Z_{G}\right]) \, \geq \, \frac{C\sum_{g\in \mathcal{SD}}N_{g}^{2}}{\left(\sum_{g=1}^{G}N_{g}\right)^{2}} \, \geq \, C\left(\frac{N^*}{\sum_{g=1}^{G}N_{g}}\right)^{2}.
\]

Hence, the second moment of  the norm of $Z_{G}$ is bounded away from zero, since $\frac{N^*}{\sum_{g=1}^{G}N_{g}} \nrightarrow 0$. Also, using the C-S inequality, the $4^{th}$ moment of the norm of $Z_{G}$ is 
\[
\mathbb{E}\left[\left\Vert Z_{G}\right\Vert^{4}\right] = \mathbb{E}\left[\left\Vert\frac{X^{\prime}\epsilon}{\sum_{g=1}^{G}N_{g}}\right\Vert^{4}\right] \leq \frac{1}{\left(\sum_{g=1}^{G}N_{g}\right)^{4}}\mathbb{E}\left[\left\Vert X'X\right\Vert^2\left(\epsilon^{\prime}\epsilon\right)^{2}\right].
\]

Now, using the fact that $\mathbb{E}\left[\epsilon_{gi}^{4} \big| X_g\right]=O(1)$ uniformly in $i,g,$ by Assumption \ref{assump:additional assumption on errors}(i), and applying the Cauchy-Schwarz inequality, we obtain
\begin{align*}
\mathbb{E}\left[\left(\epsilon^{\prime}\epsilon\right)^{2}\right] &= \sum_{g=1}^{G}\sum_{g_1=1}^{G}\sum_{i=1}^{N_{g}}\sum_{j=1}^{N_{g_1}}\mathbb{E}\left[\epsilon_{gi}^{2}\epsilon_{g_1j}^{2}\right]\\
 &\leq \sum_{g=1}^{G}\sum_{g_1=1}^{G}\sum_{i=1}^{N_{g}}\sum_{j=1}^{N_{g_1}}\sqrt{\mathbb{E}\left[\epsilon_{gi}^{4}\right]\mathbb{E}\left[\epsilon_{g_1j}^{4}\right]}\\
 &= O\left(\left(\sum_{g=1}^{G}N_{g}\right)^{2}\right).
\end{align*}
Note that using the independence across clusters by Assumption \ref{assump:error independence}, using Assumption \ref{assump: Xg'Xg=O(Ng)}(ii) with $q=2,$ and applying Lemma \ref{lemma:Exact order summability}, we obtain
\begin{align*}
    \mathbb{E}\left[\left\Vert X'X\right\Vert^2\right] & = \sum_{g=1}^{G} \mathbb{E}\left[\left\Vert X_g'X_g\right\Vert^2\right] + \sum_{g=1}^{G} \sum_{\substack{g_1=1 \\ g_1 \neq g}}^{G} \operatorname{tr}\left( \mathbb{E}\left[X_g'X_g\right] \mathbb{E}\left[X_{g_1}'X_{g_1}\right] \right) \\
    & = O\left(\left(\sum_{g=1}^{G} N_g\right)^2\right).
\end{align*}
Hence, we have $\mathbb{E}\left[\left\Vert Z_{G}\right\Vert^{4}\right]=O(1)$. Therefore, the fourth moment of the norm of $Z_{G}$ is bounded. We have also shown that
the second moment of the norm of $Z_{G}$ is bounded away from zero, that is, $Z_{G}$ does not converge to $0$ in the mean square. Since the boundedness of the $(2+\delta)^{th}$ moment for some $0<\delta<\infty$ implies uniform integrability of $\left\{ \left\Vert Z_{G}\right\Vert^{2}:G\in\mathbb{N}\right\},$ the family is uniformly integrable. So, we can claim that $Z_{G}$
does not converge to $0$ in probability, since probability
convergence and uniform integrability in $\mathscr{L}^{p}$ implies
convergence at the $p^{th}$ moment. Hence, $$Z_{G}=\frac{X^{\prime}\epsilon}{\sum_{g=1}^{G}N_{g}}\neq o_P(1),$$ which deduces the fact that $\hat{\beta}_{POLS}$ is inconsistent for $\beta.$

\hfill
\qed

\textbf{Proof of Theorem \ref{thm:ols normality}}

To show the asymptotic normality of $\hat{\beta}_{POLS}$, first note that 
\[
\hat{\beta}_{POLS}-\beta = \left(\frac{{X^{\prime}X}}{{\sum_{g=1}^{G}N_{g}}}\right)^{-1}\frac{1}{\sum_{g=1}^{G}N_{g}}\sum_{g=1}^{G}X_{g}^{\prime}\epsilon_{g},
\]
 which implies 
 $$
\begin{aligned}
\frac{\sum_{g=1}^G N_g}{\sqrt{\sum_{g=1}^{G}N_{g}^{2}}}W_{G}^{-1/2}\left(\hat{\beta}_{POLS}-\beta\right) 
&= W_{G}^{-1/2}C_{G}^{-1}\frac{1}{\sqrt{\sum_{g=1}^{G}N_{g}^{2}}}\sum_{g=1}^{G}X_{g}^{\prime}\epsilon_{g} \\
&\quad + W_{G}^{-1/2}\left[\left(\frac{X^{\prime}X}{\sum_{g=1}^{G}N_{g}}\right)^{-1}-C_{G}^{-1}\right] \\
&\quad \times D_{G}^{1/2}D_{G}^{-1/2}\frac{1}{\sqrt{\sum_{g=1}^{G}N_{g}^{2}}}\sum_{g=1}^{G}X_{g}^{\prime}\epsilon_{g}.
\end{aligned}
$$
We prove this theorem in three steps.

\textbf{Step 1:} Here, we show that $D_{G}^{-1/2}\frac{1}{\sqrt{\sum_{g=1}^{G}N_{g}^{2}}}\sum_{g=1}^{G}X_{g}^{\prime}\epsilon_{g}\xrightarrow{d}N_{k}\left(0,I_{k}\right),$ as $G\to\infty.$

To prove this asymptotic normality, it suffices to prove the Lyapunov
condition for CLT. To prove the Lyapunov condition, it is required to
show that for any fixed $z\in\mathbb{R}^{k}$ and for some $p>1,$ 
\begin{align*}
\frac{\sum_{g=1}^{G}\mathbb{E} \left|  z^{\prime}D_{G}^{-1/2}\frac{1}{\sqrt{\sum_{g=1}^{G}N_{g}^{2}}}X_{g}^{\prime}\epsilon_{g} \right| ^{2p}}{\left(\sum_{g=1}^{G}\mathbb{E}\left[z^{\prime}D_{G}^{-1/2}\frac{1}{\sqrt{\sum_{g=1}^{G}N_{g}^{2}}}X_{g}^{\prime}\epsilon_{g}\right]^{2}\right)^{p}} & \to0,\text{ as }G\to\infty\\
\implies\frac{\sum_{g=1}^{G}\mathbb{E} \left|  z^{\prime}D_{G}^{-1/2}X_{g}^{\prime}\epsilon_{g} \right| ^{2p}}{\left(\sum_{g=1}^{G}\mathbb{E}\left[z^{\prime}D_{G}^{-1/2}X_{g}^{\prime}\epsilon_{g}\right]^{2}\right)^{p}} & \to0,\text{ as }G\to\infty.
\end{align*}
For the numerator, applying Result \ref{result:trace inequality} in the second and third lines yields
\begin{eqnarray}
& & \sum_{g=1}^{G}\mathbb{E} \left|  z^{\prime}D_{G}^{-1/2}X_{g}^{\prime}\epsilon_{g} \right| ^{2p} \nonumber\\
 & = & \sum_{g=1}^{G}\mathbb{E}\mathbb{E}\left[\left(z^{\prime}D_{G}^{-1/2}X_{g}^{\prime}\epsilon_{g}\epsilon_{g}^{\prime}X_{g}D_{G}^{-1/2}z\right)^{p} \big|  X_g\right]\nonumber\\
 & \leq & \sum_{g=1}^{G}\mathbb{E}\mathbb{E}\left[\left(z^{\prime}D_{G}^{-1}z\right)^{p}\left[\operatorname{tr}\left(X_{g}^{\prime}\epsilon_{g}\epsilon_{g}^{\prime}X_{g}\right)\right]^{p} \big|  X_g\right]\nonumber\\
 & \leq & \left(z^{\prime}D_{G}^{-1}z\right)^{p}\sum_{g=1}^{G}\mathbb{E}\left[\left(\operatorname{tr}\left(X_{g}^{\prime}X_{g}\right)\right)^{p}\mathbb{E}\left[\left(\epsilon_{g}^{\prime}\epsilon_{g}\right)^{p} \big|  X_g\right]\right]. \label{eq:normalOLS1}
\end{eqnarray}

Now, using H\"{o}lder's inequality and Assumption \ref{assump:additional assumption on errors}(i),
\[
\mathbb{E}\left[\left(\epsilon_{g}^{\prime}\epsilon_{g}\right)^{p} \big| X_g\right]\leq \mathbb{E}\left[\left(\sum_{i=1}^{N_{g}}1^{\frac{p}{p-1}}\right)^{p-1}\sum_{i=1}^{N_{g}} \left| \epsilon_{gi} \right| ^{2p} \Bigg| X_g\right]=O\left(N_{g}^{p}\right), \text{ uniformly in } g.
\]

Also, for the denominator, we have
\begin{eqnarray*}
\left(\sum_{g=1}^{G}\mathbb{E}\left[z^{\prime}D_{G}^{-1/2}X_{g}^{\prime}\epsilon_{g}\right]^{2}\right)^{p} & = & \left(\sum_{g=1}^{G}\mathbb{E}\mathbb{E}\left[z^{\prime}D_{G}^{-1/2}X_{g}^{\prime}\epsilon_{g}\epsilon_{g}^{\prime}X_{g}D_{G}^{-1/2}z \big|  X_g\right]\right)^{p}\\
 & = & \left(\sum_{g=1}^{G}\mathbb{E}\left[z^{\prime}D_{G}^{-1/2}X_{g}^{\prime}\,\mathbb{E}\left[\epsilon_{g}\epsilon_{g}^{\prime} \big|  X_g\right]X_{g}D_{G}^{-1/2}z\right]\right)^{p}\\
 & = & \left(z^{\prime}D_{G}^{-1/2}\sum_{g=1}^{G}\mathbb{E}\left[X_{g}^{\prime}\Omega_{g}X_{g}\right]D_{G}^{-1/2}z\right)^{p}\\
 & = & \left(\sum_{g=1}^{G}N_{g}^{2}\right)^{p}\left(z^{\prime}D_{G}^{-1/2}D_{G}D_{G}^{-1/2}z\right)^{p}\\
 & = & \left(\sum_{g=1}^{G}N_{g}^{2}\right)^{p}\left(z^{\prime}z\right)^{p}.
\end{eqnarray*}

Note that by Assumption \ref{assump:lamba_max=lambda_min for OLS}, we have 
\begin{align*}
\frac{z^{\prime}D_{G}^{-1}z}{z^{\prime}z}\leq\lambda_{\max}\left(D_{G}^{-1}\right) & =\frac{1}{\lambda_{\min}\left(D_{G}\right)}\\
& = \begin{cases}
O\left(1\right), & \text{for strong dependence,}\\
O\left(\frac{\sum_{g=1}^{G}N_{g}^{2}}{\sum_{g=1}^{G}N_{g}h(N_{g})}\right), & \text{for semi-strong dependence,}\\
O\left(\frac{\sum_{g=1}^{G}N_{g}^{2}}{\sum_{g=1}^{G}N_{g}}\right), & \text{for weak dependence.}
\end{cases}
\end{align*}

Given the uniform moment bound $\mathbb{E}\left[\left(\operatorname{tr}\left(X_{g}^{\prime}X_{g}\right)\right)^{p}\right] = O\left(N_g^p \right)$ uniformly in $g$ by Assumption \ref{assump: Xg'Xg=O(Ng)}(ii), and applying Lemma \ref{lemma:Exact order summability} in  \eqref{eq:normalOLS1}, the Lyapunov condition reduces to 
\begin{align*}
& \quad \frac{\sum_{g=1}^{G}\mathbb{E} \left| z^{\prime}D_{G}^{-1/2}X_{g}^{\prime}\epsilon_{g} \right|^{2p}}{\left(\sum_{g=1}^{G}\mathbb{E}\left[z^{\prime}D_{G}^{-1/2}X_{g}^{\prime}\epsilon_{g}\right]^{2}\right)^{p}} \\
&\leq \frac{1}{\left(\sum_{g=1}^{G}N_{g}^{2}\right)^{p}}\frac{\left(z^{\prime}D_{G}^{-1}z\right)^{p}}{\left(z^{\prime}z\right)^{p}}\sum_{g=1}^{G}\mathbb{E}\left[\left(\operatorname{tr}\left(X_{g}^{\prime}X_{g}\right)\right)^{p}\mathbb{E}\left[\left(\epsilon_{g}^{\prime}\epsilon_{g}\right)^{p} \big|  X_g\right]\right]\\
&= \frac{\sum_{g=1}^{G}O\left(N_{g}^{2p}\right)}{\left(\sum_{g=1}^{G}N_{g}^{2}\right)^{p}} 
\begin{cases}
O\left(1\right), & \text{for strong dependence,}\\
O\left(\left(\frac{\sum_{g=1}^{G}N_{g}^{2}}{\sum_{g=1}^{G}N_{g}h(N_{g})}\right)^{p}\right), & \text{for semi-strong dependence,}\\
O\left(\left(\frac{\sum_{g=1}^{G}N_{g}^{2}}{\sum_{g=1}^{G}N_{g}}\right)^{p}\right), & \text{for weak dependence,}
\end{cases}\\
&= \begin{cases}
O\left(\frac{\sum_{g=1}^{G}N_{g}^{2p}}{\left(\sum_{g=1}^{G}N_{g}^{2}\right)^{p}}\right), & \text{for strong dependence,}\\
O\left(\frac{\sum_{g=1}^{G}N_{g}^{2p}}{\left(\sum_{g=1}^{G}N_{g}h(N_{g})\right)^{p}}\right), & \text{for semi-strong dependence,}\\
O\left(\frac{\sum_{g=1}^{G}N_{g}^{2p}}{\left(\sum_{g=1}^{G}N_{g}\right)^{p}}\right), & \text{for weak dependence.}
\end{cases}
\end{align*}
Hence, the Lyapunov condition for CLT holds under the restrictive conditions on the cluster sizes.

\textbf{Step 2:} We show that
\[
W_{G}^{-1/2}\left[\left(\frac{X^{\prime}X}{\sum_{g=1}^{G}N_{g}}\right)^{-1}-C_{G}^{-1}\right]D_{G}^{1/2}D_{G}^{-1/2}\frac{1}{\sqrt{\sum_{g=1}^{G}N_{g}^{2}}}\sum_{g=1}^{G}X_{g}^{\prime}\epsilon_{g}=o_{P}(1).
\]
Under Assumption \ref{assump:lamba_max=lambda_min for OLS}, for strong dependence, we have $D_{G}$ is finite and positive definite,
for semi-strong dependence, we have $\frac{\sum_{g=1}^{G}N_{g}^{2}}{\sum_{g=1}^{G}N_{g}h(N_{g})}D_{G}$ is finite and positive definite,
and for weak dependence, we have $\frac{\sum_{g=1}^{G}N_{g}^{2}}{\sum_{g=1}^{G}N_{g}}D_{G}$ is finite and positive definite
This also implies that for strong dependence, $W_{G}$ is finite and positive definite for semi-strong dependence, we have $\frac{\sum_{g=1}^{G}N_{g}^{2}}{\sum_{g=1}^{G}N_{g}h(N_{g})}W_{G}$ is finite and positive definite,
and for weak dependence, we have $\frac{\sum_{g=1}^{G}N_{g}^{2}}{\sum_{g=1}^{G}N_{g}}W_{G}$ is finite and positive definite,
since $C_{G}$ is uniformly positive definite, by Assumption \ref{assump: Xg'Xg=O(Ng)}. Now, using $\left(\frac{X^{\prime}X}{\sum_{g=1}^{G}N_{g}}\right)^{-1}-C_{G}^{-1}=o_P(1),$ and $D_{G}^{-1/2}\frac{1}{\sqrt{\sum_{g=1}^{G}N_{g}^{2}}}\sum_{g=1}^{G}X_{g}^{\prime}\epsilon_{g}=O_{P}(1),$ Step 2 holds by Lemma \ref{lemma:big-OP sum and multiplication}. 

\textbf{Step 3:} We show that $W_{G}^{-1/2}C_{G}^{-1}\frac{1}{\sqrt{\sum_{g=1}^{G}N_{g}^{2}}}\sum_{g=1}^{G}X_{g}^{\prime}\epsilon_{g}\xrightarrow{d}N_{k}\left(0,I_{k}\right).$ 

Since $\left\{ D_{G}\right\} $ and $\left\{ D_{G}^{-1}\right\} $
are finite and positive definite, by Assumption \ref{assump:lamba_max=lambda_min for OLS}, $\left\{ C_{G}\right\} $
is finite and uniformly positive definite by Assumption \ref{assump: Xg'Xg=O(Ng)}
and $D_{G}^{-1/2}\frac{1}{\sqrt{\sum_{g=1}^{G}N_{g}^{2}}}\sum_{g=1}^{G}X_{g}^{\prime}\epsilon_{g}\xrightarrow{d}N_{k}\left(0,I_{k}\right),$
from Step 1, it immediately follows from Corollary 4.24 of \cite{white1984asymptotic} that
\begin{align*}
    W_{G}^{-1/2}C_{G}^{-1}\frac{1}{\sqrt{\sum_{g=1}^{G}N_{g}^{2}}}\sum_{g=1}^{G}X_{g}^{\prime}\epsilon_{g} & = W_{G}^{-1/2}C_{G}^{-1}D_G^{1/2}D_G^{-1/2}\frac{1}{\sqrt{\sum_{g=1}^{G}N_{g}^{2}}}\sum_{g=1}^{G}X_{g}^{\prime}\epsilon_{g} \\
    & \xrightarrow{d}N_{k}\left(0,I_{k}\right),
\end{align*}
since $W_{G}^{-1/2}C_{G}^{-1}D_G^{1/2}D_G^{1/2}C_G^{-1/2}W_G^{-1/2}=W_G^{-1/2}W_G W_G^{-1/2}=I_k.$

From the above three steps, the asymptotic normality of the POLS estimator holds (i) for strong dependence if $\frac{\sum_{g=1}^G N_g^{2p}}{\left(\sum_{g=1}^G N_g^2\right)^p}\to 0$, (ii) for semi-strong dependence if $\frac{\sum_{g=1}^G N_g^{2p}}{\left(\sum_{g=1}^G N_g h(N_g)\right)^p}\to 0$, (iii) for weak dependence if $\frac{\sum_{g=1}^G N_g^{2p}}{\left(\sum_{g=1}^G N_g\right)^p}\to 0$. 

\hfill
\qed

For the sake of completeness, we provide the following lemma.
\begin{lemma}\label{lemma:lambda_min lambda_max inequality}
Suppose that $A$ is a $m\times n$ matrix with
full row rank $m(\leq n),$ where $m$ is finite. Then, for any $n\times n$
positive definite matrix $B,$  $\lambda_{\min}\left(ABA^{\prime}\right)\ge\lambda_{\min}\left(AA^{\prime}\right)\lambda_{\min}\left(B\right)$,
and $\lambda_{\max}\left(ABA^{\prime}\right)\leq\lambda_{\max}\left(AA^{\prime}\right)\lambda_{\max}\left(B\right).$
\end{lemma}

\textbf{Proof of Lemma \ref{lemma:lambda_min lambda_max inequality}}

Note that 
\[
\lambda_{\min}\left(ABA^{\prime}\right)=\underset{z\in\mathbb{R}^{m}}{min}\frac{z^{\prime}ABA^{\prime}z}{z^{\prime}z}\geq\underset{z\in\mathbb{R}^{m}}{min}\frac{z^{\prime}AA^{\prime}z}{z^{\prime}z}\lambda_{\min}\left(B\right)\ge\lambda_{\min}\left(AA^{\prime}\right)\lambda_{\min}\left(B\right),
\]

and
\[
\lambda_{\max}\left(ABA^{\prime}\right)=\underset{z\in\mathbb{R}^{m}}{max}\frac{z^{\prime}ABA^{\prime}z}{z^{\prime}z}\leq\underset{z\in\mathbb{R}^{m}}{max}\frac{z^{\prime}AA^{\prime}z}{z^{\prime}z}\lambda_{\max}\left(B\right)\leq\lambda_{\max}\left(AA^{\prime}\right)\lambda_{\max}\left(B\right).
\]

\hfill
\qed

\begin{lemma}\label{lemma:order of beta ols -beta}
For both (i) $\left\Vert\frac{X^{\prime}\epsilon}{\sum_{g=1}^{G}N_{g}}\right\Vert$ and (ii) $\left\Vert\hat{\beta}_{POLS}-\beta\right\Vert$, the asymptotic order is given by:
\[
\begin{cases}
O_{P}\left(\frac{\sqrt{\sum_{g=1}^{G}N_{g}^{2}}}{\sum_{g=1}^{G}N_{g}}\right), & \text{for strong dependence,}\\[2ex]
O_{P}\left(\frac{\sqrt{\sum_{g=1}^{G}N_{g}h(N_{g})}}{\sum_{g=1}^{G}N_{g}}\right), & \text{for semi-strong dependence,}\\[2ex]
O_{P}\left(\frac{1}{\sqrt{\sum_{g=1}^{G}N_{g}}}\right), & \text{for weak dependence.}
\end{cases}
\]
\end{lemma}
 
\textbf{Proof of Lemma \ref{lemma:order of beta ols -beta}}

(i) Since the variance-covariance matrix of $X^{\prime}\epsilon$ is finite dimensional, it follows from \eqref{eq: Variance order of X'epsilon term 2} given in the proof of Theorem \ref{thm:ols consistency} and Assumption \ref{assump:error independence} that
\[
\mathbb{E}\left\Vert\frac{X^{\prime}\epsilon}{\sum_{g=1}^{G}N_{g}}\right\Vert^{2} = \begin{cases}
O\left(\frac{\sum_{g=1}^{G}N_{g}^{2}}{\left(\sum_{g=1}^{G}N_{g}\right)^{2}}\right), & \text{for strong dependence,}\\
O\left(\frac{\sum_{g=1}^{G}N_{g}h(N_{g})}{\left(\sum_{g=1}^{G}N_{g}\right)^{2}}\right), & \text{for semi-strong dependence,}\\
O\left(\frac{1}{\sum_{g=1}^{G}N_{g}}\right), & \text{for weak dependence,}
\end{cases}
\]
which proves the first part of this lemma.

(ii) Note that using the norm inequality for multiplication
in the third line, the triangular inequality in the fourth line, and applying  Lemma \ref{lemma:big-OP sum and multiplication} and Lemma \ref{lemma:order of beta ols -beta}(i) in the last line,
\begin{align*}
\left\Vert\hat{\beta}_{POLS}-\beta\right\Vert & =\left\Vert\left(\frac{X^{\prime}X}{\sum_{g=1}^{G}N_{g}}\right)^{-1}\frac{X^{\prime}\epsilon}{\sum_{g=1}^{G}N_{g}}\right\Vert\\
 & =\left\Vert\left(C_{G}^{-1}+\left(\frac{X^{\prime}X}{\sum_{g=1}^{G}N_{g}}\right)^{-1}-C_{G}^{-1}\right)\frac{X^{\prime}\epsilon}{\sum_{g=1}^{G}N_{g}}\right\Vert\\
 & \leq\left\Vert C_{G}^{-1}+\left(\frac{X^{\prime}X}{\sum_{g=1}^{G}N_{g}}\right)^{-1}-C_{G}^{-1}\right\Vert\left\Vert\frac{X^{\prime}\epsilon}{\sum_{g=1}^{G}N_{g}}\right\Vert\\
 & \leq\left[\left\Vert C_{G}^{-1}\right\Vert+\left\Vert\left(\frac{X^{\prime}X}{\sum_{g=1}^{G}N_{g}}\right)^{-1}-C_{G}^{-1}\right\Vert\right]\left\Vert\frac{X^{\prime}\epsilon}{\sum_{g=1}^{G}N_{g}}\right\Vert\\
 & =\left[O(1)+o_{P}(1)\right]\left\Vert\frac{X^{\prime}\epsilon}{\sum_{g=1}^{G}N_{g}}\right\Vert\\
 & =\begin{cases}
O_{P}\left(\frac{\sqrt{\sum_{g=1}^{G}N_{g}^{2}}}{\sum_{g=1}^{G}N_{g}}\right), & \text{for strong dependence,}\\
O_{P}\left(\frac{\sqrt{\sum_{g=1}^{G}N_{g}h(N_{g})}}{\sum_{g=1}^{G}N_{g}}\right), & \text{for semi-strong dependence,}\\
O_{P}\left(\frac{1}{\sqrt{\sum_{g=1}^{G}N_{g}}}\right), & \text{for weak dependence.}
\end{cases}
\end{align*}

\hfill
\qed

\begin{lemma}\label{lemma:order of Xg epsilon g OLS}
$\left\Vert\epsilon_{g}^{\prime}X_{g}\right\Vert=\begin{cases}
O_{P}\left(N_{g}\right), & \text{for strong dependence,}\\
O_{P}\left(\sqrt{N_{g}h(N_{g})}\right), & \text{for semi-strong dependence,}\\
O_{P}\left(\sqrt{N_{g}}\right), & \text{for weak dependence,}
\end{cases}$ \\
uniformly in $g.$
\end{lemma}

\textbf{Proof of Lemma \ref{lemma:order of Xg epsilon g OLS}}

Note that 
\[
\mathbb{E}\left\Vert\epsilon_{g}^{\prime}X_{g}\right\Vert^{2} = \mathbb{E}\mathbb{E}\left[\operatorname{tr}\left(X_{g}^{\prime}\epsilon_{g}\epsilon_{g}^{\prime}X_{g}\right) \big|  X_g\right] = \mathbb{E}\left[\operatorname{tr}\left(X_{g}^{\prime}\Omega_{g}X_{g}\right)\right].
\]

By Result \ref{result:trace inequality}, $\mathbb{E}\left[\operatorname{tr}\left({X}'_{g}\Omega_g{X}_{g}\right)\right] \leq \mathbb{E}\left[\operatorname{tr}\left({X}'_{g}{X}_{g}\right)\right] \lambda_{\max}(\Omega_g)$. Combined with the condition that $\mathbb{E}\left[\operatorname{tr}\left(X_{g}^{\prime}X_{g}\right)\right]=O\left(N_{g}\right)$ uniformly in $g$ (Assumption \ref{assump: Xg'Xg=O(Ng)}), it follows from Assumption \ref{assump:error independence} that
\[
\mathbb{E}\left\Vert\epsilon_{g}^{\prime}X_{g}\right\Vert^{2} = 
\begin{cases}
O\left(N_{g}^{2}\right), & \text{for strong dependence,}\\
O\left(N_{g}h(N_{g})\right), & \text{for semi-strong dependence,}\\
O\left(N_{g}\right), & \text{for weak dependence,}
\end{cases}
\]
uniformly in $g.$ Hence, the lemma holds.

\hfill
\qed

\textbf{Proof of Theorem \ref{thm:Vols_consistent}} 

To establish the consistency of $\hat{W}_{G}$ for $W_{G}$, it suffices to show that $\frac{1}{\left\Vert W_{G}\right\Vert}\left\Vert\hat{W}_{G}-W_{G}\right\Vert=o_{P}(1).$ We first show that $\frac{1}{\left\Vert D_{G}\right\Vert}\left\Vert\hat{D}_{G}-D_{G}\right\Vert=o_{P}(1).$
Note that $u_{g}=Y_{g}-X_{g}\hat{\beta}_{POLS}=\epsilon_{g}-X_{g}\left(\hat{\beta}_{POLS}-\beta\right)$, which implies
\begin{align*}
u_{g}u_{g}^{\prime} &= \epsilon_{g}\epsilon_{g}^{\prime} + X_{g}\left(\hat{\beta}_{POLS}-\beta\right)\left(\hat{\beta}_{POLS}-\beta\right)^{\prime}X_{g}^{\prime} \\
&\quad - X_{g}\left(\hat{\beta}_{POLS}-\beta\right)\epsilon_{g}^{\prime} - \epsilon_{g}\left(\hat{\beta}_{POLS}-\beta\right)^{\prime}X_{g}^{\prime}.
\end{align*}
Consequently, we can decompose the difference $\hat{D}_{G}-D_{G}$ as
\begin{align*}
\hat{D}_{G}-D_{G} &= \frac{1}{\sum_{g=1}^{G}N_{g}^{2}}\sum_{g=1}^{G}X_{g}^{\prime}u_{g}u_{g}^{\prime}X_{g}-\mathbb{E}\left[\frac{1}{\sum_{g=1}^{G}N_{g}^{2}}\sum_{g=1}^{G}X_{g}^{\prime}\Omega_{g}X_{g}\right] \\
 &= T_{1} + T_{2} + T_{3} + T_{4},
\end{align*}
where the individual terms are defined as follows:
\begin{align*}
T_{1} &= \frac{1}{\sum_{g=1}^{G}N_{g}^{2}}\sum_{g=1}^{G}X_{g}^{\prime}\epsilon_{g}\epsilon_{g}^{\prime}X_{g}-\mathbb{E}\left[\frac{1}{\sum_{g=1}^{G}N_{g}^{2}}\sum_{g=1}^{G}X_{g}^{\prime}\Omega_{g}X_{g}\right], \\
T_{2} &= \frac{1}{\sum_{g=1}^{G}N_{g}^{2}}\sum_{g=1}^{G}X_{g}^{\prime}X_{g}\left(\hat{\beta}_{POLS}-\beta\right)\left(\hat{\beta}_{POLS}-\beta\right)^{\prime}X_{g}^{\prime}X_{g}, \\
T_{3} &= -\frac{1}{\sum_{g=1}^{G}N_{g}^{2}}\sum_{g=1}^{G}X_{g}^{\prime}X_{g}\left(\hat{\beta}_{POLS}-\beta\right)\epsilon_{g}^{\prime}X_{g}, \\
T_{4} &= -\frac{1}{\sum_{g=1}^{G}N_{g}^{2}}\sum_{g=1}^{G}X_{g}^{\prime}\epsilon_{g}\left(\hat{\beta}_{POLS}-\beta\right)^{\prime}X_{g}^{\prime}X_{g}.
\end{align*}

Now, using Assumption \ref{assump:error independence}, by the law of iterated expectations conditioning on $X$, we have
\[
\mathbb{E}\left[T_{1}\right]=\mathbb{E}\left[\frac{1}{\sum_{g=1}^{G}N_{g}^{2}}\sum_{g=1}^{G}X_{g}^{\prime}\mathbb{E}\left[\epsilon_{g}\epsilon_{g}^{\prime} |  X_g\right]X_{g}-\frac{1}{\sum_{g=1}^{G}N_{g}^{2}}\sum_{g=1}^{G}X_{g}^{\prime}\Omega_{g}X_{g}\right]=0,
\]

and using the independence across clusters in the third line and result
\ref{result:trace inequality} in the fourth line, we have
\begin{eqnarray*}
\mathbb{E}\left\Vert T_{1}\right\Vert^{2} & = & \frac{1}{\left(\sum_{g=1}^{G}N_{g}^{2}\right)^{2}}\mathbb{E}\left\Vert\sum_{g=1}^{G}X_{g}^{\prime}\left(\epsilon_{g}\epsilon_{g}^{\prime}-\Omega_{g}\right)X_{g}\right\Vert^{2}\\
 & = & \frac{1}{\left(\sum_{g=1}^{G}N_{g}^{2}\right)^{2}}\mathbb{E}\left[\left\Vert\sum_{g=1}^{G}X_{g}^{\prime}\left(\epsilon_{g}\epsilon_{g}^{\prime}-\Omega_{g}\right)X_{g}\right\Vert^{2}\right]\\
 & = & \frac{1}{\left(\sum_{g=1}^{G}N_{g}^{2}\right)^{2}}\mathbb{E}\left[\sum_{g=1}^{G}\left\Vert X_{g}^{\prime}\left(\epsilon_{g}\epsilon_{g}^{\prime}-\Omega_{g}\right)X_{g}\right\Vert^{2} \right]\\
 & \leq & \frac{1}{\left(\sum_{g=1}^{G}N_{g}^{2}\right)^{2}}\sum_{g=1}^{G}\mathbb{E}\left[\operatorname{tr}^2\left(X_{g}^{\prime}X_{g}\right)\left\Vert\left(\epsilon_{g}\epsilon_{g}^{\prime}-\Omega_{g}\right)\right\Vert^{2}\right]\\
 & = & {\left(\sum_{g=1}^{G}N_{g}^{2}\right)^{2}}\sum_{g=1}^{G}\mathbb{E}\left[\operatorname{tr}^2\left(X_{g}^{\prime}X_{g}\right)\mathbb{E}\left[\left\Vert\left(\epsilon_{g}\epsilon_{g}^{\prime}-\Omega_{g}\right)\right\Vert^{2}\big| X_g\right]\right].
\end{eqnarray*}

Using the fact that $\mathbb{E}\left[\Vert\epsilon_{g}\epsilon_{g}^{\prime}\Vert^{2} \big|  X_g\right]=O\left(N_{g}^{2}\right)$ uniformly in $g,$ from the proof of Theorem \ref{thm:ols normality} with $p=2$, the following holds uniformly in $g$:
\[
\mathbb{E}\left[\left\Vert\left(\epsilon_{g}\epsilon_{g}^{\prime}-\Omega_{g}\right)\right\Vert^{2} \Big|  X_g\right]\leq2\left(\mathbb{E}\left[\left\Vert\epsilon_{g}\epsilon_{g}^{\prime}\right\Vert^{2} \Big|  X_g\right]+\left\Vert\Omega_{g}\right\Vert^{2}\right)=O\left(N_{g}^{2}\right).
\] 
Since $\mathbb{E}\left[\operatorname{tr}^2\left(X_{g}^{\prime}X_{g}\right)\right]=O\left(N_{g}^{2}\right)$ uniformly in $g$ by Assumption \ref{assump: Xg'Xg=O(Ng)}(ii) with $q=2$, applying Lemma \ref{lemma:Exact order summability}, we have $\mathbb{E}\left\Vert T_{1}\right\Vert^{2}=O\left(\frac{\sum_{g=1}^{G}N_{g}^{4}}{\left(\sum_{g=1}^{G}N_{g}^{2}\right)^{2}}\right)$, which further deduces that
\[
\left\Vert T_{1}\right\Vert=O_{P}\left(\frac{\sqrt{\sum_{g=1}^{G}N_{g}^{4}}}{\sum_{g=1}^{G}N_{g}^{2}}\right).
\]

Now, using the triangular inequality in the first line and the norm inequality in the second line,  we have 
\begin{eqnarray*}
\left\Vert T_{2} \right\Vert & \leq & \frac{1}{\sum_{g=1}^{G}N_{g}^{2}}\sum_{g=1}^{G}\left\Vert X_{g}^{\prime}X_{g}\left(\hat{\beta}_{POLS}-\beta\right)\left(\hat{\beta}_{POLS}-\beta\right)^{\prime}X_{g}^{\prime}X_{g}\right\Vert\\
& \leq & \frac{1}{\sum_{g=1}^{G}N_{g}^{2}}\sum_{g=1}^{G}\left\Vert X_{g}^{\prime}X_{g}\right\Vert^{2}\left\Vert\left(\hat{\beta}_{POLS}-\beta\right)\left(\hat{\beta}_{POLS}-\beta\right)^{\prime}\right\Vert\\
& = &  \left\Vert\hat{\beta}_{POLS}-\beta\right\Vert^2 \frac{1}{\sum_{g=1}^{G}N_{g}^{2}}\sum_{g=1}^{G}\left\Vert X_{g}^{\prime}X_{g}\right\Vert^{2}.
\end{eqnarray*}

Since $\mathbb{E}\left[\left\Vert X_{g}^{\prime}X_{g}\right\Vert^{2}\right]= O\left(N_g^2\right)$ uniformly in $g$ by Assumption \ref{assump: Xg'Xg=O(Ng)}(ii) with $q=2$, it follows from Lemma \ref{lemma:big-OP sum and multiplication} and Lemma \ref{lemma:order of beta ols -beta} that
\begin{align*}
\left\Vert T_{2} \right\Vert &= \begin{cases}
O_{P}\left(\frac{\sum_{g_{1}=1}^{G}N_{g_{1}}^{2}}{\left(\sum_{g_{1}=1}^{G}N_{g_{1}}\right)^{2}}\frac{1}{\sum_{g=1}^{G}N_{g}^{2}}\sum_{g=1}^{G} N_{g}^{2} \right), & \text{for strong dependence,}\\
O_{P}\left(\frac{\sum_{g_{1}=1}^{G}N_{g_{1}}h(N_{g_{1}})}{\left(\sum_{g_{1}=1}^{G}N_{g_{1}}\right)^{2}}\frac{1}{\sum_{g=1}^{G}N_{g}^{2}}\sum_{g=1}^{G} N_{g}^{2} \right), & \text{for semi-strong dependence,}\\
O_{P}\left(\frac{\sum_{g_{1}=1}^{G}N_{g_{1}}}{\left(\sum_{g_{1}=1}^{G}N_{g_{1}}\right)^{2}}\frac{1}{\sum_{g=1}^{G}N_{g}^{2}}\sum_{g=1}^{G}N_{g}^{2} \right), & \text{for weak dependence,}
\end{cases} \\
&= \begin{cases}
O_{P}\left(\frac{\sum_{g=1}^{G}N_{g}^{2}}{\left(\sum_{g=1}^{G}N_{g}\right)^{2}}\right), & \text{for strong dependence,}\\[2ex]
O_{P}\left(\frac{\sum_{g=1}^{G}N_{g}h(N_{g})}{\left(\sum_{g=1}^{G}N_{g}\right)^{2}}\right), & \text{for semi-strong dependence,}\\[2ex]
O_{P}\left(\frac{1}{\sum_{g=1}^{G}N_{g}}\right), & \text{for weak dependence.}
\end{cases}
\end{align*}

Using the triangular inequality in the first line and the norm inequality in the second line, we have 
\begin{eqnarray*}
\left\Vert T_{3} \right\Vert & \leq & \frac{1}{\sum_{g=1}^{G}N_{g}^{2}}\sum_{g=1}^{G}\left\Vert X_{g}^{\prime}X_{g}\left(\hat{\beta}_{POLS}-\beta\right)\epsilon_{g}^{\prime}X_{g}\right\Vert\\
& \leq & \frac{1}{\sum_{g=1}^{G}N_{g}^{2}}\sum_{g=1}^{G}\left\Vert X_{g}^{\prime}X_{g}\right\Vert\left\Vert\hat{\beta}_{POLS}-\beta\right\Vert\left\Vert\epsilon_{g}^{\prime}X_{g}\right\Vert\\
 & = & \left\Vert\hat{\beta}_{POLS}-\beta\right\Vert \frac{1}{\sum_{g=1}^{G}N_{g}^{2}}\sum_{g=1}^{G}\left\Vert X_{g}^{\prime}X_{g}\right\Vert\left\Vert\epsilon_{g}^{\prime}X_{g}\right\Vert.
\end{eqnarray*}

Using the C-S inequality in the first line and applying Assumption \ref{assump: Xg'Xg=O(Ng)}, Lemma \ref{lemma:big-OP sum and multiplication} and Lemma \ref{lemma:order of Xg epsilon g OLS} in the second line, for some $0<a,b,c<\infty,$
\begin{eqnarray*}
    \mathbb{E}\left[\left\Vert X_{g}^{\prime}X_{g}\right\Vert\left\Vert\epsilon_{g}^{\prime}X_{g}\right\Vert\right] & \leq & \sqrt{\mathbb{E}\left[\left\Vert X_{g}^{\prime}X_{g}\right\Vert^2\right]} \sqrt{\mathbb{E}\left[\left\Vert\epsilon_{g}^{\prime}X_{g}\right\Vert^2\right]} \\
    & \leq & \begin{cases}
aN^2_g, & \text{for strong dependence,}\\
bN_{g}^{3/2}\sqrt{h(N_{g})}, & \text{for semi-strong dependence,}\\
cN_{g}^{3/2}, & \text{for weak dependence,}
\end{cases} 
\end{eqnarray*}
for all $g.$ Therefore, applying Assumption \ref{assump: Xg'Xg=O(Ng)}, Lemma \ref{lemma:big-OP sum and multiplication}, Lemma \ref{lemma:order of beta ols -beta} and Lemma \ref{lemma:order of Xg epsilon g OLS}, it follows that
\[
\Vert T_3 \Vert = 
\begin{cases}
O_{P}\left(\frac{\sqrt{\sum_{g_{1}=1}^{G}N_{g_{1}}^{2}}}{\sum_{g_{1}=1}^{G}N_{g_{1}}}\frac{\sum_{g=1}^{G}N_{g}^{2}}{\sum_{g=1}^{G}N_{g}^{2}}\right), & \text{for strong dependence,}\\
O_{P}\left(\frac{\sqrt{\sum_{g_{1}=1}^{G}N_{g_{1}}h(N_{g_{1}})}}{\sum_{g_{1}=1}^{G}N_{g_{1}}}\frac{\sum_{g=1}^{G}N_{g}^{3/2}\sqrt{h(N_{g})}}{\sum_{g=1}^{G}N_{g}^{2}}\right), & \text{for semi-strong dependence,}\\
O_{P}\left(\frac{\sqrt{\sum_{g_{1}=1}^{G}N_{g_{1}}}}{\sum_{g_{1}=1}^{G}N_{g_{1}}}\frac{\sum_{g=1}^{G}N_{g}^{3/2}}{\sum_{g=1}^{G}N_{g}^{2}}\right), & \text{for weak dependence.}
\end{cases}
\]
Since $\sum_{g=1}^G N_g \sqrt{N_gh(N_g)}\leq \sqrt{\sum_{g=1}^G N_g^2}\sqrt{\sum_{g=1}^G N_g h(N_g)},$ and $\sum_{g=1}^G N_g^{3/2}\leq \sqrt{\sum_{g=1}^G N_g^2}\sqrt{\sum_{g=1}^G N_g}$ hold by the Cauchy-Schwarz inequality, it follows that
\[
\Vert T_3 \Vert \,=\,
\begin{cases}
O_{P}\left(\frac{\sqrt{\sum_{g=1}^{G}N_{g}^{2}}}{\sum_{g=1}^{G}N_{g}}\right), & \text{for strong dependence,}\\
O_{P}\left(\frac{\sum_{g=1}^{G}N_{g}h(N_{g})}{\sum_{g=1}^{G}N_{g}\sqrt{\sum_{g=1}^G N_g^2}} \right), & \text{for semi-strong dependence,}\\
O_{P}\left(\frac{1}{\sqrt{\sum_{g=1}^G N_g^2}}\right), & \text{for weak dependence.}
\end{cases}
\]
Also, $\Vert T_4\Vert = \Vert T_3\Vert.$ Since $\left\Vert T_{1} \right\Vert$ grows stochastically faster than $\left\Vert T_{2} \right\Vert,\left\Vert T_{3} \right\Vert$ and
$\left\Vert T_{4} \right\Vert$, for any kind of dependence, we obtain by the triangular inequality that
\[
\left\Vert\hat{D}_{G}-D_{G}\right\Vert \leq \left\Vert T_{1} \right\Vert+\left\Vert T_{2} \right\Vert+\left\Vert T_{3} \right\Vert+\left\Vert T_{4} \right\Vert=O_{P}\left(\frac{\sqrt{\sum_{g=1}^G N_g^4}}{\sum_{g=1}^G N_g^2}\right).
\]
Since $\left\Vert D_{G}\right\Vert=\begin{cases}
O_{e}\left(1\right), & \text{for strong dependence,}\\
O_{e}\left(\frac{\sum_{g=1}^{G}N_{g}h(N_{g})}{\sum_{g=1}^{G}N_{g}^{2}}\right), & \text{for semi-strong dependence,}\\
O_{e}\left(\frac{\sum_{g=1}^{G}N_{g}}{\sum_{g=1}^{G}N_{g}^{2}}\right), & \text{for weak dependence,}
\end{cases}$
by Assumption \ref{assump:lamba_max=lambda_min for OLS}, it follows that
\[
\frac{1}{\left\Vert D_{G}\right\Vert}\left\Vert\hat{D}_{G}-D_{G}\right\Vert=
\begin{cases}
O_{P}\left(\frac{\sqrt{\sum_{g=1}^{G}N_{g}^{4}}}{\sum_{g=1}^{G}N_{g}^2}\right), & \text{for strong dependence,}\\
O_{P}\left(\frac{\sqrt{\sum_{g=1}^G N_g^4}}{\sum_{g=1}^G N_g h(N_g)}\right), & \text{for semi-strong dependence,}\\
O_{P}\left(\frac{\sqrt{\sum_{g=1}^G N_g^4}}{\sum_{g=1}^G N_g}\right), & \text{for weak dependence.}
\end{cases}
\]
Hence, $\frac{1}{\left\Vert D_{G}\right\Vert}\left\Vert\hat{D}_{G}-D_{G}\right\Vert = o_P(1)$  for strong dependence if $\frac{\sum_{g=1}^G N_g^{4}}{\left(\sum_{g=1}^G N_g^2\right)^2}\to 0$,  for semi-strong dependence if $\frac{\sum_{g=1}^G N_g^{4}}{\left(\sum_{g=1}^G N_g h(N_g)\right)^2}\to 0$, and for weak dependence if $\frac{\sum_{g=1}^G N_g^{4}}{\left(\sum_{g=1}^G N_g\right)^2}\to 0$. 
Since $W_{G}=C_{G}^{-1}D_{G}C_{G}^{-1},$ it follows from Lemma \ref{lemma:lambda_min lambda_max inequality}
and Assumption \ref{assump: Xg'Xg=O(Ng)} that 
\begin{align*}
\lambda_{\min}\left(W_{G}\right) & \ge\lambda_{\min}\left(C_{G}^{-2}\right)\lambda_{\min}\left(D_{G}\right)=O_{e}(1)\lambda_{\min}\left(D_{G}\right) \\ \lambda_{\max}\left(W_{G}\right) & 
\leq\lambda_{\max}\left(C_{G}^{-2}\right)\lambda_{\max}\left(D_{G}\right)=O_{e}(1)\lambda_{\max}\left(D_{G}\right).
\end{align*} 
Now, using the fact that $O_{e}\left(\lambda_{\min}\left(D_{G}\right)\right)=O_{e}\left(\lambda_{\max}\left(D_{G}\right)\right)$
by Assumption \ref{assump:lamba_max=lambda_min for OLS}, we have $\left\Vert W_{G}\right\Vert=O_{e}\left(\left\Vert D_{G}\right\Vert\right).$
Assuming $\hat{C}_G=\frac{X^\prime X}{\sum_{g=1}^G N_g},$ we can write
\begin{align}
\hat{C}_G^{-1}\hat{D}_{G}\hat{C}_G^{-1}-C_{G}^{-1}D_{G}C_{G}^{-1}
= & \left[\hat{C}_G^{-1}-C_G^{-1}\right]D_{G}C_G^{-1} \nonumber\\
  & + \left[\hat{C}_G^{-1}-C_G^{-1}\right]\left[\hat{D}_G-D_{G}\right]C_G^{-1} \nonumber\\
  & + C_G^{-1}\left[\hat{D}_G-D_{G}\right]C_G^{-1} + C_G^{-1}D_{G}\left[\hat{C}_G^{-1}-C_G^{-1}\right] \nonumber\\
  & + \left[\hat{C}_G^{-1}-C_G^{-1}\right]D_{G}\left[\hat{C}_G^{-1}-C_G^{-1}\right] \nonumber\\
  & + \left[\hat{C}_G^{-1}-C_G^{-1}\right]\left[\hat{D}_G-D_{G}\right]\left[\hat{C}_G^{-1}-C_G^{-1}\right] \nonumber\\
  & + C_G^{-1}\left[\hat{D}_G-D_{G}\right]\left[\hat{C}_G^{-1}-C_G^{-1}\right]. \label{eq:splitting of hat(ABA)-ABA}
\end{align}
Using the triangular inequality for norms and the norm inequality in the above equation, it follows that
\[
\frac{1}{\left\Vert D_G \right\Vert}\left\Vert \hat{C}_G^{-1}\hat{D}_{G}\hat{C}_G^{-1}-C_{G}^{-1}D_{G}C_{G}^{-1}\right\Vert=o_P
(1),\]
since $\left\Vert\hat{C}_G^{-1}-C_G^{-1}\right\Vert=o_P(1)$ follows from Assumption \ref{assump: Xg'Xg=O(Ng)} by the continuity theorem of probability , $\left\Vert C_{G}^{-1}\right\Vert=O_e(1),$ $\frac{D_G}{\left\Vert D_G\right\Vert}$ is finite and uniformly positive definite by Assumption \ref{assump:lamba_max=lambda_min for OLS}, and  $\frac{1}{\left\Vert D_G\right\Vert} \left\Vert \hat{D}_G-D_{G}\right\Vert=o_P(1).$ Hence, we have
\begin{align*}
& \quad \frac{1}{\left\Vert W_{G}\right\Vert}\left\Vert\hat{W}_{G}-W_{G}\right\Vert \\
&= \frac{1}{O_{e}\left(\left\Vert D_{G}\right\Vert\right)}\left\Vert\left(\frac{X^{\prime}X}{\sum_{g=1}^{G}N_{g}}\right)^{-1}\hat{D}_{G}\left(\frac{X^{\prime}X}{\sum_{g=1}^{G}N_{g}}\right)^{-1}-C_{G}^{-1}D_{G}C_{G}^{-1}\right\Vert \\
&= o_{P}(1),
\end{align*}
 which deduces that $\hat{W}_{G}$ is consistent for $W_{G}.$ This completes the proof of Theorem \ref{thm:Vols_consistent}.

\hfill
\qed

\textbf{Proof of Theorem \ref{thm:testing for OLS}}

We prove this theorem in two steps.

\textbf{Step 1} : In this step, we show that under $H_{0},$
\[
\frac{\sum_{g=1}^{G}N_{g}}{\sqrt{\sum_{g=1}^{G}N_{g}^{2}}}\Pi_{G}^{-1/2}\left(R\hat{\beta}_{POLS}-r\right)\xrightarrow{d}N_{l}(0,I_{l}), \quad \text{ as } G\to\infty.
\]
To show this, first note that under $H_{0},$
\begin{align*}
    & \quad \frac{\sum_{g=1}^{G}N_{g}}{\sqrt{\sum_{g=1}^{G}N_{g}^{2}}}\Pi_{G}^{-1/2}\left(R\hat{\beta}_{POLS}-r\right) \\
    &  = \frac{\sum_{g=1}^{G}N_{g}}{\sqrt{\sum_{g=1}^{G}N_{g}^{2}}}\Pi_{G}^{-1/2}RW_{G}^{1/2}W_{G}^{-1/2}\left(\hat{\beta}_{POLS}-\beta\right).
\end{align*}
Observe that $\frac{\sum_{g=1}^{G}N_{g}}{\sqrt{\sum_{g=1}^{G}N_{g}^{2}}}W_{G}^{-1/2}\left(\hat{\beta}_{POLS}-\beta\right)\xrightarrow{d}N_{k}(0,I_{k}),$ under the given assumptions from Theorem \ref{thm:ols normality} with $p=2$.
Since $R$ is a finite matrix of dimension $l\times k$ with full
row rank $l,$ and $W_{G}$ is finite and positive definite, it follows
from Corollary 4.24 of \cite{white1984asymptotic} that as $G\rightarrow\infty,$
\[
\frac{\sum_{g=1}^{G}N_{g}}{\sqrt{\sum_{g=1}^{G}N_{g}^{2}}}\Pi_{G}^{-1/2}RW_{G}^{1/2}W_{G}^{-1/2}\left(\hat{\beta}_{POLS}-\beta\right)\xrightarrow{d}N_{l}(0,I_{l}).
\]

\textbf{Step 2} : In this step, we show that under $H_{0},$
\[
\frac{\sum_{g=1}^{G}N_{g}}{\sqrt{\sum_{g=1}^{G}N_{g}^{2}}}\hat{\Pi}_{G}^{-1/2}\left(R\hat{\beta}_{POLS}-r\right)\xrightarrow{d}N_{l}(0,I_{l}), \quad \text{ as } G\to\infty.
\]
To show this, note that using the norm inequality,
\begin{eqnarray*}
& &  \left\Vert\frac{\sum_{g=1}^{G}N_{g}}{\sqrt{\sum_{g=1}^{G}N_{g}^{2}}}\hat{\Pi}_{G}^{-1/2}\left(R\hat{\beta}_{POLS}-r\right)-\frac{\sum_{g=1}^{G}N_{g}}{\sqrt{\sum_{g=1}^{G}N_{g}^{2}}}\Pi_{G}^{-1/2}\left(R\hat{\beta}_{POLS}-r\right)\right\Vert\\
& \leq & \left\Vert\hat{\Pi}_{G}^{-1/2}\Pi_{G}^{1/2}-I_{l}\right\Vert\left\Vert\frac{\sum_{g=1}^{G}N_{g}}{\sqrt{\sum_{g=1}^{G}N_{g}^{2}}}\Pi_{G}^{-1/2}\left(R\hat{\beta}_{POLS}-r\right)\right\Vert.
\end{eqnarray*}
 From Step 1, we have shown that under $H_{0},$ $\frac{\sum_{g=1}^{G}N_{g}}{\sqrt{\sum_{g=1}^{G}N_{g}^{2}}}\Pi_{G}^{-1/2}\left(R\hat{\beta}_{POLS}-r\right)=O_{P}(1).$
It is enough to show that $\left\Vert\hat{\Pi}_{G}^{-1/2}\Pi_{G}^{1/2}-I_{l}\right\Vert=o_{P}(1).$
Since $\Pi_{G}=RW_{G}R^{\prime},$ it follows from Lemma \ref{lemma:lambda_min lambda_max inequality} that 
\begin{align*}
\lambda_{\min}\left(\Pi_{G}\right) & \ge\lambda_{\min}\left(RR^{\prime}\right)\lambda_{\min}\left(W_{G}\right)=O_{e}(1)\lambda_{\min}\left(W_{G}\right) \\
\lambda_{\max}\left(\Pi_{G}\right) & \leq\lambda_{\max}\left(RR^{\prime}\right)\lambda_{\max}\left(W_{G}\right)=O_{e}(1)\lambda_{\max}\left(W_{G}\right).
\end{align*}
Now, using $O_{e}\left(\lambda_{\min}\left(D_{G}\right)\right)=O_{e}\left(\lambda_{\max}\left(D_{G}\right)\right)$
by Assumption \ref{assump:lamba_max=lambda_min for OLS}, and $\left\Vert W_{G}\right\Vert=O_{e}\left(\left\Vert D_{G}\right\Vert\right)$
from the proof of Theorem \ref{thm:Vols_consistent}, we have 
\[
\left\Vert\Pi_{G}\right\Vert=O_{e}\left(\left\Vert W_{G}\right\Vert\right).
\]
 In the proof of Theorem \ref{thm:Vols_consistent}, we have shown that
$\frac{1}{\left\Vert W_{G}\right\Vert}\left\Vert\hat{W}_{G}-W_{G}\right\Vert=o_{P}(1),$
which, by the continuity theorem of probability convergence implies that $\frac{1}{O_{e}\left(\left\Vert W_{G}\right\Vert\right)}\left\Vert R\hat{W}_{G}R^{\prime}-RW_{G}R^{\prime}\right\Vert=o_{P}(1).$ Therefore,  $\frac{1}{\left\Vert\Pi_{G}\right\Vert}\left\Vert\hat{\Pi}_{G}-\Pi_{G}\right\Vert=o_{P}(1).$
By Assumption \ref{assump:lamba_max=lambda_min for OLS}, we deduce that $\frac{W_{G}}{\left\Vert W_{G}\right\Vert}$
is uniformly positive definite, and it subsequently follows that $\frac{\Pi_{G}}{\left\Vert\Pi_{G}\right\Vert}$
is also uniformly positive definite. So, by the continuity theorem for
probability convergence, taking $g(Z)=Z^{-1/2}\left(\frac{\Pi_{G}}{\left\Vert\Pi_{G}\right\Vert}\right)^{1/2}$ as a continuous function of $Z$, putting $Z=\frac{\hat{\Pi}_{G}}{\left\Vert\Pi_{G}\right\Vert},$ we have $\left\Vert\hat{\Pi}_{G}^{-1/2}\Pi_{G}^{1/2}-I_{l}\right\Vert=o_{P}(1).$

Hence, from Step 1 and Step 2, it follows that under $H_{0},$
\[
\frac{\sum_{g=1}^{G}N_{g}}{\sqrt{\sum_{g=1}^{G}N_{g}^{2}}}\hat{\Pi}_{G}^{-1/2}\left(R\hat{\beta}_{POLS}-r\right)\xrightarrow{d}N_{l}(0,I_{l}), \quad \text{ as } G\to\infty.
\]
Consequently, by the continuous mapping theorem, it follows that:
\[
\frac{\left(\sum_{g=1}^{G}N_{g}\right)^{2}}{\sum_{g=1}^{G}N_{g}^{2}}\left(R\hat{\beta}_{POLS}-r\right)^{\prime}\hat{\Pi}_{G}^{-1}\left(R\hat{\beta}_{POLS}-r\right)\xrightarrow{d}\chi_{l}^{2}, \quad \text{ as } G\to\infty
\]
for strong dependence if $\frac{\sum_{g=1}^G N_g^{4}}{\left(\sum_{g=1}^G N_g^2\right)^2}\to 0$,  for semi-strong dependence if $\frac{\sum_{g=1}^G N_g^{4}}{\left(\sum_{g=1}^G N_g h(N_g)\right)^2}\to 0$, and for weak dependence if $\frac{\sum_{g=1}^G N_g^{4}}{\left(\sum_{g=1}^G N_g\right)^2}\to 0$. 

\hfill
\qed

{\bf Proof of Theorem \ref{thm:consistent}} 

To prove the consistency of $\hat{\beta}_A,$ first, note that $$\hat{\beta}_A - \beta = \left(\frac{\bar{X}'\bar{X}}{G}\right)^{-1} \frac{\bar{X}'\bar{Y}}{G} - \beta  = \left(\frac{\bar{X}'\bar{X}}{G}\right)^{-1} \frac{\bar{X}'\bar{\epsilon}}{G}.$$

Note that $\left(\frac{\bar{X}^{\prime}\bar{X}}{G}\right)^{-1}-A_{G}^{-1}=o_P(1),$ which follows from Assumption \ref{assump: Xbar'Xbar=O(G)}(i), where $A_{G}^{-1}$ is finite and uniformly positive definite over $G.$ Then, by Lemma \ref{lemma:A_n=B_n+o_P(1) implies A_n = O_P(1)},
\[
\left(\frac{\bar{X}^{\prime}\bar{X}}{G}\right)^{-1}=O_P(1).
\]
Here, $\left(\frac{\bar{X}^{\prime}\bar{X}}{G}\right)^{-1}=O_P(1)$ means that $\left|\left[\left(\frac{\bar{X}^{\prime}\bar{X}}{G}\right)^{-1}\right]_{j,j_1}\right|=O_P(1),$ for all $j,j_1=1,2,\ldots,k.$
Now, using Assumption \ref{assump:exogeneity}, note that 
\[
\mathbb{E}\left[\frac{\bar{X}^{\prime}\bar{\epsilon}}{G}\right] = \frac{1}{G}\sum_{g=1}^{G}\mathbb{E}\left[\bar{X}_g^{\prime}\,\mathbb{E}\left[\bar{\epsilon}_g | X_g\right]\right]=0.
\]
 Also, the variance-covariance matrix of $\frac{\bar{X}^{\prime}\bar{\epsilon}}{G}$ is 
 \begin{equation}
 \label{eq: var-cov expansion for beta hat}
     \frac{1}{G^2}\sum_{g=1}^G \operatorname{Var}\left( \bar{X}_g^{\prime}\bar{\epsilon}_g\right) + \frac{1}{G^2}\sum_{\underset{g\neq g_{1}}{g,g_{1}=1}}^{G}\operatorname{Cov}\left(\bar{X}_{g}^{\prime}\bar{\epsilon}_{g},\bar{X}_{g_{1}}^{\prime}\bar{\epsilon}_{g_{1}}\right).
 \end{equation}

Since $\operatorname{Var}\left( \bar{X}_g^{\prime}\bar{\epsilon}_g\right) = \operatorname{Var}\left( \mathbb{E}\left[\bar{X}_g^{\prime}\bar{\epsilon}_g\mid X_g\right]\right) + \mathbb{E}\left[\operatorname{Var}\left(\bar{X}_g^{\prime}\bar{\epsilon}_g\right)\mid X_g\right],$ applying Assumption \ref{assump:exogeneity} the first term in \eqref{eq: var-cov expansion for beta hat} reduces to $\frac{1}{G^2}\sum_{g=1}^{G} \mathbb{E}\left[\bar{X}_{g}^{\prime}\bar{X}_{g}\right]\sigma_{g}^{2},$ where $\sigma_{g}^{2}=\frac{\mathbf{1}^\prime\Omega_{g}\mathbf{1}}{N_{g}^{2}}.$

Also, using the fact that $\operatorname{Cov}\left( \bar{X}_{g}^{\prime}\bar{\epsilon}_{g},\bar{X}_{g_{1}}^{\prime}\bar{\epsilon}_{g_{1}} \right) = 0$ by Assumptions \ref{assump:exogeneity} and \ref{assump:error independence}(i), the second term in \eqref{eq: var-cov expansion for beta hat} vanishes. Therefore, we obtain
\begin{equation} 
\label{eq: variance of X_bar epsilon_bar/G}
    \operatorname{Var}\left(\frac{\bar{X}^{\prime}\bar{\epsilon}}{G}\right) = \frac{1}{G^2}\sum_{g=1}^{G} \mathbb{E}\left[\bar{X}_{g}^{\prime}\bar{X}_{g}\right]\sigma_{g}^{2}.
\end{equation}

Since for all $g$, the diagonals of $\Omega_g$ are uniformly bounded, $\sigma_{g}^{2} = \frac{\mathbf{1}'\Omega_g\mathbf{1}}{N_g^2} \leq \frac{\operatorname{tr}(\Omega_g)}{N_g} \leq c$ holds for all $g$ for some $0<c<\infty$. Observe that for all $G\geq 1,$ the matrix $\frac{1}{G^2}\sum_{g=1}^{G} \mathbb{E}\left[\bar{X}_{g}^{\prime}\bar{X}_{g}\right]\left(c-\sigma_{g}^{2}\right) $ is a nonnegative definite matrix. This follows because for any $z\in\mathbb{R}^k,$ $z^\prime \left( \frac{1}{G^2}\sum_{g=1}^{G} \mathbb{E}\left[\bar{X}_{g}^{\prime}\bar{X}_{g}\right]c \right) z \geq z^\prime \left( \frac{1}{G^2}\sum_{g=1}^{G} \mathbb{E}\left[\bar{X}_{g}^{\prime}\bar{X}_{g}\right]\sigma_{g}^{2} \right) z.$ Therefore, we have
\[
 \frac{1}{G^2}\sum_{g=1}^{G} \mathbb{E}\left[\bar{X}_{g}^{\prime}\bar{X}_{g}\right]\sigma_{g}^{2} \leq \frac{1}{G^2}\sum_{g=1}^{G} \mathbb{E}\left[\bar{X}_{g}^{\prime}\bar{X}_{g}\right]c = \frac{c}{G} A_G.
\]
Since both the matrices $\frac{1}{G^2}\sum_{g=1}^{G} \mathbb{E}\left[\bar{X}_{g}^{\prime}\bar{X}_{g}\right]\sigma_{g}^{2}$ and $\frac{c}{G} A_G$ are positive semi-definite, the norms are also in the same direction by Result \ref{result:trace inequality}. Hence,
\[
    \left\Vert \operatorname{Var}\left(\frac{\bar{X}^{\prime}\bar{\epsilon}}{G}\right)\right\Vert = \left\Vert \frac{1}{G^2}\sum_{g=1}^{G} \mathbb{E}\left[\bar{X}_{g}^{\prime}\bar{X}_{g}\right]\sigma_{g}^{2}\right\Vert  \leq \frac{c}{G} \left\Vert A_G\right\Vert.
\]
Therefore, it follows that 
\[
    \left\Vert \operatorname{Var}\left(\frac{\bar{X}^{\prime}\bar{\epsilon}}{G}\right)\right\Vert \;= \; O\left(\frac{1}{G}\right)\to 0, \text{ as } G\to\infty,
\]
since $A_G=\mathbb{E}\left[\frac{\bar{X}^{\prime}\bar{X}}{G}\right]$
is finite and uniformly positive definite over $G,$ by Assumption \ref{assump: Xbar'Xbar=O(G)}(i). Therefore, 
\[
   \frac{\bar{X}^{\prime}\bar{\epsilon}}{G} = o_P(1). 
\]
Now, using the facts that $\left(\frac{\bar{X}^{\prime}\bar{X}}{G}\right)^{-1}=O_P(1)$ and $\frac{\bar{X}^{\prime}\bar{\epsilon}}{G} = o_P(1),$ we have from Lemma \ref{lemma:big-OP sum and multiplication},
\[
\hat{\beta}_A-\beta=O_P(1)\cdot o_P(1)=o_P(1),
\]
and subsequently $\hat{\beta}_{A}$ is consistent for any kind of
dependence of $\epsilon_{g}$ and for any cluster sizes.

\hfill 
\qed

\textbf{Proof of Lemma \ref{lemma: sum h^2/N^2 to 0}} 

Note that under semi-strong dependence,
\[\frac{\sum_{g=1}^{G}\frac{h^{2}(N_{g})}{N_{g}^{2}}}{\left(\sum_{g=1}^{G}\frac{h(N_{g})}{N_{g}}\right)^{2}}=\frac{1}{1+\frac{\sum_{g_{1}\neq g_{2}}\frac{h(N_{g_{1}})h(N_{g_{2}})}{N_{g_{1}}N_{g_{2}}}}{\sum_{g}\frac{h^{2}(N_{g})}{N_{g}^{2}}}}=\frac{1}{1+a_{G}}, \text{ say. }
\] 

(i) All the clusters are nearly balanced such that $N_{g}\simeq N,$
for all $g.$ Then, 
\[
a_{G} \simeq \frac{G(G-1)\frac{h^{2}(N)}{N^{2}}}{G\frac{h^{2}(N)}{N^{2}}}=G-1=O(G).
\]

(ii) Suppose that $L$ clusters are extremely large as
compared to the other clusters such that $N_{1},\ldots,N_{L}\to\infty$
and $N_{g}\simeq N\to\infty,$ for $g=L+1,\ldots,G,$ but $\frac{N}{N_{j}}=o(1),$
for $j=1,2,\ldots,L.$ Let $\mathcal{L}$ be the set of extremely
large clusters and without loss of generality, assume that $\frac{h(N_{1})}{N_{1}}<\frac{h(N_{2})}{N_{2}}<\cdots<\frac{h(N_{L})}{N_{L}}.$
Then,
\begin{align*}
 & \quad a_{G}\\
 & = \frac{\sum_{g_{1},g_{2}\in\mathcal{L}}\frac{h(N_{g_{1}})h(N_{g_{2}})}{N_{g_{1}}N_{g_{2}}}+\sum_{g_{1}\in\mathcal{L},g_{2}\in\mathcal{L}^{c}}\frac{h(N_{g_{1}})h(N_{g_{2}})}{N_{g_{1}}N_{g_{2}}}+(G-L)(G-L-1)\frac{h^{2}(N)}{N^{2}}}{\sum_{g\in\mathcal{L}}\frac{h^{2}(N_{g})}{N_{g}^{2}}+(G-L)\frac{h^{2}(N)}{N^{2}}}\\
 & \geq \frac{L(L-1)\frac{h(N_{1})h(N_{2})}{N_{1}N_{2}}+2L(G-L)\frac{h(N_{1})h(N)}{N_{1}N}+(G-L)(G-L-1)\frac{h^{2}(N)}{N^{2}}}{L\frac{h^{2}(N_{L})}{N_{L}^{2}}+(G-L)\frac{h^{2}(N)}{N^{2}}}\\
 & = \frac{\frac{L(L-1)}{(G-L)}\frac{h(N_{1})h(N_{2})N^{2}}{N_{1}N_{2}h^{2}(N)}+2L\frac{h(N_{1})N}{N_{1}h(N)}+(G-L-1)}{\frac{L}{G-L}\frac{h^{2}(N_{L})N^{2}}{N_{L}^{2}h^{2}(N)}+1},\text{ dividing by }(G-L)\frac{h^{2}(N)}{N^{2}}.
\end{align*}

Clearly, the numerator is bigger than $G-L-1$ and the denominator
is $o\left(\frac{L}{G-L}\right)+1,$ due to the fact that $\frac{h^{2}(N_{L})N^{2}}{N_{L}^{2}h^{2}(N)}\to0.$
Hence, if $L$ is finite, then $a_{G}=O(G).$ Also, note that $L$
may be $O(G)$ additionally with $G-L=O(G),$ for example, $L=G/2.$
Then also $a_{G}=O(G).$ So, for case (i) and case (ii), $\frac{\sum_{g=1}^{G}\frac{h^{2}(N_{g})}{N_{g}^{2}}}{\left(\sum_{g=1}^{G}\frac{h(N_{g})}{N_{g}}\right)^{2}}\to 0,$ as $G\to\infty.$

Putting $h(N_{g})=1,$ we can have
$\frac{\sum_{g=1}^{G}\frac{1}{N_{g}^{2}}}{\left(\sum_{g=1}^{G}\frac{1}{N_{g}}\right)^{2}}\to0$ as $G\to\infty$, under weak dependence, for the similar cases described above.
This concludes the proof of Lemma \ref{lemma: sum h^2/N^2 to 0}.

\hfill
\qed

{\bf Proof of Theorem \ref{thm:normal}}

To show the asymptotic normality of $\hat{\beta}_A,$  note that  $\sqrt{G}\left(\hat{\beta}_{A}-\beta\right)=\left(\frac{\bar{X}^{\prime}\bar{X}}{G}\right)^{-1}\frac{1}{\sqrt{G}}\sum_{g=1}^{G}\bar{X}_{g}^{\prime}\bar{\epsilon}_{g}.$
So, 
\begin{align*}
    & \quad \sqrt{G}\left(\hat{\beta}_{A}-\beta\right)-A_{G}^{-1}\frac{1}{\sqrt{G}}\sum_{g=1}^{G}\bar{X}_{g}^{\prime}\bar{\epsilon}_{g} \\
    & = \left[\left(\frac{\bar{X}^{\prime}\bar{X}}{G}\right)^{-1}-A_{G}^{-1}\right]B_{G}^{1/2}B_{G}^{-1/2}\frac{1}{\sqrt{G}}\sum_{g=1}^{G}\bar{X}_{g}^{\prime}\bar{\epsilon}_{g},
\end{align*}
which implies 
\begin{equation}\label{eq:asympnorm}
\begin{aligned}
V_{G}^{-1/2}\sqrt{G}\left(\hat{\beta}_{A}-\beta\right) 
&= V_{G}^{-1/2}A_{G}^{-1}\frac{1}{\sqrt{G}}\sum_{g=1}^{G}\bar{X}_{g}^{\prime}\bar{\epsilon}_{g} \\
&\quad + V_{G}^{-1/2}\left[\left(\frac{\bar{X}^{\prime}\bar{X}}{G}\right)^{-1}-A_{G}^{-1}\right] \\
&\quad \times B_{G}^{1/2}B_{G}^{-1/2}\frac{1}{\sqrt{G}}\sum_{g=1}^{G}\bar{X}_{g}^{\prime}\bar{\epsilon}_{g}.
\end{aligned}
\end{equation}
We will then complete the proof in three steps.

\textbf{Step 1:} Here, we will show that $B_{G}^{-1/2}\frac{1}{\sqrt{G}}\sum_{g=1}^{G}\bar{X}_{g}^{\prime}\bar{\epsilon}_{g}\xrightarrow{d}N_{k}\left(0,I_{k}\right),$ as $G\to\infty.$

To prove the asymptotic normality of $B_{G}^{-1/2}\frac{1}{\sqrt{G}}\sum_{g=1}^{G}\bar{X}_{g}^{\prime}\bar{\epsilon}_{g},$
it suffices to prove the Lyapunov condition for CLT. To prove the Lyapunov
condition, it is required to show that for any fixed $z\in\mathbb{R}^{k},$
and for some $p>1,$ 
\[
\frac{\sum_{g=1}^{G}\mathbb{E}\left| z^{\prime}B_{G}^{-1/2}\frac{1}{\sqrt{G}}\bar{X}_{g}^{\prime}\bar{\epsilon}_{g}\right|^{2p}}{\left[\sum_{g=1}^{G}\mathbb{E}\left(z^{\prime}B_{G}^{-1/2}\frac{1}{\sqrt{G}}\bar{X}_{g}^{\prime}\bar{\epsilon}_{g}\right)^{2}\right]^{p}}\to0,\text{ as }G\to\infty.
\] 
For simplicity, we take $p=2.$ Then, using Result \ref{result:trace inequality} in the second line, 
\begin{eqnarray}
& & \sum_{g=1}^{G}\mathbb{E} \left|  z^{\prime}B_{G}^{-1/2}\frac{1}{\sqrt{G}}\bar{X}_{g}^{\prime}\bar{\epsilon}_{g} \right|^{4} \nonumber \\
 & = & \frac{1}{G^{2}}\sum_{g=1}^{G}\mathbb{E}\left[\left(z^{\prime}B_{G}^{-1/2}\bar{X}_{g}^{\prime}\bar{X}_{g}B_{G}^{-1/2}z\right)^{2}| \bar{\epsilon}_{g} | ^{4}\right]\nonumber\\
 & \leq & \frac{1}{G^{2}}\sum_{g=1}^{G}\mathbb{E}\left[\left(z^{\prime}B_{G}^{-1}z\right)^{2}\left(\operatorname{tr}\left(\bar{X}_{g}^{\prime}\bar{X}_{g}\right)\right)^{2}| \bar{\epsilon}_{g} | ^{4}\right]\nonumber\\
 & = & \frac{1}{G^{2}}\left(z^{\prime}B_{G}^{-1}z\right)^{2}\sum_{g=1}^{G}\mathbb{E}\left[\left(\operatorname{tr}\left(\bar{X}_{g}^{\prime}\bar{X}_{g}\right)\right)^{2}\mathbb{E}\left[ | \bar{\epsilon}_{g} | ^{4} \big|  X_g \right]\right].\label{eq:asympnorm2}
\end{eqnarray}

Now, the $4^{th}$ absolute moment of $\bar{\epsilon}_{g}$ conditional
on $X_g$ is
\begin{align*}
\mathbb{E}\left[|\bar{\epsilon}_g|^4 \big| X_g \right]
&= \frac{1}{N_g^4}\mathbb{E}\Big[
\sum_i \epsilon_{gi}^4
+ 4\sum_{i\neq j}\epsilon_{gi}^3\epsilon_{gj}
+ 3\sum_{i\neq j}\epsilon_{gi}^2\epsilon_{gj}^2 \\
&\qquad\qquad
+ 6\sum_{i\neq j\neq s}\epsilon_{gi}^2\epsilon_{gj}\epsilon_{gs}
+ \sum_{i\neq j\neq s\neq t}\epsilon_{gi}\epsilon_{gj}\epsilon_{gs}\epsilon_{gt}
\;\Big|\, X_g\Big].
\end{align*}

Then, it follows from Assumption \ref{assump:additional assumption on errors} that the conditional fourth moment of $\bar{\epsilon}_{g}$ is
\[
\mathbb{E}\left[ | \bar{\epsilon}_{g} | ^{4}\vert X_g \right] = \begin{cases}
O(1), & \text{for strong dependence,}\\
O\left(\frac{h^{2}(N_{g})}{N_{g}^{2}}\right), & \text{for semi-strong dependence,}\\
O\left(\frac{1}{N_{g}^{2}}\right), & \text{for weak dependence.}
\end{cases}
\]

uniformly in $g.$
Also, for the square of the variance, we have
\begin{eqnarray*}
\left(\sum_{g=1}^{G}\mathbb{E}\left[z^{\prime}B_{G}^{-1/2}\frac{1}{\sqrt{G}}\bar{X}_{g}^{\prime}\bar{\epsilon}_{g}\right]^{2}\right)^{2} & = & \frac{1}{G^{2}}\left(\sum_{g=1}^{G}\mathbb{E}\mathbb{E}\left[z^{\prime}B_{G}^{-1/2}\bar{X}_{g}^{\prime}\bar{X}_{g}\bar{\epsilon}_{g}^{2}B_{G}^{-1/2}z \big|  X_g \right]\right)^{2}\\
 & = & \frac{1}{G^{2}}\left(\sum_{g=1}^{G}\mathbb{E}\left[z^{\prime}B_{G}^{-1/2}\bar{X}_{g}^{\prime}\bar{X}_{g}\sigma_{g}^{2}B_{G}^{-1/2}z\right]\right)^{2}\\
 & = & \left(z^{\prime}B_{G}^{-1/2}\,\mathbb{E}\left[\frac{1}{G}\sum_{g=1}^{G}\bar{X}_{g}^{\prime}\bar{X}_{g}\sigma_{g}^{2}\right]B_{G}^{-1/2}z\right)^{2}\\
 & = & \left(z^{\prime}B_{G}^{-1/2}B_{G}B_{G}^{-1/2}z\right)^{2} \\
 & = & \left(z^{\prime}z\right)^{2}.
\end{eqnarray*}

Note that by Assumption \ref{assump:lamba_max=lambda_min}, we have 
\begin{align*}
\frac{z^{\prime}B_{G}^{-1}z}{z^{\prime}z} 
&\leq \lambda_{\max}\left(B_{G}^{-1}\right) = \frac{1}{\lambda_{\min}\left(B_{G}\right)} \\
&= \begin{cases}
O(1), & \text{for strong dependence,}\\[2ex]
O\left(\left(\frac{1}{G}\sum_{g=1}^{G}\frac{h(N_{g})}{N_{g}}\right)^{-1}\right), & \text{for semi-strong dependence,}\\[2ex]
O\left(\left(\frac{1}{G}\sum_{g=1}^{G}N_{g}^{-1}\right)^{-1}\right), & \text{for weak dependence.}
\end{cases}
\end{align*}

From the identity \eqref{eq:asympnorm2}, using the above facts along
with the fact that $\mathbb{E}\left[\left(\operatorname{tr}\left(\bar{X}_{g}^{\prime}\bar{X}_{g}\right)\right)^{2}\right]=O(1)$ uniformly in $g$ by Assumption \ref{assump: Xbar'Xbar=O(G)}(ii), and applying Lemma \ref{lemma:Exact order summability}, the Lyapunov condition reduces to 
{\allowdisplaybreaks
\begin{eqnarray}
 &  & \frac{\sum_{g=1}^{G}\mathbb{E} \left|  z^{\prime}B_{G}^{-1/2}\frac{1}{\sqrt{G}}\bar{X}_{g}^{\prime}\bar{\epsilon}_{g} \right| ^{4}}{\left(\sum_{g=1}^{G}\mathbb{E}\left[z^{\prime}B_{G}^{-1/2}\frac{1}{\sqrt{G}}\bar{X}_{g}^{\prime}\bar{\epsilon}_{g}\right]^{2}\right)^{2}}\nonumber\\
 & \leq & \frac{1}{G^{2}}\frac{\left(z^{\prime}B_{G}^{-1}z\right)^{2}}{\left(z^{\prime}z\right)^{2}}\sum_{g=1}^{G}\mathbb{E}\left[\left(\operatorname{tr}\left(\bar{X}_{g}^{\prime}\bar{X}_{g}\right)\right)^{2}\mathbb{E}\left[ | \bar{\epsilon}_{g} | ^{4} \big|  X_g \right]\right]\nonumber\\
 & = & \begin{cases}
\frac{1}{G^{2}}O\left(1\right)\sum_{g=1}^{G}O(1), & \text{for strong dependence,}\\
\frac{1}{G^{2}}O\left(\frac{1}{\left(\frac{1}{G}\sum_{g=1}^{G}\frac{h(N_{g})}{N_{g}}\right)^{2}}\right)\sum_{g=1}^{G}O\left(\frac{h^{2}(N_{g})}{N_{g}^{2}}\right), & \text{for semi-strong dependence,}\nonumber\\
\frac{1}{G^{2}}O\left(\frac{1}{\left(\frac{1}{G}\sum_{g=1}^{G}\frac{1}{N_{g}}\right)^{2}}\right)\sum_{g=1}^{G}O\left(\frac{1}{N_{g}^{2}}\right), & \text{for weak dependence,}
\end{cases}\nonumber\\
 & = & \begin{cases}
O\left(G^{-1}\right), & \text{for strong dependence,}\\
O\left(\frac{\sum_{g=1}^{G}\frac{h^{2}(N_{g})}{N_{g}^{2}}}{\left(\sum_{g=1}^{G}\frac{h(N_{g})}{N_{g}}\right)^{2}}\right), & \text{for semi-strong dependence,}\\
O\left(\frac{\sum_{g=1}^{G}\frac{1}{N_{g}^{2}}}{\left(\sum_{g=1}^{G}\frac{1}{N_{g}}\right)^{2}}\right), & \text{for weak dependence.}
\end{cases} \label{eq: Lyapunov condition orders for beta hat}
\end{eqnarray}}

From Lemma \ref{lemma: sum h^2/N^2 to 0}, $\frac{\sum_{g=1}^{G}\frac{h^{2}(N_{g})}{N_{g}^{2}}}{\left(\sum_{g=1}^{G}\frac{h(N_{g})}{N_{g}}\right)^{2}}\to 0$ and  $\frac{\sum_{g=1}^{G}\frac{1}{N_{g}^{2}}}{\left(\sum_{g=1}^{G}\frac{1}{N_{g}}\right)^{2}}\to 0,$
as $ G\to\infty$, for the nearly balanced case, and also for the unbalanced case with $O(G)$ nearly balanced clusters. Consequently, the CLT applies for $B_{G}^{-1/2}\frac{1}{\sqrt{G}}\sum_{g=1}^{G}\bar{X}_{g}^{\prime}\bar{\epsilon}_{g}$
under strong dependence without further restrictions, and under semi-strong or weak dependence for the cases mentioned above.

\textbf{Step 2:} We will show that 
\[
V_{G}^{-1/2}\left[\left(\frac{\bar{X}^{\prime}\bar{X}}{G}\right)^{-1}-A_{G}^{-1}\right]B_{G}^{1/2}B_{G}^{-1/2}\frac{1}{\sqrt{G}}\sum_{g=1}^{G}\bar{X}_{g}^{\prime}\bar{\epsilon}_{g}=o_{P}(1).
\]
By Assumption \ref{assump:lamba_max=lambda_min}, we have 
$\lambda_{\min}(B_{G})$ and $\lambda_{\max}(B_{G})$ are of the exact order:
\[
\begin{cases}
O_{e}\left(1\right), & \text{for strong dependence,}\\
O_{e}\left(\frac{1}{G}\sum_{g=1}^{G}\frac{h(N_{g})}{N_{g}}\right), & \text{for semi-strong dependence,}\\
O_{e}\left(\frac{1}{G}\sum_{g=1}^{G}\frac{1}{N_{g}}\right), & \text{for weak dependence.}
\end{cases}
\]
The above condition, along with Lemma \ref{lemma:lambda_min lambda_max inequality} and Assumption \ref{assump: Xbar'Xbar=O(G)}, also implies that for strong, semi-strong, and weak dependence, respectively, we have:
\begin{align*}
    O_e\left(\lambda_{\min}(V_{G})\right) &= O_e\left(\lambda_{\max}(V_{G})\right) = O_e(1), \\
    O_e\left(\lambda_{\min}\left(\frac{G}{\sum_{g=1}^{G}\frac{h(N_{g})}{N_{g}}}V_{G}\right)\right) &= O_e\left(\lambda_{\max}\left(\frac{G}{\sum_{g=1}^{G}\frac{h(N_{g})}{N_{g}}}V_{G}\right)\right) = O_e(1),  \\
    O_e\left(\lambda_{\min}\left(\frac{G}{\sum_{g=1}^{G}\frac{1}{N_{g}}}V_{G}\right)\right) &= O_e\left(\lambda_{\max}\left(\frac{G}{\sum_{g=1}^{G}\frac{1}{N_{g}}}V_{G}\right)\right) = O_e(1).
\end{align*}

Using $\left[\left(\frac{\bar{X}^{\prime}\bar{X}}{G}\right)^{-1}-A_{G}^{-1}\right]=o_{P}(1)$ by Assumption \ref{assump: Xbar'Xbar=O(G)}, and $B_{G}^{-1/2}\frac{1}{\sqrt{G}}\sum_{g=1}^{G}\bar{X}_{g}^{\prime}\bar{\epsilon}_{g}=O_{P}(1)$ from Step 1, we have by Lemma \ref{lemma:big-OP sum and multiplication}, $V_{G}^{-1/2}\left[\left(\frac{\bar{X}^{\prime}\bar{X}}{G}\right)^{-1}-A_{G}^{-1}\right]B_{G}^{1/2}B_{G}^{-1/2}\frac{1}{\sqrt{G}}\sum_{g=1}^{G}\bar{X}_{g}^{\prime}\bar{\epsilon}_{g}=o_{P}(1).$ 

\textbf{Step 3:} We will show that $V_{G}^{-1/2}A_{G}^{-1}\frac{1}{\sqrt{G}}\sum_{g=1}^{G}\bar{X}_{g}^{\prime}\bar{\epsilon}_{g}\xrightarrow{d}N_{k}\left(0,I_{k}\right),$  as $G\to\infty.$ 

Since $\left\{ B_{G}\right\} $ and $\left\{ B_{G}^{-1}\right\} $
are finite and positive definite by Assumption \ref{assump:lamba_max=lambda_min}, $\left\{ A_{G}\right\} $ is finite and uniformly positive definite by Assumption \ref{assump: Xbar'Xbar=O(G)}, and $B_{G}^{-1/2}\frac{1}{\sqrt{G}}\sum_{g=1}^{G}\bar{X}_{g}^{\prime}\bar{\epsilon}_{g}\xrightarrow{d}N_{k}\left(0,I_{k}\right)$ from Step 1, it immediately follows from Corollary 4.24 of \cite{white1984asymptotic} that 
\[
V_{G}^{-1/2}A_{G}^{-1}\frac{1}{\sqrt{G}}\sum_{g=1}^{G}\bar{X}_{g}^{\prime}\bar{\epsilon}_{g}=V_{G}^{-1/2}A_{G}^{-1}B_{G}^{1/2}B_{G}^{-1/2}\frac{1}{\sqrt{G}}\sum_{g=1}^{G}\bar{X}_{g}^{\prime}\bar{\epsilon}_{g}\xrightarrow{d}N_{k}\left(0,I_{k}\right),
\]
since $V_{G}^{-1/2}A_{G}^{-1}B_{G}^{1/2}I_{k}B_{G}^{1/2}A_{G}^{-1}V_{G}^{-1/2}=V_{G}^{-1/2}V_{G}V_{G}^{-1/2}=I_{k}.$

From equation \eqref{eq:asympnorm}, the three steps altogether lead to the fact that $V_{G}^{-1/2}\sqrt{G}\left(\hat{\beta}_{A}-\beta\right)\xrightarrow{d}N_{k}\left(0,I_{k}\right).$
This completes the proof of Theorem \ref{thm:normal}.

\hfill
\qed

\begin{lemma}\label{lemma:order of beta hat proposed - beta}

For both (i) $\left\Vert\frac{\bar{X}^{\prime}\bar{\epsilon}}{G}\right\Vert$ and (ii) $\left\Vert\hat{\beta}_{A}-\beta\right\Vert$, the asymptotic order is given by:
\[
\begin{cases}
O_{P}\left(\frac{1}{\sqrt{G}}\right), & \text{for strong dependence,}\\
O_{P}\left(\frac{1}{G}\sqrt{\sum_{g=1}^{G}\frac{h(N_{g})}{N_{g}}}\right), & \text{for semi-strong dependence,}\\
O_{P}\left(\frac{1}{G}\sqrt{\sum_{g=1}^{G}\frac{1}{N_{g}}}\right), & \text{for weak dependence.}
\end{cases}
\]
\end{lemma}

\textbf{Proof of Lemma \ref{lemma:order of beta hat proposed - beta}}

(i)  Using Assumption \ref{assump:exogeneity}, note that
\[
\mathbb{E}\left[\frac{\bar{X}^{\prime}\bar{\epsilon}}{G}\right]=\frac{1}{G}\sum_{g=1}^G \mathbb{E}\left[\bar{X}_g^{\prime}\,\mathbb{E}\left[\bar{\epsilon}_g \mid  X_g \right]\right]=0.
\]

Also, utilizing \eqref{eq: variance of X_bar epsilon_bar/G} in the proof of Theorem \ref{thm:consistent}, we obtain
\[
\mathbb{E}\left\Vert\frac{\bar{X}^{\prime}\bar{\epsilon}}{G}\right\Vert^{2}  = \operatorname{tr} \left(\operatorname{Var}\left[\frac{\bar{X}^{\prime}\bar{\epsilon}}{G}\right]\right) = \frac{1}{G^{2}}\sum_{g=1}^{G}\mathbb{E}\left[\bar{X}_{g}\bar{X}_{g}^{\prime}\sigma_{g}^{2}\right].
\]

Furthermore, by Assumption \ref{assump: Xbar'Xbar=O(G)}(ii), we have $\mathbb{E}\left[\bar{X}_{g}\bar{X}_{g}^{\prime}\right]=O(1)$ uniformly in $g$. Combining this with the order of $\sigma_g^2$ given in \eqref{eq:order of sigma_g^2} of Remark \ref{rem:order of sigma_g^2}, and applying Lemma \ref{lemma:Exact order summability}, it follows that
\[
\mathbb{E}\left\Vert\frac{\bar{X}^{\prime}\bar{\epsilon}}{G}\right\Vert^{2} = \begin{cases}
O\left(\frac{1}{G}\right), & \text{for strong dependence,}\\
O\left(\frac{1}{G^{2}}\sum_{g=1}^{G}\frac{h(N_{g})}{N_{g}}\right), & \text{for semi-strong dependence,}\\
O\left(\frac{1}{G^{2}}\sum_{g=1}^{G}\frac{1}{N_{g}}\right), & \text{for weak dependence.}
\end{cases}
\]
This proves part (i) of the lemma.

(ii) Applying the norm inequality for multiplication in the third line, the triangular inequality for norms in the fourth line, part (i) of this lemma, and Lemma \ref{lemma:big-OP sum and multiplication} in the last line, we obtain 
\allowdisplaybreaks{
\begin{align*}
\left\Vert\hat{\beta}_{A}-\beta\right\Vert & =\left\Vert\left(\frac{\bar{X}^{\prime}\bar{X}}{G}\right)^{-1}\frac{\bar{X}^{\prime}\bar{\epsilon}}{G}\right\Vert\\
 & =\left\Vert\left(A_{G}^{-1}+\left(\frac{\bar{X}^{\prime}\bar{X}}{G}\right)^{-1}-A_{G}^{-1}\right)\frac{\bar{X}^{\prime}\bar{\epsilon}}{G}\right\Vert\\
 & \leq\left\Vert A_{G}^{-1}+\left(\frac{\bar{X}^{\prime}\bar{X}}{G}\right)^{-1}-A_{G}^{-1}\right\Vert\left\Vert\frac{\bar{X}^{\prime}\bar{\epsilon}}{G}\right\Vert\\
 & \leq\left[\left\Vert A_{G}^{-1}\right\Vert+\left\Vert\left(\frac{\bar{X}^{\prime}\bar{X}}{G}\right)^{-1}-A_{G}^{-1}\right\Vert\right]\left\Vert\frac{\bar{X}^{\prime}\bar{\epsilon}}{G}\right\Vert\\
 & =\left[O(1)+o_{P}(1)\right]\left\Vert\frac{\bar{X}^{\prime}\bar{\epsilon}}{G}\right\Vert\\
 & =\begin{cases}
O_{P}\left(\frac{1}{\sqrt{G}}\right), & \text{for strong dependence,}\\
O_{P}\left(\frac{1}{G}\sqrt{\sum_{g=1}^{G}\frac{h(N_{g})}{N_{g}}}\right), & \text{for semi-strong dependence,}\\
O_{P}\left(\frac{1}{G}\sqrt{\sum_{g=1}^{G}\frac{1}{N_{g}}}\right), & \text{for weak dependence.}
\end{cases}
\end{align*}
}
Thus, the lemma is proved.

\hfill
\qed

\begin{lemma}\label{lemma:order of epsilon bar X bar}
$\left\Vert\bar{\epsilon}_{g}\bar{X}_{g}\right\Vert=\begin{cases}
O_{P}\left(1\right), & \text{for strong dependence,}\\
O_{P}\left(\sqrt{\frac{h(N_{g})}{N_{g}}}\right), & \text{for semi-strong dependence,}\\
O_{P}\left(\sqrt{\frac{1}{N_{g}}}\right), & \text{for weak dependence,}
\end{cases}$ 

uniformly in $g.$
\end{lemma}

\textbf{Proof of Lemma \ref{lemma:order of epsilon bar X bar}} 

Using the fact that $\mathbb{E}\left[\bar{X}_{g}\bar{X}_{g}^{\prime}\right]=O(1)$ uniformly in $g$ by Assumption \ref{assump: Xbar'Xbar=O(G)}(ii), and using the order of $\sigma_g^2$ given in \eqref{eq:order of sigma_g^2} of Remark \ref{rem:order of sigma_g^2},  we obtain
\begin{align*}
\mathbb{E}\left\Vert\bar{\epsilon}_{g}\bar{X}_{g}\right\Vert^{2} & =\mathbb{E}\left[\left\Vert\bar{X}_{g}\right\Vert^{2}\mathbb{E}\left[\bar{\epsilon}_{g}^{2} \big|  X_g \right]\right]\\
 & =\mathbb{E}\left[\bar{X}_{g}\bar{X}_{g}^{\prime}\sigma_{g}^{2}\right]\\
 & =\begin{cases}
O\left(1\right), & \text{for strong dependence,}\\
O\left(\frac{h(N_{g})}{N_{g}}\right), & \text{for semi-strong dependence,}\\
O\left(\frac{1}{N_{g}}\right), & \text{for weak dependence,}
\end{cases}
\end{align*}
uniformly in $g$. Therefore, the lemma holds.

\hfill
\qed

{\bf Proof of Theorem \ref{thm:consistency of V}}

To show that $\hat{V}_{G}$ is consistent for
$V_{G},$ it suffices to show $\frac{1}{\left\Vert V_{G}\right\Vert}\left\Vert\hat{V}_{G}-V_{G}\right\Vert=o_{P}(1).$
First we will show that $\frac{1}{\left\Vert B_{G}\right\Vert}\left\Vert\hat{B}_{G}-B_{G}\right\Vert=o_{P}(1).$
To show this, note that $e_{g}=\bar{Y}_{g}-\bar{X}_{g}\hat{\beta}_{A}=\bar{\epsilon}_{g}-\bar{X}_{g}\left(\hat{\beta}_{A}-\beta\right),$
which implies $e_{g}^{2}=\bar{\epsilon}_{g}^{2}+\left(\hat{\beta}_{A}-\beta\right)^{\prime}\bar{X}_{g}^{\prime}\bar{X}_{g}\left(\hat{\beta}_{A}-\beta\right)-2\bar{\epsilon}_{g}\bar{X}_{g}\left(\hat{\beta}_{A}-\beta\right).$
Now, 
\begin{align*}
\hat{B}_{G}-B_{G} & =\frac{1}{G}\sum_{g=1}^{G}\bar{X_{g}}^{\prime}\bar{X_{g}}e_{g}^{2}-\mathbb{E}\left[\frac{1}{G}\sum_{g=1}^{G}\bar{X_{g}}^{\prime}\bar{X_{g}}\sigma_{g}^{2}\right]\\
 & =T_{1}+T_{2}+T_{3},\text{ say,}
\end{align*}
where 
\begin{align*}
T_{1} &= \frac{1}{G}\sum_{g=1}^{G}\bar{X}_{g}^{\prime}\bar{X}_{g}\bar{\epsilon}_{g}^{2}-\mathbb{E}\left[\frac{1}{G}\sum_{g=1}^{G}\bar{X}_{g}^{\prime}\bar{X}_{g}\sigma_{g}^{2}\right], \\
T_{2} &= \frac{1}{G}\sum_{g=1}^{G}\bar{X}_{g}^{\prime}\bar{X}_{g}\left(\hat{\beta}_{A}-\beta\right)^{\prime}\bar{X}_{g}^{\prime}\bar{X}_{g}\left(\hat{\beta}_{A}-\beta\right), \\
T_{3} &= -\frac{2}{G}\sum_{g=1}^{G}\bar{X}_{g}^{\prime}\bar{X}_{g}\bar{\epsilon}_{g}\bar{X}_{g}\left(\hat{\beta}_{A}-\beta\right).
\end{align*}
Now, using Assumption \ref{assump:error independence}, conditional on $X,$
\[
\mathbb{E}\left[T_{1}\right] = \mathbb{E}\left[\frac{1}{G}\sum_{g=1}^{G}\bar{X_{g}}^{\prime}\bar{X_{g}}\mathbb{E}\left[\bar{\epsilon}_{g}^{2} \mid  X_g\right]-\frac{1}{G}\sum_{g=1}^{G}\bar{X_{g}}^{\prime}\bar{X_{g}}\sigma_{g}^{2}\right]=0,
\]

and using the independence across clusters by Assumption \ref{assump:error independence}, we have
\begin{eqnarray*}
\mathbb{E}\left\Vert T_{1}\right\Vert^{2} & = & \frac{1}{G^{2}}\mathbb{E}\left\Vert\sum_{g=1}^{G}\bar{X_{g}}^{\prime}\bar{X_{g}}\left(\bar{\epsilon}_{g}^{2}-\sigma_{g}^{2}\right)\right\Vert^{2}\\
 & = & \frac{1}{G^{2}}\mathbb{E}\left[\sum_{g=1}^{G}\left\Vert\bar{X_{g}}^{\prime}\bar{X_{g}}\left(\bar{\epsilon}_{g}^{2}-\sigma_{g}^{2}\right)\right\Vert^{2}\right]\\
 & = & \frac{1}{G^{2}}\sum_{g=1}^{G}\mathbb{E}\left[\left\Vert\bar{X_{g}}^{\prime}\bar{X_{g}}\right\Vert^{2}\mathbb{E}\left[\left(\bar{\epsilon}_{g}^{2}-\sigma_{g}^{2}\right)^{2} \big|  X_g\right]\right].
\end{eqnarray*}

Using the fact that 
\[
\mathbb{E}\left[ | \bar{\epsilon}_{g} | ^{4}\mid X_g\right]=\begin{cases}
O(1), & \text{for strong dependence,}\\
O\left(\frac{h^{2}(N_{g})}{N_{g}^{2}}\right), & \text{for semi-strong dependence,}\\
O\left(\frac{1}{N_{g}^{2}}\right), & \text{for weak dependence,}
\end{cases}
\]
uniformly in $g$ from the proof of Theorem \ref{thm:normal}, and $\sigma_g^4=O\left(1\right),O\left(\frac{h^2(N_g)}{N^2_g}\right)$ and $O\left(\frac{1}{N^2_g}\right)$ uniformly in $g,$ for strong, semi-strong and weak dependence, respectively, by the equation \eqref{eq:order of sigma_g^2} of Remark \ref{rem:order of sigma_g^2}, we have 
\begin{align*}
\mathbb{E}\left[\left(\bar{\epsilon}_{g}^{2}-\sigma_{g}^{2}\right)^{2} |  X_g \right] & \leq \mathbb{E}\left[2\left( | \bar{\epsilon}_{g} | ^{4}+\sigma_{g}^{4}\right)\vert X_g \right] \\
& =\begin{cases}
O(1), & \text{for strong dependence,}\\
O\left(\frac{h^{2}(N_{g})}{N_{g}^{2}}\right), & \text{for semi-strong dependence,}\\
O\left(\frac{1}{N_{g}^{2}}\right), & \text{for weak dependence,}
\end{cases}
\end{align*}
uniformly in $g.$  Since $\mathbb{E}\left[\left\Vert\bar{X_{g}}^{\prime}\bar{X_{g}}\right\Vert^{2}\right]=O(1)$ uniformly in $g,$ we have from Lemma \ref{lemma:Exact order summability} that
$$\mathbb{E}\left\Vert T_{1}\right\Vert^{2}=\begin{cases}
O\left(\frac{1}{G}\right), & \text{for strong dependence,}\\
O\left(\frac{\sum_{g=1}^{G}\frac{h^{2}(N_{g})}{N_{g}^{2}}}{G^{2}}\right), & \text{for semi-strong dependence,}\\
O\left(\frac{\sum_{g=1}^{G}\frac{1}{N_{g}^{2}}}{G^{2}}\right), & \text{for weak dependence,}
\end{cases}$$ 
which deduces that 
\[
\left\Vert T_{1} \right\Vert=\begin{cases}
O_{P}\left(\frac{1}{\sqrt{G}}\right), & \text{for strong dependence,}\\
O_{P}\left(\frac{\sqrt{\sum_{g=1}^{G}\frac{h^{2}(N_{g})}{N_{g}^{2}}}}{G}\right), & \text{for semi-strong dependence,}\\
O_{P}\left(\frac{\sqrt{\sum_{g=1}^{G}\frac{1}{N_{g}^{2}}}}{G}\right), & \text{for weak dependence.}
\end{cases}
\]

Using the triangular inequality for norms in the first line, and Result \ref{result:trace inequality} in the second and third line, we have 
\begin{eqnarray*}
\left\Vert T_{2} \right\Vert & \leq & \frac{1}{G} \sum_{g=1}^G\left\Vert\bar{X}_{g}^{\prime}\bar{X}_{g}\left(\hat{\beta}_{A}-\beta\right)^{\prime}\bar{X}_{g}^{\prime}\bar{X}_{g}\left(\hat{\beta}_{A}-\beta\right)\right\Vert \\
& \leq & \frac{1}{G} \sum_{g=1}^G\left\Vert\bar{X}_{g}^{\prime}\bar{X}_{g}\right\Vert\left\Vert\left(\hat{\beta}_{A}-\beta\right)^{\prime}\bar{X}_{g}^{\prime}\bar{X}_{g}\left(\hat{\beta}_{A}-\beta\right)\right\Vert\\
 & \leq & \frac{1}{G} \sum_{g=1}^G\left\Vert\bar{X}_{g}^{\prime}\bar{X}_{g}\right\Vert^{2}\left\Vert\left(\hat{\beta}_{A}-\beta \right) \left(\hat{\beta}_{A}-\beta\right)^{\prime}\right\Vert\\
 & = & \left\Vert\hat{\beta}_{A}-\beta\right\Vert^2 \frac{1}{G} \sum_{g=1}^G\left\Vert\bar{X}_{g}^{\prime}\bar{X}_{g}\right\Vert^{2}.
\end{eqnarray*}

Since $\mathbb{E}\left\Vert\bar{X}_{g}^{\prime}\bar{X}_{g}\right\Vert^{2}<c_0$ uniformly in $g$ by Assumption \ref{assump: Xbar'Xbar=O(G)}, we have from Lemma \ref{lemma:big-OP sum and multiplication} and Lemma \ref{lemma:order of beta hat proposed - beta}
\[
\left\Vert T_{2} \right\Vert \; = \; \begin{cases}
O_{P}\left(\frac{1}{G}\right), & \text{for strong dependence,}\\
O_{P}\left(\frac{1}{G^{2}}\sum_{g=1}^{G}\frac{h(N_{g})}{N_{g}}\right), & \text{for semi-strong dependence,}\\
O_{P}\left(\frac{1}{G^{2}}\sum_{g=1}^{G}\frac{1}{N_{g}}\right), & \text{for weak dependence.}
\end{cases}
\]
For $T_3,$ using the triangular inequality for norms in the first line, and Result \ref{result:trace inequality} in the second line,
\begin{eqnarray*}
\left\Vert T_{3} \right\Vert & \leq & \frac{1}{G} \sum_{g=1}^G \left\Vert 2\bar{X}_{g}^{\prime}\bar{X}_{g}\bar{\epsilon}_{g}\bar{X}_{g}\left(\hat{\beta}_{A}-\beta\right)\right\Vert\\ 
& \leq & \frac{1}{G} \sum_{g=1}^G 2\left\Vert\bar{X}_{g}^{\prime}\bar{X}_{g}\right\Vert\left\Vert\bar{\epsilon}_{g}\bar{X}_{g}\right\Vert\left\Vert\hat{\beta}_{A}-\beta\right\Vert\\
& = & 2\left\Vert\hat{\beta}_{A}-\beta\right\Vert \frac{1}{G} \sum_{g=1}^G \left\Vert\bar{X}_{g}^{\prime}\bar{X}_{g}\right\Vert\left\Vert\bar{\epsilon}_{g}\bar{X}_{g}\right\Vert.
\end{eqnarray*}

Applying the C-S inequality in the first line, and using Assumption \ref{assump: Xbar'Xbar=O(G)} and Lemma \ref{lemma:order of epsilon bar X bar} in the second line, for some $0<a,b,c<\infty,$
\begin{align*}
\mathbb{E}\left[\left\Vert\bar{X}_{g}^{\prime}\bar{X}_{g}\right\Vert\left\Vert\bar{\epsilon}_{g}\bar{X}_{g}\right\Vert\right] & \leq \sqrt{\mathbb{E}\left[\left\Vert\bar{X}_{g}^{\prime}\bar{X}_{g}\right\Vert^2\right]} \sqrt{\mathbb{E}\left[\left\Vert\bar{\epsilon}_{g}\bar{X}_{g}\right\Vert^2\right]} \\
& \leq \begin{cases}
a, & \text{for strong dependence,}\\
b \sqrt{\frac{h(N_{g})}{N_{g}}}, & \text{for semi-strong dependence,}\\
c \sqrt{\frac{1}{N_{g}}}, & \text{for weak dependence,}
\end{cases}
\end{align*}
uniformly in $g.$ Hence, applying Assumption \ref{assump: Xbar'Xbar=O(G)}, Lemma \ref{lemma:big-OP sum and multiplication}, Lemma \ref{lemma:order of beta hat proposed - beta}, and Lemma \ref{lemma:order of epsilon bar X bar} in the first line, and the C-S inequality $\sum_{g=1}^{G}\sqrt{\frac{h(N_{g})}{N_{g}}}\ensuremath{\leq\sqrt{\sum_{g=1}^{G}\frac{h(N_{g})}{N_{g}}}\sqrt{G}}$ in the last line, it follows that
\begin{eqnarray*}
 \left\Vert T_{3} \right\Vert  & = & \begin{cases}
O_{P}\left(\frac{1}{\sqrt{G}}\right), & \text{for strong dependence,}\\
O_{P}\left(\frac{1}{G^{2}}\sum_{g=1}^{G}\sqrt{\frac{h(N_{g})}{N_{g}}}\sqrt{\sum_{g_{1}=1}^{G}\frac{h(N_{g_{1}})}{N_{g_{1}}}}\right), & \text{for semi-strong dependence,}\\
O_{P}\left(\frac{1}{G^{2}}\sum_{g=1}^{G}\sqrt{\frac{1}{N_{g}}}\sqrt{\sum_{g_{1}=1}^{G}\frac{1}{N_{g_{1}}}}\right), & \text{for weak dependence,}
\end{cases}\\
 & = & \begin{cases}
O_{P}\left(\frac{1}{\sqrt{G}}\right), & \text{for strong dependence,}\\
O_{P}\left(\frac{1}{G\sqrt{G}}\sum_{g=1}^{G}\frac{h(N_{g})}{N_{g}}\right), & \text{for semi-strong dependence,}\\
O_{P}\left(\frac{1}{G\sqrt{G}}\sum_{g=1}^{G}\frac{1}{N_{g}}\right), & \text{for weak dependence.}
\end{cases}
\end{eqnarray*}
Now, it is easy to observe that $\left\Vert T_{3} \right\Vert$ grows stochastically faster than $\left\Vert T_{2} \right\Vert.$ Also, using the C-S inequality: $\frac{1}{\sqrt{G}}\sum_{g=1}^{G}\frac{h(N_{g})}{N_{g}}\leq\sqrt{\sum_{g=1}^{G}\frac{h^{2}(N_{g})}{N_{g}^{2}}},$ note that $\left\Vert T_{1} \right\Vert$ grows stochastically faster than $\left\Vert T_{3} \right\Vert.$ So, by the triangular inequality for norms, we have the upper bound of $\Vert \hat{B}_{G}-B_{G}\Vert$ is
\[
\left\Vert T_{1} \right\Vert+\left\Vert T_{2} \right\Vert+\left\Vert T_{3} \right\Vert=\begin{cases}
O_{P}\left(\frac{1}{\sqrt{G}}\right), & \text{for strong dependence,}\\
O_{P}\left(\frac{\sqrt{\sum_{g=1}^{G}\frac{h^{2}(N_{g})}{N_{g}^{2}}}}{G}\right), & \text{for semi-strong dependence,}\\
O_{P}\left(\frac{\sqrt{\sum_{g=1}^{G}\frac{1}{N_{g}^{2}}}}{G}\right), & \text{for weak dependence.}
\end{cases}
\]

Since by Assumption \ref{assump:lamba_max=lambda_min}, 
\[
\left\Vert B_{G}\right\Vert=\begin{cases}
O_{e}\left(1\right), & \text{for strong dependence,}\\
O_{e}\left(\frac{1}{G}\sum_{g=1}^{G}\frac{h(N_{g})}{N_{g}}\right), & \text{for semi-strong dependence,}\\
O_{e}\left(\frac{1}{G}\sum_{g=1}^{G}\frac{1}{N_{g}}\right), & \text{for weak dependence,}
\end{cases}
\]
we have 
\begin{equation}
\label{eq: consistency of V orders}
    \frac{1}{\left\Vert B_{G}\right\Vert}\left\Vert \hat{B}_{G}-B_{G} \right\Vert=\begin{cases}
O_{P}\left(\frac{1}{\sqrt{G}}\right), & \text{for strong dependence,}\\
O_{P}\left(\frac{\sqrt{\sum_{g=1}^{G}\frac{h^{2}(N_{g})}{N_{g}^{2}}}}{\sum_{g=1}^{G}\frac{h(N_{g})}{N_{g}}}\right), & \text{for semi-strong dependence,}\\
O_{P}\left(\frac{\sqrt{\sum_{g=1}^{G}\frac{1}{N_{g}^{2}}}}{\sum_{g=1}^{G}\frac{1}{N_{g}}}\right), & \text{for weak dependence.}
\end{cases}
\end{equation}

From Lemma \ref{lemma: sum h^2/N^2 to 0}, we have $\frac{\sum_{g=1}^{G}\frac{h^{2}(N_{g})}{N_{g}^{2}}}{\left(\sum_{g=1}^{G}\frac{h(N_{g})}{N_{g}}\right)^{2}}\to 0$ and  $\frac{\sum_{g=1}^{G}\frac{1}{N_{g}^{2}}}{\left(\sum_{g=1}^{G}\frac{1}{N_{g}}\right)^{2}}\to 0,$
as $G\to\infty,$ for the nearly balanced case, and the unbalanced case with
$O(G)$ nearly balanced clusters. Thus, we have 
\begin{eqnarray*}
\frac{1}{\left\Vert B_{G}\right\Vert}\left\Vert \hat{B}_{G}-B_{G} \right\Vert & = & o_{P}(1).
\end{eqnarray*}

Since $V_{G}=A_{G}^{-1}B_{G}A_{G}^{-1},$ it follows from Lemma \ref{lemma:lambda_min lambda_max inequality}
and Assumption \ref{assump: Xbar'Xbar=O(G)} that
\begin{align*}
    \lambda_{\min}\left(V_{G}\right) & \ge \lambda_{\min}\left(A_{G}^{-2}\right)\lambda_{\min}\left(B_{G}\right)=O_{e}(1)\lambda_{\min}\left(B_{G}\right) \\
    \lambda_{\max}\left(V_{G}\right) & \leq\lambda_{\max}\left(A_{G}^{-2}\right) \lambda_{\max}\left(B_{G}\right)=O_{e}(1)\lambda_{\max}\left(B_{G}\right).
\end{align*}
Now, using the fact that $O_{e}\left(\lambda_{\min}\left(B_{G}\right)\right)=O_{e}\left(\lambda_{\max}\left(B_{G}\right)\right)$
by Assumption \ref{assump:lamba_max=lambda_min}, we have 
\[
\left\Vert V_{G}\right\Vert=O_{e}\left(\left\Vert B_{G}\right\Vert\right)=\begin{cases}
O_{e}\left(1\right), & \text{for strong dependence,}\\
O_{e}\left(\frac{1}{G}\sum_{g=1}^{G}\frac{h(N_{g})}{N_{g}}\right), & \text{for semi-strong dependence,}\\
O_{e}\left(\frac{1}{G}\sum_{g=1}^{G}\frac{1}{N_{g}}\right), & \text{for weak dependence.}
\end{cases}
\]

Note that assuming $\hat{A}_G=\frac{\bar{X}^{\prime}\bar{X}}{G},$ we can split $\hat{A}_G^{-1}\hat{B}_{G}\hat{A}_G^{-1}-A_{G}^{-1}B_{G}A_{G}^{-1}$ in a way similar to equation \eqref{eq:splitting of hat(ABA)-ABA} in the proof of Theorem \ref{thm:Vols_consistent}. Using the triangular inequality for the norms and the norm inequality, we obtain
\[
\frac{1}{\left\Vert B_G \right\Vert}\left\Vert \hat{A}_G^{-1}\hat{B}_{G}\hat{A}_G^{-1}-A_{G}^{-1}B_{G}A_{G}^{-1}\right\Vert=o_P
(1),\]
since $\left\Vert\hat{A}_G^{-1}-A_G^{-1}\right\Vert=o_P(1)$ follows from Assumption \ref{assump: Xbar'Xbar=O(G)} by the continuity theorem of probability, $\frac{1}{\left\Vert B_{G}\right\Vert}B_{G}$ is finite and uniformly positive definite by Assumption \ref{assump:lamba_max=lambda_min}, $\left\Vert A_{G}^{-1}\right\Vert=O_e(1),$ and $\frac{1}{\left\Vert B_G\right\Vert} \left\Vert \hat{B}_G-B_{G}\right\Vert=o_P(1).$ 

Hence, we have
\begin{align*}
    \frac{1}{\left\Vert V_{G}\right\Vert}\left\Vert\hat{V}_{G}-V_{G}\right\Vert & =\frac{1}{O_{e}\left(\left\Vert B_{G}\right\Vert\right)}\left\Vert\left(\frac{\bar{X}^{\prime}\bar{X}}{G}\right)^{-1}\hat{B}_{G}\left(\frac{\bar{X}^{\prime}\bar{X}}{G}\right)^{-1}-A_{G}^{-1}B_{G}A_{G}^{-1}\right\Vert \\
    & =o_{P}(1),
\end{align*}
 which deduces that $\hat{V}_{G}$ is consistent for $V_{G}.$
 
\hfill
\qed

{\bf Proof of Theorem \ref{thm:testing}}

We show the asymptotic distribution of the Wald-type test statistic based on $\hat{\beta}_A$ in two steps.

\textbf{Step 1} : In this step, we will show that under $H_{0},$
\[
\Delta_{G}^{-1/2}\sqrt{G}\left(R\hat{\beta}_{A}-r\right)\xrightarrow{d}N_{l}(0,I_{l}), \quad \text{ as } G\to\infty,
\] 
(i) for the nearly balanced case, (ii) for the unbalanced case and strong dependence, (iii) for the unbalanced case, and weak or semi-strong dependence, if $O(G)$ clusters are nearly balanced. 

First note that under $H_{0},$
\begin{align*}
    \Delta_{G}^{-1/2}\sqrt{G}\left(R\hat{\beta}_{A}-r\right) & =\Delta_{G}^{-1/2}\sqrt{G}R\left(\hat{\beta}_{A}-\beta\right) \\
    & = \Delta_{G}^{-1/2}RV_{G}^{1/2}V_{G}^{-1/2}\sqrt{G}\left(\hat{\beta}_{A}-\beta\right).
\end{align*}
Note that Theorem \ref{thm:normal} yields $V_{G}^{-1/2}\sqrt{G}\left(\hat{\beta}_{A}-\beta\right)\xrightarrow{d}N_{k}(0,I_{k}),$ (i) for nearly balanced case, (ii) for the unbalanced case and strong dependence, (iii) for the unbalanced case, and weak or semi-strong dependence, if $O(G)$ clusters are nearly balanced.
Since $R$ is a finite matrix of dimension $l\times k$ with full
row rank $l,$ and $V_{G}$ is finite and positive definite, it follows
from Corollary 4.24 of \cite{white1984asymptotic} that
\[
\Delta_{G}^{-1/2}RV_{G}^{1/2}V_{G}^{-1/2}\sqrt{G}\left(\hat{\beta}_{A}-\beta\right)\xrightarrow{d}N_{l}(0,I_{l}), \quad \text{ as } G\to\infty
\]
for the cases mentioned already.

\textbf{Step 2} : In this step, we will show that under $H_{0},$
\[
\hat{\Delta}_{G}^{-1/2}\sqrt{G}\left(R\hat{\beta}_{A}-r\right)\xrightarrow{d}N_{l}(0,I_{l}), \quad \text{ as } G\to\infty.
\]

To show the above, first note that using norm inequality,
\begin{align*}
    & \quad \left\Vert\hat{\Delta}_{G}^{-1/2}\sqrt{G}\left(R\hat{\beta}_{A}-r\right)-\Delta_{G}^{-1/2}\sqrt{G}\left(R\hat{\beta}_{A}-r\right)\right\Vert \\
    & \leq  \left\Vert\hat{\Delta}_{G}^{-1/2}\Delta_{G}^{1/2}-I_{l}\right\Vert\left\Vert\Delta_{G}^{-1/2}\sqrt{G}\left(R\hat{\beta}_{A}-r\right)\right\Vert.
\end{align*}
 From Step 1, we have shown that $\Delta_{G}^{-1/2}\sqrt{G}\left(R\hat{\beta}_{A}-r\right)\xrightarrow{d}N_{l}(0,I_{l}),$ i.e., $\left\Vert\Delta_{G}^{-1/2}\sqrt{G}\left(R\hat{\beta}_{A}-r\right)\right\Vert=O_P(1),$ under $H_{0}.$ Therefore, it is enough to show that $\left\Vert\hat{\Delta}_{G}^{-1/2}\Delta_{G}^{1/2}-I_{l}\right\Vert=o_{P}(1).$

Since $\Delta_{G}=RV_{G}R^{\prime},$ it follows from Lemma \ref{lemma:lambda_min lambda_max inequality} that 
\begin{align*}
\lambda_{\min}\left(\Delta_{G}\right) & \ge\lambda_{\min}\left(RR^{\prime}\right)\lambda_{\min} \left(V_{G}\right)=O_{e}(1)\lambda_{\min}\left(V_{G}\right) \\
\lambda_{\max}\left(\Delta_{G}\right) & \leq\lambda_{\max}\left(RR^{\prime}\right)\lambda_{\max}\left(V_{G}\right)=O_{e}(1)\lambda_{\max}\left(V_{G}\right).
\end{align*}

Now, using the fact that $O_{e}\left(\lambda_{\min}\left(B_{G}\right)\right)=O_{e}\left(\lambda_{\max}\left(B_{G}\right)\right)$
by Assumption \ref{assump:lamba_max=lambda_min}, and $\left\Vert V_{G}\right\Vert=O_{e}\left(\left\Vert B_{G}\right\Vert\right)$
from the proof of Theorem \ref{thm:consistency of V}, we have 
$\left\Vert\Delta_{G}\right\Vert=O_{e}\left(\left\Vert V_{G}\right\Vert\right).$
In the proof of Theorem \ref{thm:consistency of V}, we have shown that
$\frac{1}{\left\Vert V_{G}\right\Vert}\left\Vert\hat{V}_{G}-V_{G}\right\Vert=o_{P}(1),$
which implies $\frac{1}{O_{e}\left(\left\Vert V_{G}\right\Vert\right)}\left\Vert R\hat{V}_{G}R^{\prime}-RV_{G}R^{\prime}\right\Vert=o_{P}(1),$
which further implies that $\frac{1}{\left\Vert\Delta_{G}\right\Vert}\left\Vert\hat{\Delta}_{G}-\Delta_{G}\right\Vert=o_{P}(1).$
By Assumption \ref{assump:lamba_max=lambda_min}, we deduce that $\frac{V_{G}}{\left\Vert V_{G}\right\Vert}$
is uniformly positive definite, and it
subsequently follows that $\frac{\Delta_{G}}{\left\Vert\Delta_{G}\right\Vert}$
is also uniformly positive definite. So, by the continuity theorem for
probability convergence, taking $g(Z)=Z^{-1/2}\left(\frac{\Delta_G}{\left\Vert\Delta_G\right\Vert}\right)^{1/2}$ as a continuous function of $Z$ and putting $Z=\frac{\hat{\Delta}_G}{\left\Vert\Delta_G\right\Vert}$, we have $\left\Vert\hat{\Delta}_{G}^{-1/2}\Delta_{G}^{1/2}-I_{l}\right\Vert=o_{P}(1).$

Hence, from Step 1 and Step 2, under $H_{0},$ $\hat{\Delta}_{G}^{-1/2}\sqrt{G}\left(R\hat{\beta}_{A}-r\right)\xrightarrow{d}N_{l}(0,I_{l}),$
as $G\to\infty.$ Consequently, by the continuous mapping theorem, it follows that:
\[
G\left(R\hat{\beta}_{A}-r\right)^{\prime}\hat{\Delta}_{G}^{-1}\left(R\hat{\beta}_{A}-r\right)\xrightarrow{d}\chi_{l}^{2}, \quad \text{ as } G\to\infty
\]
(i) for the nearly balanced case, (ii) for the unbalanced case and strong dependence, (iii) for the unbalanced case, and weak or semi-strong dependence, if $O(G)$ clusters are nearly balanced. This proves the theorem.

\hfill
\qed

{\bf Proof of Result \ref{result:Inconsistent_ols_endogeneity}}

First note that from equation \eqref{eq:beta hat ols for measurement error}, the POLS estimator is $$\hat{\beta}_{POLS}=\beta+\left(\frac{X^{\prime}X}{\sum_{g=1}^{G}N_{g}}\right)^{-1}\frac{\sum_{g=1}^{G}X_{g}^{\prime}\epsilon_{g}}{\sum_{g=1}^{G}N_{g}}-\left(\frac{X^{\prime}X}{\sum_{g=1}^{G}N_{g}}\right)^{-1}\frac{\sum_{g=1}^{G}X_{g}^{\prime}\Gamma_{g}\beta}{\sum_{g=1}^{G}N_{g}},$$
which implies the following:
\begin{eqnarray}
    \hat{\beta}_{POLS}-\beta=\left(\frac{X^{\prime}X}{\sum_{g=1}^{G}N_{g}}\right)^{-1}\left(\frac{\sum_{g=1}^{G}X_{g}^{\prime}\epsilon_{g}}{\sum_{g=1}^{G}N_{g}}-\frac{\sum_{g=1}^{G}X_{g}^{\prime}\Gamma_{g}\beta}{\sum_{g=1}^{G}N_{g}}\right). \label{eq:beta hat ols - beta expression for measurement error}
\end{eqnarray}

Note that
\[
\frac{X^{\prime}X}{\sum_{g=1}^{G}N_{g}}=\frac{\sum_{g=1}^{G}X_{g}^{*\prime}X_{g}^{*}}{\sum_{g=1}^{G}N_{g}}+\frac{\sum_{g=1}^{G}X_{g}^{*\prime}\Gamma_{g}}{\sum_{g=1}^{G}N_{g}}+\frac{\sum_{g=1}^{G}\Gamma_{g}^{\prime}X_{g}^{*}}{\sum_{g=1}^{G}N_{g}}+\frac{\sum_{g=1}^{G}\Gamma_{g}^{\prime}\Gamma_{g}}{\sum_{g=1}^{G}N_{g}}.\]

Now, $\frac{\sum_{g=1}^{G}X_{g}^{*\prime}X_{g}^{*}}{\sum_{g=1}^{G}N_{g}}=C_{G}^{*}+o_{P}(1),$
where $C_{G}^{*}$ is finite and uniformly positive definite, by Assumption \ref{assump: Xg'Xg=O(Ng) for measurement error}.
Also, by Assumption \ref{assump:exogeneity for measurement error}, $\mathbb{E}\left[\frac{\sum_{g=1}^{G}X_{g}^{*\prime}\Gamma_{g}}{\sum_{g=1}^{G}N_{g}}\right] = \mathbb{E}\left[\frac{1}{\sum_{g=1}^{G}N_{g}}\sum_{g=1}^{G}X_{g}^{*\prime}\,\mathbb{E}\left[\Gamma_{g} | X_g^{*}\right]\right]=0,$
and 
\begin{eqnarray*}
 &  & \mathbb{E}\left\Vert\frac{\sum_{g=1}^{G}X_{g}^{*\prime}\Gamma_{g}}{\sum_{g=1}^{G}N_{g}}\right\Vert^{2}\\
 & = & \frac{1}{\left(\sum_{g=1}^{G}N_{g}\right)^{2}}\mathbb{E}\operatorname{tr}\left[\sum_{g=1}^{G}X_{g}^{*\prime}\Gamma_{g}\Gamma_{g}^{\prime}X_{g}^{*}+\sum_{\underset{g\neq g_{1}}{g,g_{1}=1}}^{G}X_{g}^{*\prime}\Gamma_{g}\Gamma_{g_{1}}^{\prime}X_{g_{1}}^{*}\right]\\
 & = & \frac{1}{\left(\sum_{g=1}^{G}N_{g}\right)^{2}}\mathbb{E}\operatorname{tr}\left[\sum_{g=1}^{G}X_{g}^{*\prime}\Gamma_{g}\Gamma_{g}^{\prime}X_{g}^{*} \right],\text{ due to independence }\\
 & = & \frac{1}{\left(\sum_{g=1}^{G}N_{g}\right)^{2}}\sum_{g=1}^{G}\mathbb{E}\left[\operatorname{tr}\left(X_{g}^{*\prime}\,\mathbb{E}\left[\Gamma_{g}\Gamma_{g}^{\prime} \big|  X_g^{*}\right]X_{g}^{*}\right)\right]\\
 & \leq & \frac{1}{\left(\sum_{g=1}^{G}N_{g}\right)^{2}}\sum_{g=1}^{G}\mathbb{E}\left[\operatorname{tr}\left(X_{g}^{*\prime}X_{g}^{*}\right)\operatorname{tr}\,\mathbb{E}\left[\Gamma_{g}\Gamma_{g}^{\prime} \big|  X_g^{*}\right]\right],\text{ by Result \ref{result:trace inequality}}\\
 & = & \frac{1}{\left(\sum_{g=1}^{G}N_{g}\right)^{2}}\sum_{g=1}^{G}\mathbb{E}\left[\operatorname{tr}\left(X_{g}^{*\prime}X_{g}^{*}\right)\sum_{i=1}^{N_{g}}\sum_{j=1}^{k}\mathbb{E}\left[\gamma_{gij}^{2} \big|  X_g^{*}\right]\right]\\
 & = & \frac{1}{\left(\sum_{g=1}^{G}N_{g}\right)^{2}}\sum_{g=1}^{G}\mathbb{E}\left[\operatorname{tr}\left(X_{g}^{*\prime}X_{g}^{*}\right)\right]O\left(N_{g}\right),\text{ by Assumption \ref{assump:error independence for measurement error}}\\
 & = & O\left(\frac{\sum_{g=1}^{G}N_{g}^{2}}{\left(\sum_{g=1}^{G}N_{g}\right)^{2}}\right),\text{ by Assumption \ref{assump: Xg'Xg=O(Ng) for measurement error} and Lemma \ref{lemma:Exact order summability}.}
\end{eqnarray*}

Following the proof of Theorem \ref{thm:ols consistency}, $O\left(\frac{\sum_{g=1}^{G}N_{g}^{2}}{\left(\sum_{g=1}^{G}N_{g}\right)^{2}}\right)\to0,$
if $\frac{\underset{g:g\leq G}{\text{ max }}N_g}{\sum_{g=1}^G N_g} \to 0$. So, in this scenario, $\frac{\sum_{g=1}^{G}X_{g}^{*\prime}\Gamma_{g}}{\sum_{g=1}^{G}N_{g}}=o_{P}(1).$ In the similar fashion, it can be shown
that $\frac{\sum_{g=1}^{G}\Gamma_{g}^{\prime}X_{g}^{*}}{\sum_{g=1}^{G}N_{g}}=o_{P}(1),$ if $\frac{\underset{g:g\leq G}{\text{ max }}N_g}{\sum_{g=1}^G N_g} \to 0$.  Now, for the last term, note that $\mathbb{E}\left[\frac{\sum_{g=1}^{G}\Gamma_{g}^{\prime}\Gamma_{g}}{\sum_{g=1}^{G}N_{g}} \bigg|  X_g^{*}\right]\to C_{0},$
which is positive definite, by Assumption \ref{assump: convergence of Gamma'Gamma}(i), and
\begin{eqnarray*}
 &  & \mathbb{E}\left\Vert\frac{\sum_{g=1}^{G}\Gamma_{g}^{\prime}\Gamma_{g}}{\sum_{g=1}^{G}N_{g}}-\mathbb{E}\left[\frac{\sum_{g=1}^{G}\Gamma_{g}^{\prime}\Gamma_{g}}{\sum_{g=1}^{G}N_{g}}\right]\right\Vert^{2}\\
 & = & \left(\frac{1}{\sum_{g=1}^{G}N_{g}}\right)^{2}\mathbb{E}\left[\operatorname{tr}\left(\sum_{g=1}^{G}\left(\Gamma_{g}^{\prime}\Gamma_{g}-\mathbb{E}\left[\Gamma_{g}^{\prime}\Gamma_{g}\right]\right)\right)^{2}\right]\\
 & = & \left(\frac{1}{\sum_{g=1}^{G}N_{g}}\right)^{2}\sum_{g=1}^{G}\mathbb{E}\mathbb{E}\left[\operatorname{tr}\left(\Gamma_{g}^{\prime}\Gamma_{g}-\mathbb{E}\left[\Gamma_{g}^{\prime}\Gamma_{g}\right]\right)^{2} \Big|  X_g^{*}\right],\text{ due to independence}\\
 & = & o(1),\text{ by Assumption \ref{assump: convergence of Gamma'Gamma}(ii).}
\end{eqnarray*}

So, we have shown that $\frac{\sum_{g=1}^{G}\Gamma_{g}^{\prime}\Gamma_{g}}{\sum_{g=1}^{G}N_{g}}=C_{0}+o_{P}(1).$
Hence, $\frac{X^{\prime}X}{\sum_{g=1}^{G}N_{g}}=C_{G}^{*}+C_{0}+o_{P}(1),$
where $C_{G}^{*}$ is uniformly positive definite and $C_{0}$ is positive definite, which implies
that $\left(\frac{X^{\prime}X}{\sum_{g=1}^{G}N_{g}}\right)^{-1} = \left(C_{G}^{*}+C_{0}\right)^{-1}+o_P(1),$
by the continuity theorem of probability convergence, where $\left(C_{G}^{*}+C_{0}\right)^{-1}$
is also uniformly positive definite. Therefore, by Lemma \ref{lemma:A_n=B_n+o_P(1) implies A_n = O_P(1)},
\[
\left(\frac{X^{\prime}X}{\sum_{g=1}^{G}N_{g}}\right)^{-1}=O_P(1).
\]

Now, note that 
\[\frac{\sum_{g=1}^{G}X_{g}^{\prime}\epsilon_{g}}{\sum_{g=1}^{G}N_{g}}=\frac{\sum_{g=1}^{G}X_{g}^{*\prime}\epsilon_{g}}{\sum_{g=1}^{G}N_{g}}+\frac{\sum_{g=1}^{G}\Gamma_{g}^{\prime}\epsilon_{g}}{\sum_{g=1}^{G}N_{g}}.\]

Following the proof of Theorem \ref{thm:ols consistency}, it can be shown that
$\frac{\sum_{g=1}^{G}X_{g}^{*\prime}\epsilon_{g}}{\sum_{g=1}^{G}N_{g}}=o_P(1),$ for the following cases: (i) nearly balanced clusters under any kind of dependence, (ii)  unbalanced clusters and weak or semi-strong dependence,   (iii) unbalanced clusters and strong dependence, if $\frac{\underset{g:g=1(1)G}{\text{ max }} N_g}{\sum_{g=1}^G N_g} \to 0.$ For the second term, since $\epsilon_{g}$ and
$\Gamma_{g}$ are independently distributed, $\mathbb{E}\left[\frac{\sum_{g=1}^{G}\Gamma_{g}^{\prime}\epsilon_{g}}{\sum_{g=1}^{G}N_{g}}\right]=0.$
Also, using the independence across clusters in the first line, the C-S inequality in the second line, applying Assumption \ref{assump:error independence for measurement error} in the third line, and using $\mathbb{E}\left[\left\Vert\Gamma_{g}^{\prime}\Gamma_{g}\right\Vert \big|  X_g^{*}\right]=O\left(N_{g}\right)$ uniformly in $g$ and $\mathbb{E}\left[\left\Vert\epsilon_{g}^{\prime}\epsilon_{g}\right\Vert \big|  X_g^{*}\right]=O\left(N_{g}\right)$ uniformly in $g$
in the last line, we have
\begin{align*}
\mathbb{E}\left\Vert\frac{\sum_{g=1}^{G}\Gamma_{g}^{\prime}\epsilon_{g}}{\sum_{g=1}^{G}N_{g}}\right\Vert^{2} & =\left(\frac{1}{\sum_{g=1}^{G}N_{g}}\right)^{2}\sum_{g=1}^{G}\mathbb{E}\left[\left\Vert\Gamma_{g}^{\prime}\epsilon_{g}\right\Vert^{2}\right]\\
 & \leq\left(\frac{1}{\sum_{g=1}^{G}N_{g}}\right)^{2}\sum_{g=1}^{G}\mathbb{E}\left[\left\Vert\Gamma_{g}^{\prime}\Gamma_{g}\right\Vert\cdot\left\Vert\epsilon_{g}^{\prime}\epsilon_{g}\right\Vert\right]\\
  & = \left(\frac{1}{\sum_{g=1}^{G}N_{g}}\right)^{2}\sum_{g=1}^{G}\mathbb{E}\left[\left\Vert\Gamma_{g}^{\prime}\Gamma_{g}\right\Vert\right]\mathbb{E}\left[\left\Vert\epsilon_{g}^{\prime}\epsilon_{g}\right\Vert\right]\\
 & =O\left(\frac{\sum_{g=1}^{G}N_{g}^{2}}{\left(\sum_{g=1}^{G}N_{g}\right)^{2}}\right), \;\text{ by Lemma \ref{lemma:Exact order summability}.}
\end{align*}
Note that $O\left(\frac{\sum_{g=1}^{G}N_{g}^{2}}{\left(\sum_{g=1}^{G}N_{g}\right)^{2}}\right)\to0,$
if $\frac{\underset{g:g\leq G}{\text{ max }}N_g}{\sum_{g=1}^G N_g} \to 0$.  So, $\frac{\sum_{g=1}^{G}\Gamma_{g}^{\prime}\epsilon_{g}}{\sum_{g=1}^{G}N_{g}}=o_P(1),$  if $\frac{\underset{g:g\leq G}{\text{ max }}N_g}{\sum_{g=1}^G N_g} \to 0$. Hence, 
\[
\frac{\sum_{g=1}^{G}X_{g}^{\prime}\epsilon_{g}}{\sum_{g=1}^{G}N_{g}}=o_P(1), \;\text{  if } \frac{\underset{g:g\leq G}{\text{ max }}N_g}{\sum_{g=1}^G N_g} \to 0.
\]

Now, for the term $\frac{\sum_{g=1}^{G}X_{g}^{\prime}\Gamma_{g}\beta}{\sum_{g=1}^{G}N_{g}},$ first note that \[
\frac{\sum_{g=1}^{G}X_{g}^{\prime}\Gamma_{g}\beta}{\sum_{g=1}^{G}N_{g}}=\frac{\sum_{g=1}^{G}X_{g}^{*\prime}\Gamma_{g}\beta}{\sum_{g=1}^{G}N_{g}}+\frac{\sum_{g=1}^{G}\Gamma_{g}^{\prime}\Gamma_{g}\beta}{\sum_{g=1}^{G}N_{g}}.\]

We already have shown that $\frac{\sum_{g=1}^{G}X_{g}^{*\prime}\Gamma_{g}}{\sum_{g=1}^{G}N_{g}}=o_{P}(1)$
and $\frac{\sum_{g=1}^{G}\Gamma_{g}^{\prime}\Gamma_{g}}{\sum_{g=1}^{G}N_{g}}=C_{0}+o_{P}(1),$
where $C_{0}$ is positive definite, which implies that $\frac{\sum_{g=1}^{G}X_{g}^{*\prime}\Gamma_{g}\beta}{\sum_{g=1}^{G}N_{g}}=o_P(1)$
and $\frac{\sum_{g=1}^{G}\Gamma_{g}^{\prime}\Gamma_{g}\beta}{\sum_{g=1}^{G}N_{g}} = C_{0}\beta + o_P(1),$ since $\beta$ is a finite-dimensional vector. Therefore, for any kind of dependence of $\gamma_{g\cdot j},$
\[
\frac{\sum_{g=1}^{G}X_{g}^{\prime}\Gamma_{g}\beta}{\sum_{g=1}^{G}N_{g}}=C_{0}\beta + o_P(1).
\]

Hence, combining all the facts, from equation \eqref{eq:beta hat ols - beta expression for measurement error}, we have 
\begin{align*}
\hat{\beta}_{POLS}-\beta & =\left(\frac{X^{\prime}X}{\sum_{g=1}^{G}N_{g}}\right)^{-1}\left(\frac{\sum_{g=1}^{G}X_{g}^{\prime}\epsilon_{g}}{\sum_{g=1}^{G}N_{g}}-\frac{\sum_{g=1}^{G}X_{g}^{\prime}\Gamma_{g}\beta}{\sum_{g=1}^{G}N_{g}}\right)\\
 & =O_{P}(1)\cdot\left(o_P(1)-\left(C_{0}\beta+o_P(1)\right)\right)\\
 & =O_{P}\left(-C_{0}\beta\right), \text{ by Lemma \ref{lemma:big-OP sum and multiplication}.}
\end{align*}

So, under any type of dependence of $\epsilon_{g}$ and $\gamma_{g\cdot j},$ we have $\hat{\beta}_{POLS}-\beta\neq o_P(1),$ unless $\beta=0.$ Hence, $\hat{\beta}_{POLS}$ is biased and inconsistent for any kind of dependence of $\epsilon_{g}$ and $\gamma_{g\cdot j}.$

\hfill
\qed

\textbf{Proof of Lemma \ref{lemma: m' error consistency N_g large condition}} 

(i) Let $\mathcal{W}_G$ be the set of indices $g$ where $N_g \not\to \infty$ (i.e., the bounded clusters). By assumption, its cardinality is $|\mathcal{W}_G| = o(G)$. Let $\mathcal{U}_G$ be the set of remaining unbounded clusters where $N_g \to \infty$. 

Furthermore, let $\mathcal{C}_1$ be the set of clusters where $f(N_g) = O_e(h(N_g)/N_g)$, which by assumption satisfies $|\mathcal{C}_1| = O(G)$. Let $\mathcal{C}_2$ be the set of remaining clusters where $f(N_g) = O_e(1/N_g)$. We can decompose the average into three parts based on boundedness and the functional form of $f(N_g)$:
\[
\frac{1}{G} \sum_{g=1}^G f(N_g) = \frac{1}{G} \sum_{g \in \mathcal{W}_G} f(N_g) + \frac{1}{G} \sum_{g \in \mathcal{U}_G \cap \mathcal{C}_1} f(N_g) + \frac{1}{G} \sum_{g \in \mathcal{U}_G \cap \mathcal{C}_2} f(N_g).
\]

For the first term, since $N_g$ is bounded for $g \in \mathcal{W}_G$, both $1/N_g$ and $h(N_g)/N_g$ evaluate to bounded values. Therefore, $f(N_g)$ is bounded by some constant $C_1 > 0$ regardless of whether $g$ belongs to $\mathcal{C}_1$ or $\mathcal{C}_2$. Thus, the first term is bounded by $C_1 \frac{|\mathcal{W}_G|}{G}$. Since $|\mathcal{W}_G| = o(G)$, this term converges to $0$ as $G \to \infty$.

For the second term, we consider unbounded clusters in $\mathcal{C}_1$. Since $N_g \to \infty$ for all $g \in \mathcal{U}_G$, the minimum cluster size in this intersection, $M_1 = \min_{g \in \mathcal{U}_G \cap \mathcal{C}_1} N_g \to \infty$ as $G \to \infty$. Because $\frac{h(x)}{x} \to 0$ as $x \to \infty$, the maximum value in this set, $\max_{g \in \mathcal{U}_G \cap \mathcal{C}_1} f(N_g)$, must also converge to $0$. The second term is bounded by $\frac{|\mathcal{C}_1|}{G} \max_{g \in \mathcal{U}_G \cap \mathcal{C}_1} f(N_g)$. Since $|\mathcal{C}_1| = O(G)$, the ratio $\frac{|\mathcal{C}_1|}{G}$ is bounded, and thus the entire second term converges to $0$.

For the third term, we evaluate unbounded clusters in $\mathcal{C}_2$, where $f(N_g) = O_e(1/N_g)$. Let $M_2 = \min_{g \in \mathcal{U}_G \cap \mathcal{C}_2} N_g$. Since $N_g \to \infty$ for these clusters, $M_2 \to \infty$ as $G \to \infty$. The maximum value of $f(N_g)$ in this set converges to $0$ proportionally to $1/M_2$. The third term is bounded by $\frac{|\mathcal{C}_2|}{G} \max_{g \in \mathcal{U}_G \cap \mathcal{C}_2} f(N_g)$. Because $|\mathcal{C}_2| \leq G$, this term also converges to $0$ as $G \to \infty$.

Since all three terms vanish as $G \to \infty$, we have $O\left(\frac{1}{G} \sum_{g=1}^G f(N_g)\right) \to 0$.

\vspace{1em}

(ii) Let $\mathcal{W}_G$ be defined as above, with $|\mathcal{W}_G| = o(G)$. Let $\mathcal{U}_G = \{1, \dots, G\} \setminus \mathcal{W}_G$ be the set of all remaining unbounded clusters where $N_g \to \infty$. We decompose the sum as:
\[
\frac{1}{G} \sum_{g=1}^G \frac{1}{N_g} = \frac{1}{G} \sum_{g \in \mathcal{W}_G} \frac{1}{N_g} + \frac{1}{G} \sum_{g \in \mathcal{U}_G} \frac{1}{N_g}
\]

Since $N_g \geq 1$ for all $g$, the first term is bounded by $\frac{|\mathcal{W}_G|}{G}$, which vanishes as $G \to \infty$ because $|\mathcal{W}_G| = o(G)$. 

For the second term, let $M_G = \min_{g \in \mathcal{U}_G} N_g$. Since $N_g \to \infty$ for all $g \in \mathcal{U}_G$, it follows that $M_G \to \infty$ as $G \to \infty$. The second term is bounded by $\frac{|\mathcal{U}_G|}{G} \frac{1}{M_G} \leq \frac{1}{M_G}$, which also vanishes as $G \to \infty$. 

The sum of these two terms converges to $0$, which completes the proof for (ii).

\hfill
\qed

{\bf Proof of Result \ref{result:consistent_our_estimate_endogeneity}}

First note that from equation \eqref{eq:beta hat for measurement error}, the proposed estimator $\hat{\beta}_{A}$ can be written as 
\begin{eqnarray}
    \hat{\beta}_{A}-\beta=\left(\frac{\bar{X}^{\prime}\bar{X}}{G}\right)^{-1}\left(\frac{\sum_{g=1}^{G}\bar{X}_{g}^{\prime}\bar{\epsilon}_{g}}{G}-\frac{\sum_{g=1}^{G}\bar{X}_{g}^{\prime}\bar{\gamma}_{g}\beta}{G}\right). \label{eq:beta hat - beta expression for meauserment error}
\end{eqnarray}

Note that 
\[
\frac{\bar{X}^{\prime}\bar{X}}{G}=\frac{\sum_{g=1}^{G}\bar{X}_{g}^{*\prime}\bar{X}_{g}^{*}}{G}+\frac{\sum_{g=1}^{G}\bar{X}_{g}^{*\prime}\bar{\gamma}_{g}}{G}+\frac{\sum_{g=1}^{G}\bar{\gamma}_{g}^{\prime}\bar{X}_{g}^{*}}{G}+\frac{\sum_{g=1}^{G}\bar{\gamma}_{g}^{\prime}\bar{\gamma}_{g}}{G}.
\] 

Now, $\frac{\sum_{g=1}^{G}\bar{X}_{g}^{*\prime}\bar{X}_{g}^{*}}{G}=A_{G}^{*}+o_{P}(1),$
where $A_{G}^{*}$ is finite and uniformly positive definite, by Assumption \ref{assump: Xbar'Xbar=O(G) for measurement error}.
Also, by Assumption \ref{assump:exogeneity for measurement error},
\[
\mathbb{E}\left[\frac{\sum_{g=1}^{G}\bar{X}_{g}^{*\prime}\bar{\gamma}_{g}}{G}\right]=\mathbb{E}\left[\frac{1}{G}\sum_{g=1}^{G}\bar{X}_{g}^{*\prime}\,\mathbb{E}\left[\bar{\gamma}_{g} \big|  X_g^{*}\right]\right]=0,
\]
and using the fact that $\mathbb{E}\left[\bar{\gamma}_{g}^{\prime}\bar{\gamma}_{g} \mid X_g^{*}\right] = \left( \frac{\mathbf{1}^\prime\Lambda_{g,jl}\mathbf{1}}{N_{g}^{2}} \right)_{j,l=1}^k$, we have
\begin{align*}
\mathbb{E}\left\Vert\frac{1}{G}\sum_{g=1}^{G}\bar{X}_{g}^{*\prime}\bar{\gamma}_{g}\right\Vert^{2} & =  \frac{1}{G^{2}}\mathbb{E}\left[\operatorname{tr}\left(\sum_{g=1}^{G}\bar{X}_{g}^{*\prime}\bar{\gamma}_{g}\sum_{g_{1}=1}^{G}\bar{\gamma}_{g_{1}}^{\prime}\bar{X}_{g_{1}}^{*}\right)\right]\\
 & =  \frac{1}{G^{2}}\mathbb{E}\left[\operatorname{tr}\left(\sum_{g=1}^{G}\bar{X}_{g}^{*\prime}\bar{\gamma}_{g}\bar{\gamma}_{g}^{\prime}\bar{X}_{g}^{*}\right)\right], \quad \text{due to independence}\\
 & =  \frac{1}{G^{2}}\sum_{g=1}^{G}\mathbb{E}\left[\operatorname{tr}\left(\bar{X}_{g}^{*\prime}\,\mathbb{E}\left[\bar{\gamma}_{g}\bar{\gamma}_{g}^{\prime} \mid  X_g^{*}\right]\bar{X}_{g}^{*}\right)\right]\\
 & \leq  \frac{1}{G^{2}}\sum_{g=1}^{G}\mathbb{E}\left[\operatorname{tr}\left(\bar{X}_{g}^{*\prime}\bar{X}_{g}^{*}\right)\operatorname{tr}\left(\mathbb{E}\left[\bar{\gamma}_{g}\bar{\gamma}_{g}^{\prime} \mid  X_g^{*}\right]\right)\right], \quad \text{by Result \ref{result:trace inequality}}\\
 & =  \frac{1}{G^{2}}\sum_{g=1}^{G}\mathbb{E}\left[\operatorname{tr}\left(\bar{X}_{g}^{*\prime}\bar{X}_{g}^{*}\right)\right]\sum_{j=1}^{k}\frac{\mathbf{1}^\prime\Lambda_{g,jj}\mathbf{1}}{N_{g}^{2}}.
\end{align*}

Now, using the fact that $\frac{\mathbf{1}^\prime\Lambda_{g,jj}\mathbf{1}}{N_{g}^{2}}=O(1)$ uniformly in $j,g$ for any kind of dependence, using Assumption \ref{assump: Xbar'Xbar=O(G) for measurement error}, and Lemma \ref{lemma:Exact order summability}, we have 
\[
\mathbb{E}\left\Vert\frac{\sum_{g=1}^{G}\bar{X}_{g}^{*\prime}\bar{\gamma}_{g}}{G}\right\Vert^{2} = O\left(\frac{k}{G}\right)  \rightarrow  0, \; \text{ as }G\rightarrow\infty.
\]
So, $\frac{\sum_{g=1}^{G}\bar{X}_{g}^{*\prime}\bar{\gamma}_{g}}{G}=o_{P}(1).$
Similarly, it is easy to note that  $\frac{\sum_{g=1}^{G}\bar{\gamma}_{g}^{\prime}\bar{X}_{g}^{*}}{G}=o_{P}(1).$
Now, for the last term, note that by the Cauchy-Schwarz inequality, the off-diagonal cross-moments are bounded by their corresponding diagonal variance terms; specifically, $\left| \mathbf{1}^\prime\Lambda_{g,jl}\mathbf{1} \right| \leq \left( \mathbf{1}^\prime\Lambda_{g,jj}\mathbf{1} \cdot \mathbf{1}^\prime\Lambda_{g,ll}\mathbf{1} \right)^{1/2}$. Consequently, the asymptotic order of the matrix is strictly determined by its diagonal entries. Thus, by equation \eqref{eq:order of gamma gj bar square} and Lemma \ref{lemma:Exact order summability}, we have
\begin{align*}
    \mathbb{E}\left[\frac{1}{G}\sum_{g=1}^{G}\bar{\gamma}_{g}^{\prime}\bar{\gamma}_{g}\right] & =  \left( \frac{1}{G} \sum_{g=1}^G\frac{\mathbf{1}^\prime\Lambda_{g,jl}\mathbf{1}}{N_{g}^{2}} \right)_{j,l=1}^k  \\ 
    & = \begin{cases}
    \left( O\left( \frac{1}{G} \sum_{g=1}^G\frac{h_\gamma(N_g)}{N_{g}} \right) \right)_{j,l=1}^k, & \text{for semi-strong dependence,}\\
    \left( O\left( \frac{1}{G} \sum_{g=1}^G\frac{1}{N_{g}} \right) \right)_{j,l=1}^k, & \text{for weak dependence.}
    \end{cases}
\end{align*}
The above orders converge to zero by Lemma~\ref{lemma: m' error consistency N_g large condition}, provided that the number of bounded clusters is $o(G)$, while $N_g \to \infty$ for all remaining clusters. Thus,
\begin{small}
\[
\mathbb{E}\left[\frac{\sum_{g=1}^{G}\bar{\gamma}_{g}^{\prime}\bar{\gamma}_{g}}{G}\right]
\xrightarrow[G\to\infty]{}
\begin{cases}
0^{k\times k}, & \text{under weak or semi-strong dependence of } \gamma_{g\cdot j}, \\[4pt]
C^{*}, & \text{under strong dependence of } \gamma_{g\cdot j}.
\end{cases}
\]
\end{small}
By Assumption~\ref{assump: convergence of Gamma bar'Gamma bar for average model}(i),
the above strong-dependence case holds, where $C^{*}$ is positive definite. Also, for weak or semi-strong dependence of $\gamma_{g\cdot j}$, 
\begin{eqnarray*}
& & \mathbb{E}\left\Vert\frac{\sum_{g=1}^{G}\bar{\gamma}_{g}^{\prime}\bar{\gamma}_{g}}{G}-\mathbb{E}\left[\frac{\sum_{g=1}^{G}\bar{\gamma}_{g}^{\prime}\bar{\gamma}_{g}}{G}\right]\right\Vert^{2}\\
& = & \mathbb{E}\left[\operatorname{tr}\left(\frac{1}{G^{2}}\sum_{g=1}^{G}\left[\bar{\gamma}_{g}^{\prime}\bar{\gamma}_{g}-\mathbb{E}\left[\bar{\gamma}_{g}^{\prime}\bar{\gamma}_{g}\right]\right]\sum_{g_1=1}^G\left[\bar{\gamma}_{g_1}^{\prime}\bar{\gamma}_{g_1}-\mathbb{E}\left[\bar{\gamma}_{g_1}^{\prime}\bar{\gamma}_{g_1}\right]\right]\right) \right]\\
 & = & \frac{1}{G^{2}}\sum_{g=1}^{G}\mathbb{E}\mathbb{E}\left[\operatorname{tr}\left(\left[\bar{\gamma}_{g}^{\prime}\bar{\gamma}_{g}-\mathbb{E}\left[\bar{\gamma}_{g}^{\prime}\bar{\gamma}_{g}\right]\right]^2\right) \big|  X_g^{*}\right],\text{ due to independence} \\
& = & \frac{1}{G^{2}}\sum_{g=1}^{G}\operatorname{tr}\left(\mathbb{E}\mathbb{E}\left[\left[\bar{\gamma}_{g}^{\prime}\bar{\gamma}_{g}-\mathbb{E}\left[\bar{\gamma}_{g}^{\prime}\bar{\gamma}_{g}\right]\right]^2 \big|  X_g^{*}\right]\right) \\
& = & \frac{1}{G^{2}}\sum_{g=1}^{G}\operatorname{tr}\left(\mathbb{E}\left[\left[\bar{\gamma}_{g}^{\prime}\bar{\gamma}_{g}\right]^2\right]-\mathbb{E}^2\left[\bar{\gamma}_{g}^{\prime}\bar{\gamma}_{g}\right]\right).
\end{eqnarray*}

Since $\bar{\gamma}_{g}^{\prime}\bar{\gamma}_{g}$ is symmetric and non-negative definite, its expectation $\mathbb{E}\left[\bar{\gamma}_{g}^{\prime}\bar{\gamma}_{g}\right]$ is also symmetric. Consequently, the square of this expectation, $\left(\mathbb{E}\left[\bar{\gamma}_{g}^{\prime}\bar{\gamma}_{g}\right]\right)^2$, is positive semi-definite, implying its trace is non-negative. By the linearity of the trace operator, it follows that
\begin{align*}
    & \quad \frac{1}{G^{2}}\sum_{g=1}^{G}\operatorname{tr}\left(\mathbb{E}\left[\left[\bar{\gamma}_{g}^{\prime}\bar{\gamma}_{g}\right]^2\right]-\mathbb{E}^2\left[\bar{\gamma}_{g}^{\prime}\bar{\gamma}_{g}\right]\right) \\
    & \leq \frac{1}{G^{2}}\sum_{g=1}^{G}\operatorname{tr}\left(\mathbb{E}\left[\left[\bar{\gamma}_{g}^{\prime}\bar{\gamma}_{g}\right]^2\right]\right)\\
    & = \frac{1}{G^{2}}\sum_{g=1}^{G}\mathbb{E}\left[\operatorname{tr}\left(\bar{\gamma}_{g}^{\prime}\bar{\gamma}_{g}\right)^{2}\right] \\
     & \leq  \frac{1}{G^{2}}\sum_{g=1}^{G}\mathbb{E}\left[\left(\operatorname{tr}\left(\bar{\gamma}_{g}^{\prime}\bar{\gamma}_{g}\right)\right)^{2}\right], \text{ by Result \ref{result:trace inequality}}\\
     & = \frac{1}{G^{2}}\sum_{g=1}^{G}\mathbb{E}\left[\left(\sum_{j=1}^{k}\bar{\gamma}_{g\cdot j}^{2}\right)^{2}\right],\text{ where }\bar{\gamma}_{g\cdot j}=\frac{1}{N_{g}}\sum_{i=1}^{N_{g}}\gamma_{gij}\\
    & \leq \frac{1}{G^{2}}\sum_{g=1}^{G}\mathbb{E}\left[k\sum_{j=1}^{k}\bar{\gamma}_{g\cdot j}^{4}\right],\text{ by C-S inequality}\\
    & =  \frac{k}{G^{2}}\sum_{g=1}^{G}\sum_{j=1}^{k}\mathbb{E}\mathbb{E}\left[\left(\frac{1}{N_{g}}\sum_{i=1}^{N_{g}}\gamma_{gij}\right)^{4} \Bigg|  X_g^{*}\right]\\
    & \leq  \frac{k}{G^{2}}\sum_{g=1}^{G}\sum_{j=1}^{k}\frac{N_{g}^{3}}{N_{g}^{4}}\sum_{i=1}^{N_{g}}\mathbb{E}\mathbb{E}\left[ | \gamma_{gij} | ^{4} \big|  X_g^{*}\right],\text{ by H\ensuremath{\ddot{o}}lder's inequality}\\
    & = O\left(\frac{k^{2}}{G}\right), \text{ by Assumption \ref{assump:existence of moment for gamma} and Lemma \ref{lemma:Exact order summability}}\\
    & \to  0,\text{ as }G\to\infty.
\end{align*}

For strong dependence, $\mathbb{E}\left\Vert\frac{\sum_{g=1}^{G}\bar{\gamma}_{g}^{\prime}\bar{\gamma}_{g}}{G}-\mathbb{E}\left[\frac{\sum_{g=1}^{G}\bar{\gamma}_{g}^{\prime}\bar{\gamma}_{g}}{G}\right]\right\Vert^{2}=o(1),$
as $G\to\infty,$ by Assumption \ref{assump: convergence of Gamma bar'Gamma bar for average model}(ii). So, 
\begin{small}
    \begin{equation}
\label{eq: avg of gamma_g_bar'gamma_g_bar order}
\frac{\sum_{g=1}^{G}\bar{\gamma}_{g}^{\prime}\bar{\gamma}_{g}}{G}=\begin{cases}
o_{P}(1), & \text{for weak or semi-strong dependence of \ensuremath{\gamma_{g\cdot j}}}\\
C^{*}+o_{P}(1), & \text{for strong dependence of \ensuremath{\gamma_{g\cdot j}},\text{ by Assumption \ref{assump: convergence of Gamma bar'Gamma bar for average model}} }
\end{cases}
\end{equation}
\end{small}
provided that the number of bounded clusters is $o(G)$, while $N_g \to \infty$ for all remaining clusters. Hence, 
\[
\frac{\bar{X}^{\prime}\bar{X}}{G}=\begin{cases}
A_{G}^{*}+o_{P}(1), & \text{for weak or semi-strong dependence of \ensuremath{\gamma_{g\cdot j}}}\\
A_{G}^{*}+C^{*}+o_{P}(1), & \text{for strong dependence of \ensuremath{\gamma_{g\cdot j}},\text{ by Assumption \ref{assump: convergence of Gamma bar'Gamma bar for average model}} }
\end{cases}
\]
where $A_{G}^{*}$ is uniformly positive definite by Assumption \ref{assump: Xbar'Xbar=O(G) for measurement error} and $C^{*}$ is positive definite by Assumption \ref{assump: convergence of Gamma bar'Gamma bar for average model}. By the continuity theorem of probability convergence,
\begin{small}
    \[
\left(\frac{\bar{X}^{\prime}\bar{X}}{G}\right)^{-1}= 
\begin{cases}
A_{G}^{*-1}+o_P(1), & \text{for weak or semi-strong dependence of \ensuremath{\gamma_{g\cdot j}}}\\
\left(A_{G}^{*}+C^{*}\right)^{-1}+o_P(1), & \text{for strong dependence of \ensuremath{\gamma_{g\cdot j}} }
\end{cases}
\] 
\end{small}
 where $A_{G}^{*-1}$
and $\left(A_{G}^{*}+C^{*}\right)^{-1}$ are also uniformly positive definite.
Therefore, by Lemma \ref{lemma:A_n=B_n+o_P(1) implies A_n = O_P(1)},
\[
\left(\frac{\bar{X}^{\prime}\bar{X}}{G}\right)^{-1}=O_{P}(1).
\]

Now, note that 
\[
\frac{\sum_{g=1}^{G}\bar{X}_{g}^{\prime}\bar{\epsilon}_{g}}{G}=\frac{\sum_{g=1}^{G}\bar{X}_{g}^{*\prime}\bar{\epsilon}_{g}}{G}+\frac{\sum_{g=1}^{G}\bar{\gamma}_{g}^{\prime}\bar{\epsilon}_{g}}{G}.
\]

Following the proof of Theorem \ref{thm:consistent}, it can be shown that
$\frac{\sum_{g=1}^{G}\bar{X}_{g}^{*\prime}\bar{\epsilon}_{g}}{G}=o_P(1),$ for any kind of dependence of $\epsilon_{g}.$ Since $\epsilon_{g}$ and $\Gamma_{g}$ are independently distributed,
$\mathbb{E}\left[\frac{\sum_{g=1}^{G}\bar{\gamma}_{g}^{\prime}\bar{\epsilon}_{g}}{G}\right]=0.$
Also, using the independence between clusters in the first line, Assumption \ref{assump:error independence for measurement error} in the second line, and using 
$\mathbb{E}\left[\left\Vert\bar{\gamma}_{g}^{\prime}\bar{\gamma}_{g}\right\Vert \big|  X_g^{*}\right]=O(1)$ uniformly in $g$ and $\mathbb{E}\left[ | \bar{\epsilon}_{g} | ^{2} \big|  X_g^{*}\right]=O(1)$ uniformly in $g,$ and Lemma \ref{lemma:Exact order summability} in the last line, we have
\begin{align*}
\mathbb{E}\left\Vert\frac{\sum_{g=1}^{G}\bar{\gamma}_{g}^{\prime}\bar{\epsilon}_{g}}{G}\right\Vert^{2} & = \frac{1}{G^{2}}\sum_{g=1}^{G}\mathbb{E}\left[\left\Vert\bar{\gamma}_{g}^{\prime}\bar{\epsilon}_{g}\right\Vert^{2}\right]\\
 & =\frac{1}{G^{2}}\sum_{g=1}^{G}\mathbb{E}\left[\left\Vert\bar{\gamma}_{g}^{\prime}\bar{\gamma}_{g}\right\Vert\right]\mathbb{E}\left[ | \bar{\epsilon}_{g} | ^{2}\right]\\
 & =O\left(\frac{1}{G}\right)\to 0,\text{ as }G\to\infty.
\end{align*}
 So, $\frac{\sum_{g=1}^{G}\bar{\gamma}_{g}^{\prime}\bar{\epsilon}_{g}}{G}=o_P(1),$ for any kind of dependence of $\epsilon_{g}$ and $\gamma_{g\cdot j}$. Therefore, for any kind of dependence of $\epsilon_{g}$ and $\gamma_{g\cdot j},$
\[
\frac{\sum_{g=1}^{G}\bar{X}_{g}^{\prime}\bar{\epsilon}_{g}}{G}=o_P(1).
\]
Now, for the term $\frac{\sum_{g=1}^{G}\bar{X}_{g}^{\prime}\bar{\gamma}_{g}\beta}{G},$ first note that 
\[
\frac{\sum_{g=1}^{G}\bar{X}_{g}^{\prime}\bar{\gamma}_{g}\beta}{G}=\frac{\sum_{g=1}^{G}\bar{X}_{g}^{*\prime}\bar{\gamma}_{g}\beta}{G}+\frac{\sum_{g=1}^{G}\bar{\gamma}_{g}^{\prime}\bar{\gamma}_{g}\beta}{G}.\]

We already have shown that $\frac{\sum_{g=1}^{G}\bar{X}_{g}^{*\prime}\bar{\gamma}_{g}}{G}=o_{P}(1).$ Therefore, $\frac{\sum_{g=1}^{G}\bar{X}_{g}^{*\prime}\bar{\gamma}_{g} \beta}{G} =o_{P}(1),$ since $\beta$ is finite-dimensional. Since $\beta$ is finite-dimensional, it follows from \eqref{eq: avg of gamma_g_bar'gamma_g_bar order} that  
\[
\frac{\sum_{g=1}^{G}\bar{\gamma}_{g}^{\prime}\bar{\gamma}_{g}\beta}{G}=\begin{cases}
o_P(1), & \text{for weak or semi-strong dependence of \ensuremath{\gamma_{g\cdot j}}}\\
C^{*}\beta + o_P(1), & \text{for strong dependence of \ensuremath{\gamma_{g\cdot j}} }
\end{cases}
\]

provided that the number of bounded clusters is $o(G)$, while $N_g \to \infty$ for all remaining clusters. Hence, for any kind of dependence of $\epsilon_g,$
\[
\frac{\sum_{g=1}^{G}\bar{X}_{g}^{\prime}\bar{\gamma}_{g}\beta}{G}=\begin{cases}
o_P(1), & \text{for weak or semi-strong dependence of \ensuremath{\gamma_{g\cdot j}}}\\
C^{*}\beta + o_P(1), & \text{for strong dependence of \ensuremath{\gamma_{g\cdot j}} }
\end{cases}
\]
Therefore, by combining these results and applying Lemma \ref{lemma:big-OP sum and multiplication} to equation \eqref{eq:beta hat - beta expression for meauserment error}, we conclude that
\begin{small}
\begin{align*}
& \quad \hat{\beta}_{A}-\beta\\
& =\begin{cases}
O_{P}(1)\cdot\left(o_P(1)-o_P(1)\right), & \text{for weak or semi-strong dependence of \ensuremath{\gamma_{g\cdot j}}}\\
O_{P}(1)\cdot\left(o_P(1)-\left(C^{*}\beta + o_P(1)\right)\right), & \text{for strong dependence of \ensuremath{\gamma_{g\cdot j}} }
\end{cases}\\
 & =\begin{cases}
o_P(1), & \text{for weak or semi-strong dependence of \ensuremath{\gamma_{g\cdot j}}}\\
O_{P}(-C^{*}\beta), & \text{for strong dependence of \ensuremath{\gamma_{g\cdot j}} }
\end{cases}
\end{align*}
\end{small}

Consequently, the estimator $\hat{\beta}_{A}$ is consistent under weak or semi-strong dependence in the measurement errors, but inconsistent under strong dependence. This property holds regardless of the underlying dependence structure of the structural error $\epsilon_{g}$.
 
\hfill
\qed

\setcitestyle{aysep={}}
\bibliography{references}

 \end{document}